\documentclass[aps,prx,reprint,preprintnumbers,superscriptaddress,nofootinbib,longbibliography,floatfix]{revtex4-2}

\usepackage[utf8]{inputenc}
\usepackage[T1]{fontenc}
\usepackage{lmodern}
\usepackage{microtype}
\usepackage{mathtools}

\usepackage{placeins}

\usepackage{amsmath}
\DeclareMathOperator*{\argmax}{argmax}
\DeclareMathOperator*{\argmin}{argmin}

\usepackage{amsmath}
\usepackage[caption=false]{subfig}
\usepackage{amsfonts}
\usepackage{graphicx}
\usepackage{hyperref}
\hypersetup{
  colorlinks=true,
  citecolor=blue,
  linkcolor=blue,
  urlcolor=blue
}

\newcommand{\thetaup}{\theta_A}
\newcommand{\thetadown}{\theta_B}
\newcommand{\thetanaught}{\theta_0}
\newcommand{\thetadata}{\theta_{\rm data}}
\newcommand{\thetaCL}{\theta_{\rm CL}}
\newcommand{\thetaML}{\theta_{\rm ML}}

\newcommand{\pup}{p_A}
\newcommand{\pdown}{p_B}

\usepackage{color}
\definecolor{darkred}{rgb}{1.0,0.1,0.1}
\definecolor{darkgreen}{rgb}{0.1,0.7,0.1}
\definecolor{darkblue}{rgb}{0.1,0.1,1.0}

\DeclareRobustCommand{\Sec}[1]{Sec.~\ref{sec:#1}}

\DeclareRobustCommand{\App}[1]{App.~\ref{app:#1}}

\DeclareRobustCommand{\Tab}[1]{Table~\ref{tab:#1}}

\DeclareRobustCommand{\Fig}[1]{Fig.~\ref{fig:#1}}
\DeclareRobustCommand{\Figs}[2]{Figs.~\ref{fig:#1} and \ref{fig:#2}}
\DeclareRobustCommand{\Eq}[1]{Eq.~(\ref{eq:#1})}
\DeclareRobustCommand{\Eqs}[2]{Eqs.~(\ref{eq:#1}) and (\ref{eq:#2})}
\DeclareRobustCommand{\Ref}[1]{Ref.~\cite{#1}}
\DeclareRobustCommand{\Refs}[1]{Refs.~\cite{#1}}

\begin{document}

\preprint{MIT-CTP 5271}

\title{E Pluribus Unum Ex Machina:\\ Learning from Many Collider Events at Once}

\author{Benjamin Nachman}
\email{bpnachman@lbl.gov}
\affiliation{Physics Division, Lawrence Berkeley National Laboratory, Berkeley, CA 94720, USA}
\affiliation{Berkeley Institute for Data Science, University of California, Berkeley, CA 94720, USA}

\author{Jesse Thaler}
\email{jthaler@mit.edu}
\affiliation{Center for Theoretical Physics, Massachusetts Institute of Technology, Cambridge, MA 02139, USA}
\affiliation{The NSF AI Institute for Artificial Intelligence and Fundamental Interactions}

\begin{abstract}
There have been a number of recent proposals to enhance the performance of machine learning strategies for collider physics by combining many distinct events into a single ensemble feature.
To evaluate the efficacy of these proposals, we study the connection between single-event classifiers and multi-event classifiers under the assumption that collider events are independent and identically distributed (IID).
We show how one can build optimal multi-event classifiers from single-event classifiers, and we also show how to construct multi-event classifiers such that they produce optimal single-event classifiers.
This is illustrated for a Gaussian example as well as for classification tasks relevant for searches and measurements at the Large Hadron Collider.
We extend our discussion to regression tasks by showing how they can be phrased in terms of parametrized classifiers.
Empirically, we find that training a single-event (per-instance) classifier is more effective than training a multi-event (per-ensemble) classifier, as least for the cases we studied, and we relate this fact to properties of the loss function gradient in the two cases.
While we did not identify a clear benefit from using multi-event classifiers in the collider context, we speculate on the potential value of these methods in cases involving only approximate independence, as relevant for jet substructure studies.
\end{abstract}

%\date{\today}
\maketitle

%%%%%%%%%%%%%%%%%%%%%%%%%%%%%%%%%%%%%%%%%%%%%%%%%%%%%%%

{\small
\tableofcontents
}

\section{Introduction}

Modern machine learning techniques are being widely applied to enhance or replace existing analysis techniques across collider physics~\cite{Larkoski:2017jix,Guest:2018yhq,Albertsson:2018maf,Radovic:2018dip,Bourilkov:2019yoi,hepmllivingreview}.
These approaches hold great promise for new particle searches, for Standard Model measurements, and for high-energy nuclear physics investigations.
A subset of these proposals have advocated for a multi-event strategy whereby a machine-learned function acts on multiple collision events at the same time~\cite{Lai:2018ixk,Khosa:2019kxd,Du:2019civ,Mullin:2019mmh,Chang:2020rtc,Flesher:2020kuy,Lazzarin:2020uvv,Lai:2020byl}.
This multi-event (per-ensemble) strategy contrasts with more typical single-event (per-instance) machine learning methods that process one event at a time, although both strategies make use of many events during the training process.

Intuitively, an ensemble approach might seem like a more promising learning strategy because there is more information contained in $N>1$ collision events than in one single event.
There is, however, an important distinction between the amount of information contained in a data set and the amount of information needed to encode a machine-learned function.
For this reason, there need not be a gain from using multi-event strategies over single-event strategies in the context of machine learning.

In this paper, we show that when directly compared on the same task, there is indeed no informational benefit from training a function that processes multiple events simultaneously compared to training a function that processes only a single event at a time.
This fact can be easily understood from the statistical structure of collision data.
To test for a practical benefit, we perform empirical comparisons of per-ensemble and per-instance methods on benchmark tasks relevant for the Large Hadron Collider (LHC), finding that single-event (per-instance) methods are more effective for the cases we studied.

To an excellent approximation, collider events are statistically independent and identically distributed (IID).
In simulation, this is exactly true up to deficiencies in random number generators.
In data, there are some small time-dependent effects from changing conditions and there are also some correlations between events introduced by detector effects with timescales longer than a typical bunch crossing.
These event-to-event correlations, however, are truly negligible when considering the set of events typically used for physics analysis that are selected by triggers.
The probability for two events next to each other in time to be saved by the triggers is effectively zero, since triggers save only a tiny fraction of events.
The IID nature of collision events therefore ensures that the information content is the same for ensembles of events and for single events drawn from an ensemble.

In equations, the probability to observe $N$ events $x_i$ is
\begin{equation}
\label{eq:iid_prob_relation}
p(\{x_1, \ldots, x_N\} | \theta) = \prod_{i = 1}^N p(x_i | \theta),
\end{equation}
where $\theta$ represents possible parameters of the generative model, such as the physics process being studied or the values of coupling constants.
The optimal classifier to distinguish whether events have been generated via $\thetaup$ or via $\thetadown$ depends only on the per-ensemble likelihood ratio~\cite{neyman1933ix}:
\begin{equation}
\label{eq:iid_lr_relation}
\frac{p(\{x_1, \ldots, x_N \}| \thetaup)}{p(\{x_1, \ldots, x_N \}| \thetadown)} = \prod_{i = 1}^N \frac{p(x_i | \thetaup)}{p(x_i | \thetadown)},
\end{equation}
which by the IID assumption only depends on knowing the per-instance likelihood ratio $p(x_i | \thetaup)/p(x_i | \thetadown)$.
This equality explains the informational equivalence of per-ensemble and per-event learning.

Given the simplicity of \Eq{iid_lr_relation}, why are we writing a whole paper on this topic (apart from the opportunity to invoke a gratuitously Latinate paper title that incorporates an aspiration for national unity)?
The studies in \Refs{Lai:2018ixk,Khosa:2019kxd,Du:2019civ,Mullin:2019mmh,Chang:2020rtc,Flesher:2020kuy,Lazzarin:2020uvv,Lai:2020byl} find that per-ensemble learning is effective for their respective tasks, in some cases arguing why per-instance learning is deficient.
It is certainly true that a set of events $\{x_1, \ldots, x_N\}$ contains more information than a single event $x_i$ drawn from this set.
What we will show in this paper is that if one carefully combines the per-instance information, one can recover the per-ensemble benefit, with the potential for a substantially reduced training cost.
We emphasize that our analysis does not contradict the studies in \Refs{Lai:2018ixk,Khosa:2019kxd,Du:2019civ,Mullin:2019mmh,Chang:2020rtc,Flesher:2020kuy,Lazzarin:2020uvv,Lai:2020byl}; rather this work suggests the possibility of achieving the same or better results by replacing per-ensemble learning with per-instance learning.
There may be specialized contexts where per-ensemble learning is superior, particularly if the training procedure itself can be made simpler, such as in the linear regression approach of \Ref{Flesher:2020kuy}.
This paper also gives us a chance to mention some facts about loss functions that are well known in the statistics literature but might not be as well appreciated in collider physics.
Moving away from the IID case, we speculate on the relevance of our analysis for jet substructure tasks where there is a notion of approximate independence of emissions.

The remainder of this paper is organized as follows.
In \Sec{statistics}, we provide the formal statistical basis for building multi-event classifiers from single-event classifiers, and vice versa, under the IID assumption.
We also explain how regression tasks can be translated into the language of per-instance parametrized classification.
In \Sec{empirical}, we present empirical studies that corroborate these analytic results. 
Our conclusions are given in \Sec{conclusions}.

\section{The Statistics of Per-Ensemble Learning}
\label{sec:statistics}

\subsection{Review of Per-Instance Learning}
\label{sec:per-instance}

Suppose that a collider event is represented by features in $\mathbb{E}=\mathbb{R}^M$ and we are trying to train a binary classifier to learn a target in $[0,1]$.
Let $c:\mathbb{E} \rightarrow [0,1]$ be a function that processes a single event, with the goal of distinguishing events being generated by $\thetaup$ ($c \to 1$) versus those generated by $\thetadown$ ($c \to 0$).
Such a function can be obtained by minimizing an appropriate loss functional, such as the binary cross entropy:
\begin{align}
\nonumber   L_{\rm BCE}[c]  =-\int dx\, \Big( & p(x|\thetaup) \log c(x) \\
&~ + p(x | \thetadown)  \log (1-c(x)) \Big),
\end{align}
where $p(x|\theta)$ is the probability density of $x\in\mathbb{E}$ given class $\theta$.
Here and throughout this discussion, we consider the infinite statistics limit such that we can replace sums over events by integrals.  
We have also dropped the prior factors $p(\theta_i)$, assuming that one has equal numbers of examples from the two hypotheses during training.
While this is often true in practice, it is not strictly necessary for our main conclusions, though it does simplify the notation.
It is well-known~\cite{hastie01statisticallearning,sugiyama_suzuki_kanamori_2012} (also in high-energy physics~\cite{2010.03569,1907.08209,Stoye:2018ovl,Hollingsworth:2020kjg,Brehmer:2018kdj,Brehmer:2018eca,Brehmer:2019xox,Brehmer:2018hga,Cranmer:2015bka,Badiali:2020wal,Andreassen:2020nkr,Andreassen:2019cjw,Fischer-ACAT2019}) that an optimally trained $c$ will have the following property:
\begin{align}
\label{eq:c_to_p}
    \frac{c(x)}{1-c(x)}=\frac{p(x|\thetaup)}{p(x|\thetadown)},
\end{align}
such that one learns the per-instance likelihood ratio.
By the Neyman–Pearson lemma~\cite{neyman1933ix}, this defines the optimal single-event classifier.

\begin{table*}
    \centering
    \def\arraystretch{1.8}
    \begin{tabular}{c @{$\quad$} c @{$\quad$} c @{$\quad$} c @{$\quad$} c @{$\quad$} c }
    \hline\hline
       Loss Name & $A(f)$  & $B(f)$ & $\argmin_f L[f]$ &  Integrand of $- \min_f L[f]$ & Related Divergence/Distance  \\
       \hline
        Binary Cross Entropy & $\log f$ & $\log (1-f)$ & $\frac{\pup}{\pup+\pdown}$ & $\pup \log \frac{\pup}{\pup+\pdown} + (A \leftrightarrow B)$ & $2 \big( \text{Jensen-Shannon} - \log 2 \big)$ \\
       Mean Squared Error & $-(1-f)^2$ & $-f^2$ & $\frac{\pup}{\pup+\pdown}$ & $- \frac{\pup \pdown}{\pup+\pdown}$ & $\frac{1}{2} \big(\text{Triangular} - 1 \big)$  \\ 
       Square Root & $\frac{-1}{\sqrt{f}}$ & $-\sqrt{f}$ &  $\frac{\pup}{\pdown}$ & $-2 \sqrt{\pup \pdown}$ & $2\big(\text{Hellinger}^2 - 1 \big)$ \\
       Maximum Likelihood Cl. & $\log f$ & $1-f$ & $\frac{\pup}{\pdown}$ & $\pup \log \frac{\pup}{\pdown}$ & Kullback–Leibler\\
    \hline\hline
    \end{tabular}
    \caption{Examples of loss functionals in the form of \Eq{AB_loss_func}, with the associated location and value of the loss minimum, using the shorthand $p_i \equiv p(x|\theta_i)$.
    We have used the symbol $f$ in all cases to denote the classifier, but some choices require explicit constraints on $f$ to be either non-negative or in the range $[0,1]$.
    In the last column, we indicate the relation of the loss minimum to statistical divergences and distances, up to an overall scaling and offset.  See \Ref{2005math.....10521N} for additional relations.}
    \label{tab:loss_to_f}
\end{table*}

There are many loss functionals that satisfy this property.
Consider a more general loss functional that depends on a learnable function $f:\mathbb{E} \rightarrow \mathbb{R}$ (which unlike $c$ may or may not map to $[0,1]$) as well as fixed rescaling functions $A: \mathbb{R} \to \mathbb{R}$ and $B: \mathbb{R} \to \mathbb{R}$:
\begin{align}
\label{eq:AB_loss_func}
 L[f]  =-\int dx\, &\Big( p(x | \thetaup) \, A(f(x)) + p(x|\thetadown) \ B(f(x)) \Big).
\end{align}
Taking the functional derivative with respect to $f(x)$, the extremum of $L[f]$ satisfies the property:
\begin{align}
\label{eq:AB_learned_func}
- \frac{B'(f(x))}{A'(f(x))} = \frac{p(x|\thetaup)}{p(x|\thetadown)}.
\end{align}
As long as $-B'(f)/A'(f)$ is a monotonic rescaling of $f$ and the overall loss functional is convex, then the function $f(x)$ learned by minimizing \Eq{AB_loss_func} defines an optimal classifier.
In many cases, the minimum value of $L[f]$ itself is interesting in the context of statistical divergences and distances~\cite{2005math.....10521N}, and a few examples are shown in \Tab{loss_to_f}.

To simplify the following discussion, we will focus on the ``maximum likelihood classifier'' (MLC) loss:
\begin{align}
\nonumber L_\text{MLC}[f] = - \int dx \, \Big(& p(x|\thetaup) \log f(x) \\
&~ + p(x|\thetadown) \, (1 -f(x)) \Big).
\label{eq:maxlikeloss}
\end{align}
This is of the general form in \Eq{AB_loss_func} with $A(f) = \log f$ and $B(f) = 1-f$.
To our knowledge, the MLC was first introduced in the collider physics context in \Refs{DAgnolo:2018cun,DAgnolo:2019vbw}, although with an exponential parametrization of $f(x)$.  
We call \Eq{maxlikeloss} the MLC loss to distinguish it from the related maximum likelihood loss that is often used to fit generative models~\cite{Andreassen:2018apy,brehmer2020flows,Nachman:2020lpy}.
Using \Eq{AB_learned_func}, the minimum of this loss functional yields directly the likelihood ratio:
\begin{equation}
\label{eq:maxlikeloss_argmin}
\argmin_f L_\text{MLC}[f] = \frac{p(x|\thetaup)}{p(x|\thetadown)},
\end{equation}
which will be useful to simplify later analyses.%  
\footnote{
A variation of \Eq{maxlikeloss_argmin} holds for $A(f)=\log C(f)$ and $B(f)=1-C(f)$, where $C(f)$ is any monotonically increasing function with range that covers $(0,\infty)$.  In this case, $C(\argmin_f L[f])=p(x|\thetaup)/p(x|\thetadown)$.
This can be useful in practice if $C(f)$ is everywhere positive, since $f$ can take on negative values and still yield a valid likelihood ratio.
See \Fig{MLC} for an empirical study of $C(f) = \exp f$.
\label{footnote:modMLC}
}
The MLC loss functional value at the minimum is
\begin{equation}
\label{eq:KL_divergence}
- \min_f L_\text{MLC}[f] = \int dx \, p(x|\thetaup) \log \frac{p(x|\thetaup)}{p(x|\thetadown)},
\end{equation}
which is the Kullback–Leibler (KL) divergence, also known as the relative entropy from $p(x|\thetadown)$ to $p(x|\thetaup)$.
See \App{maxlikeloss} for an intuitive derivation of \Eq{maxlikeloss}.

\subsection{Per-Ensemble Binary Classification}
\label{sec:per_ensemble_classification}

To move from single-event classification to multi-event classification, we want to learn a classification function $f_N$ that can process $N$ events simultaneously.
Here, we are using $f_N:\mathbb{E}^N\rightarrow \mathbb{R}$ instead of $c_N:\mathbb{E}^N\rightarrow [0,1]$ to avoid algebraic manipulations like \Eq{c_to_p}.
We will use the vector notation
\begin{equation}
    \vec{x} = \{x_1, \dots, x_N\}
\end{equation}
to represent an element of $\mathbb{E}^N$.
Our goal is to distinguish whether $\vec{x}$ is drawn from $p(\vec{x}|\thetaup)$ ($f_N \to \infty$) or from $p(\vec{x}|\thetadown)$ ($f_N \to 0$).
Note that we are trying to classify a pure event ensemble as coming from either $\thetaup$ or $\thetadown$, which is a different question than trying to determine the proportion of events drawn from each class in a mixed event ensemble.
For $N = 1$, $f_1$ is the same as $f$ discussed in \Eq{AB_loss_func}.

If $f_N$ is trained optimally, then the classification performance of $f_N$ evaluated on $N > 1$ events will be better than the performance of $f_1$ evaluated on a single event, as relevant to the discussions in  \Refs{Lai:2018ixk,Khosa:2019kxd,Du:2019civ,Mullin:2019mmh,Chang:2020rtc,Flesher:2020kuy,Lazzarin:2020uvv,Lai:2020byl}.
The key point of this paper is that one can construct a classifier $f_{1 \to N}$ that is built only from $f_1$, acts on $N$ events, and has the same asymptotic performance as $f_N$.

Using the MLC loss in \Eq{maxlikeloss}, but now applied to $N$ events, we have
\begin{align}
\nonumber L_\text{MLC}[f_N] = - \int d^N x \, \Big( & p(\vec{x}|\thetaup) \, \log f_N(\vec{x}) \\
& ~ + p(\vec{x}|\thetadown) \, (1 -f_N(\vec{x})) \Big),\label{eq:maxlikeloss_N}
\end{align}
whose minimum is the per-ensemble likelihood ratio:
\begin{equation}
\argmin_{f_N} L_\text{MLC}[f_N]= \frac{p(\vec{x}|\thetaup)}{p(\vec{x}|\thetadown)}.   
\end{equation}
By the Neyman–Pearson lemma, this yields the optimal per-ensemble classifier.

On the other hand, once we have trained a single-event classifier $f_1$ using \Eq{maxlikeloss}, we can build a multi-event classifier $f_{1\to N}$ without any additional training:
\begin{equation}
\label{eq:f_1_to_N}
f_{1 \to N}(\vec{x}) \equiv \prod_{i = 1}^N f_1(x_i) \quad \rightarrow \quad \frac{p(\vec{x}|\thetaup)}{p(\vec{x}|\thetadown)},
\end{equation}
where in the last step we have combined the solution found in \Eq{maxlikeloss_argmin} with the IID condition in \Eq{iid_lr_relation}.
Whereas minimizing \Eq{maxlikeloss_N} requires sampling over $\mathbb{E}^N$, constructing $f_{1 \to N}$ only requires sampling over $\mathbb{E}$, which is a considerable reduction in computational burden for large $N$.
The technical details of carrying out this procedure are explained in \Sec{manyfromone}.

Going in the converse direction, we can learn a single-event classifier $f_{N \to 1}$ starting from a constrained multi-event classifier $\tilde{f}_N$.
Using weight sharing, we can minimize \Eq{maxlikeloss_N} subject to the constraint that $\tilde{f}_N$ takes the functional form:
\begin{equation}
\label{eq:tildef_N}
\tilde{f}_N(\{x_1, \dots, x_N\}) = \prod_{i = 1}^N f_{N \to 1}(x_i),
\end{equation}
where $f_{N \to 1}(x)$ is a learnable function.
Under the IID assumption, $\tilde{f}_N$ can still learn the per-ensemble likelihood ratio, but the learned $f_{N \to 1}(x)$ will now be the per-instance likelihood ratio, at least asymptotically.%
\footnote{In the case that the two samples are composed of mixtures of two categories, then the learned $f_{N \to 1}(x)$ will be the ratio of the mixed sample likelihoods, which is monotonically related to the optimal pure sample classifier, as discussed in \Ref{Metodiev:2017vrx}.}
An examination of this converse construction is presented in \Sec{onefrommany}.

\subsection{Comparing the Loss Gradients}
\label{sec:gradients}

We have shown that the per-ensemble classifier $f_N$ and the composite per-event classifier $f_{1 \to N}$ have the same asymptotic information content, but one might wonder if there is nevertheless a practical performance gain to be had using per-ensemble learning.

Under the IID assumption, the optimal $f_N$ takes the form of $\tilde{f}_N$ in \Eq{tildef_N}, and in our empirical studies, we found no benefit to letting $f_N$ have more functional freedom.
Therefore, to get a sense of the efficacy of per-ensemble versus per-instance training, we can compare the effective loss functions for $f_{N \to 1}$ and $f_1$.
Since the inputs and outputs of these functions are the same (i.e.\ $\mathbb{E} \to \mathbb{R}$), we can do an apples-to-apples comparison of their behavior under gradient descent.
The following analysis assumes that the neural network training occurs in the vicinity of the global minimum of the loss function.

For the per-ensemble case, plugging \Eq{tildef_N} into \Eq{maxlikeloss_N} and using the IID relation in \Eq{iid_prob_relation}, we find the effective loss functional:
\begin{align}
\nonumber L_{\rm MLC}[f_{N \to 1}] + 1 &= - N \int dx \, p(x|\thetaup) \, \log f_{N \to 1}(x) \\
& \quad +  \left( \int dx \, p(x|\thetadown) f_{N \to 1}(x) \right)^N.
\end{align}
This is to be contrasted with the per-instance loss functional from \Eq{maxlikeloss}, repeated for convenience with the $f_1$ notation and typeset to be parallel to the above:
\begin{align}
\nonumber
L_\text{MLC}[f_1] + 1 &= - \int dx \, p(x|\thetaup) \, \log f_1(x) \\
& \quad + \int dx \, p(x|\thetadown) \, f_1(x).
\label{eq:maxlikeloss_alt}
\end{align}
To understand the loss gradients, we can Taylor expand the learned functions about the optimal solution:
\begin{align}
f_{N \to 1}(x) & = \frac{p(x|\thetaup)}{p(x|\thetadown)} + \epsilon(x),\\
f_{1}(x) & = \frac{p(x|\thetaup)}{p(x|\thetadown)} + \epsilon(x).
\end{align}
Plugging these into their respective loss functionals and looking at the leading-order variations, we have:
\begin{align}
\nonumber \frac{\delta L_{\rm MLC}[f_{N \to 1}]}{N} &=  \int dx \, \frac{\big( p(x|\thetadown) \, \epsilon(x) \big)^2}{2 \, p(x|\thetaup)} \\
& \quad + \frac{N-1}{2} \left( \int dx \, p(x|\thetadown) \, \epsilon(x)  \right)^2,\\
\delta L_{\rm MLC}[f_{1}] &= \int dx \, \frac{\big( p(x|\thetadown) \, \epsilon(x) \big)^2}{2 \, p(x|\thetaup)}.
\end{align}
These expressions are quadratic in $\epsilon(x)$, which means that we are expanding around the correct minimum.

The expression for $\delta L_{\rm MLC}[f_{1}]$ involves a single integral over $x$, so under gradient descent, the value of $\epsilon(x)$ can be independently adjusted at each point in phase space to find the minimum.
By contrast, $\delta L_{\rm MLC}[f_{N \to 1}]$ has an additional piece involving an integral squared, so even if at a given point in phase space $x_0$ we have achieved $\epsilon(x_0) = 0$, gradient descent will tend to push $\epsilon(x_0)$ away from the correct value until $\epsilon(x) = 0$ everywhere.
This correlated structure explains the slower convergence of $L_{\rm MLC}[f_{N \to 1}]$ compared to $L_{\rm MLC}[f_{1}]$ in our empirical studies.
While we focused on the MLC loss to simplify the algebra, the appearance of these (typically counterproductive) correlations in the loss gradient appears to be a generic feature of per-ensemble learning.

\subsection{Per-Ensemble Regression}
\label{sec:regression}

While the discussion above focused on binary classification, the same basic idea applies to regression problems as well.
The goal of regression is to infer parameters $\theta$ from the data $\vec{x}$.
There are a variety of approaches that can be used for this task, and each can be connected to parametrized per-instance classification.

\subsubsection{Maximum Likelihood}
Maximum likelihood is the most common strategy for inference in collider physics.
Symbolically, we are trying to find
\begin{equation}
    \thetaML=\argmax_\theta p(\vec{x}|\theta).
\end{equation}
One way to determine $\thetaML$ is with a two-step approach.
First, one can train a parametrized classifier $f(x,\theta)$~\cite{Cranmer:2015bka,Baldi:2016fzo} using, e.g., the per-instance MLC loss:
\begin{align}\nonumber
    L_{\rm MLC}[f]=-\int dx\, \Big(& p(x|\theta)\, p(\theta) \log f(x,\theta)\\
    &~ + p(x|\thetanaught) \, p(\theta) \, (1- f(x,\theta)) \Big).
    \label{eq:paramclass}
\end{align}    
The top line corresponds to a synthetic dataset where every event is generated from $p(x|\theta)$ with different $\theta$ values drawn from the probability density $p(\theta)$.
The bottom line corresponds to a synthetic dataset where every event is generated using the same $p(x|\thetanaught)$ for fixed $\thetanaught$ and then augmented with a value $\theta$ that follows from $p(\theta)$ independently of $x$.
Minimizing \Eq{paramclass} with respect to $f(x,\theta)$, the asymptotic solution is the likelihood ratio:
\begin{equation}
\label{eq:paramed}
f(x,\theta) = \frac{p(x|\theta)}{p(x|\thetanaught)},
\end{equation}
where the factors of $p(\theta)$ have canceled out.
Second, one can estimate $\thetaML$ by using the IID properties of the event ensemble to relate likelihoods to the classifier output $f(x,\theta)$:
    \begin{align}
        \thetaML&=\argmin_\theta\left\{- \sum_{i=1}^N\log p(x_i|\theta)\right\} \nonumber\\
        &= \argmin_\theta\left\{- \sum_{i=1}^N \log  \frac{p(x_i|\theta)}{p(x_i|\thetanaught)} \right\} \nonumber\\
        &\approx \argmin_{\theta}\left\{-\sum_{i=1}^N \log f(x_i,\theta)\right\}. \label{eq:theta_star}
    \end{align}
Thus, even though maximum likelihood regression uses information from the full event ensemble, only a parametrized per-instance classifier is required for this procedure.

\subsubsection{Classifier Loss}
Two recent proposals for parameter estimation are explicitly built on classifiers for regression~\cite{2010.03569,1907.08209}.
For any classifier, one can perform the following optimization:%
\footnote{Note that \Ref{2010.03569} used the (non-differentiable) area under the curve instead of the classifier loss, as it is not sensitive to differences in the prior $p(\theta)$ between the two data sets.}
\begin{align}
    \thetaCL =\argmax_{\theta'}\left\{\begin{matrix}\text{Loss of a classifier trained}\cr\text{to distinguish $\theta'$ from $\thetadata$}\end{matrix}\right\}.
    \label{eq:thetaCL}
\end{align}
Here, we are imagining that the $\theta'$ samples come from synthetic data sets.
The appearance of a maximum instead of minimum in \Eq{thetaCL} is because, as highlighted in \Tab{loss_to_f}, it is negative loss functions that correspond to statistical divergences and distances.

In general, the $\thetaCL$ that minimizes the classifier loss will be different from the $\thetaML$ that maximizes the likelihood.
For the special case of the MLC loss, though, they are the same in the asymptotic limit if we set $\thetaup = \thetadata$ and $\thetadown = \theta'$.
To see this, recall from \Eq{KL_divergence} that after training, the value of the MLC loss is related to the KL divergence: 
\begin{align}\nonumber
&\argmax_{\theta'}\{ \min_f L_\text{MLC}[f] \}\\\nonumber
&\hspace{2mm}=\argmax_{\theta'}\left\{ -\int dx \, p(x|\thetadata) \log \frac{p(x|\thetadata)}{p(x|\theta')}\right\}  \\\nonumber
&\hspace{2mm}\approx\argmax_{\theta'}\left\{ \sum_{i=1}^N \log \frac{p(x_i|\theta')}{p(x_i | \thetadata)} \right\}\\\nonumber
&\hspace{2mm}=\argmin_{\theta'}\left\{ -\sum_{i=1}^N \log p(x_i|\theta') \right\}\\
&\hspace{2mm}=\thetaML\,,
\label{eq:KL_divergence_alt}
\end{align}
where the sum is over data events.

\subsubsection{Direct Regression}
In terms of information content, a regression model trained in the usual way can be built from a parametrized classification model.
Suppose that $\theta\in\mathbb{R}^Q$ and $g_N:\mathbb{E}^N\rightarrow \mathbb{R}^Q$ is a regression model trained with the mean squared error loss:
\begin{equation}
\label{eq:directregressionMSE}
L_{\rm MSE}[g_N]  =-\int d^n x\, p(\vec{x},\theta) \Big(g_N(\vec{x}) - \theta \Big)^2
\end{equation}
It is well known that the optimally trained $g_N$ will be related to the expectation value of $\theta$:
\begin{equation}
g_N(\vec{x})=\mathbb{E}[\theta|\vec{x}] =\int d\theta\,\theta\, p(\theta|\vec{x}).
\label{eq:directregression}
\end{equation}
Other loss functions approximate other statistics, as discussed in \Ref{1910.03773}.
For example, the mean absolute error loss approximates the median of $\theta$.
Ultimately, all direct regression methods are functionals of $p(\theta|\vec{x})$.

We can relate $p(\theta|\vec{x})$ to a parametrized classifier $f_N(\vec{x},\theta)$ trained to distinguish $\theta$ from a baseline $\thetanaught$:
\begin{align}
\nonumber    p(\theta|\vec{x}) = \frac{p(\vec{x} | \theta) \, p(\theta)}{p(\vec{x})} &= \frac{p(\vec{x} | \theta) \, p(\theta)}{\int d \theta' \, p(\vec{x} | \theta') \, p(\theta')}\\
\nonumber     &  = \frac{\frac{p(\vec{x} | \theta)}{p(\vec{x} | \thetanaught)} \, p(\theta)}{\int d \theta' \, \frac{p(\vec{x} | \theta')}{p(\vec{x} | \thetanaught)} \, p(\theta')}\\
    & = \frac{f_N(\vec{x} , \theta) \, p(\theta)}{\int d \theta' \, f_N(\vec{x} , \theta') \, p(\theta')}, \label{eq:conditionalprobabilitytoclassifier}
\end{align}
where $p(\theta)$ is the probability density of $\theta$ used during the training of $g_N$.
Following the same logic as \Sec{per_ensemble_classification},  the per-ensemble classifier $f_N(\vec{x} , \theta)$ can be related to a per-instance classifier $f_1(x,\theta)$.
 Therefore, even though $g_N$ acts on $N$ events, it has the same information content as a parametrized classifier that acts on single events.

Performing regression via \Eqs{directregression}{conditionalprobabilitytoclassifier} is straightforward but tedious.
In practice, one would train a parametrized per-instance classifier $f_1(x,\theta)$ as in \Eq{paramed}, multiply it to construct $f_N(\vec{x},\theta)=\prod_{i=1}^N f_1(x_i,\theta)$, and then sample over values of $\theta$ to approximate the integrals.
We show examples of the above regression strategies in \Sec{regression_study}

\subsection{Beyond Regression}
\label{sec:beyond}

In addition to classification and regression, a standard machine learning task is density estimation.
While some classical machine learning methods like $k$-nearest neighbors~\cite{knn1,knn2} do require multi-instance information at prediction time, many of the standard deep learning solutions to implicit or explicit generative modeling are built on per-instance functions.
Such methods include generative adversarial networks~\cite{NIPS2014_5ca3e9b1},%
\footnote{
In the context of adversarial training, it may be beneficial to use per-ensemble information in the discriminator to mitigate mode collapse, as utilized in \Ref{Lai:2020byl}.
This is also the philosophy behind mini-batch discrimination~\cite{2016arXiv160603498S}.
}
variational autoencoders~\cite{kingma2014autoencoding}, and normalizing flows~\cite{pmlr-v37-rezende15}.

One reason for computing explicit densities is to estimate the distance to a reference density.
A common set of tools for this task are the $f$-divergences mentioned earlier.
As discussed in \Ref{2005math.....10521N} and highlighted in \Tab{loss_to_f}, there is a direct mapping between the loss value of a per-instance classification task and a corresponding $f$-divergence between the underlying probability densities.

A related quantity is the mutual information between two random variables $X$ and $Y$: 
\begin{align}
\label{eq:MI}
I(X,Y)=\int dx\, dy\, p(x,y)\log \frac{p(x,y)}{p(x) \, p(y)}.
\end{align}
For example, $Y$ could be binary (a class label) and then $I(X,Y)$ would encode how much information (in units of nats) is available in $X$ for doing classification.
This can be helpful in the context of ranking input features, and was studied in the context of quark/gluon jet classification in \Ref{Larkoski:2014pca}.

Naively, \Eq{MI} might seem like it requires estimating the densities $p(x)$, $p(y)$, and $p(x,y)$, which in turn may require ensemble information (see e.g.~\Ref{carrara2019estimation} for a study in the context of HEP).
On the other hand, \Eq{MI} takes the same form as the KL divergence in \Eq{KL_divergence}.
Therefore, this quantity can be estimated using a similar strategy as in earlier sections, by training a classifier to distinguish data following $p(x,y)$ from data following $p(x)\, p(y)$ using the MLC loss.
The value of the loss at the minimum will be an estimate of the mutual information.
A simple example of this will be studied in \Sec{beyondregression}.

\section{Empirical Studies}
\label{sec:empirical}

We now present empirical studies comparing per-instance and per-ensemble data analysis strategies to highlight the points made in \Sec{statistics}.
Our analyses are based on three case studies: a simple two Gaussian example, searching for dijet resonances, and measuring the top quark mass.

\subsection{Classifiers: Multi-Event from Single-Event}
\label{sec:manyfromone}

As argued in \Sec{per_ensemble_classification}, under the IID assumption we can build multi-event classifiers from single-event classifiers.
We now demonstrate how to construct $f_{1\rightarrow N}$ defined in \Eq{f_1_to_N}, comparing its performance to $f_N$.

\subsubsection{Two Gaussian Example}
\label{sec:manyfromone_gaussian}

 \begin{figure*}[t]
 \centering
\subfloat[]{\includegraphics[width=0.91\columnwidth]{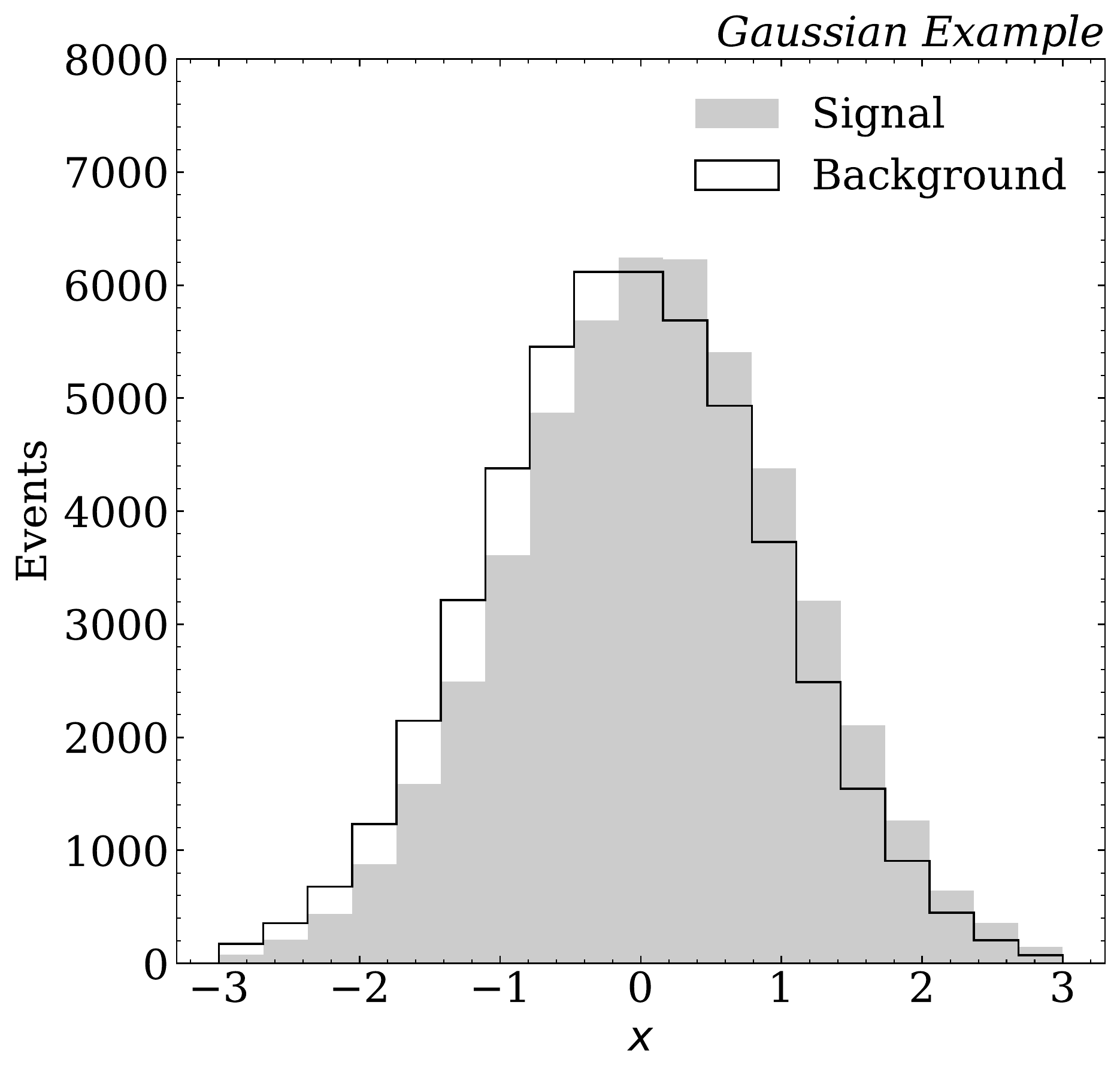}  \label{fig:Gaussian-a}} $\qquad$
 \subfloat[]{\includegraphics[width=0.91\columnwidth]{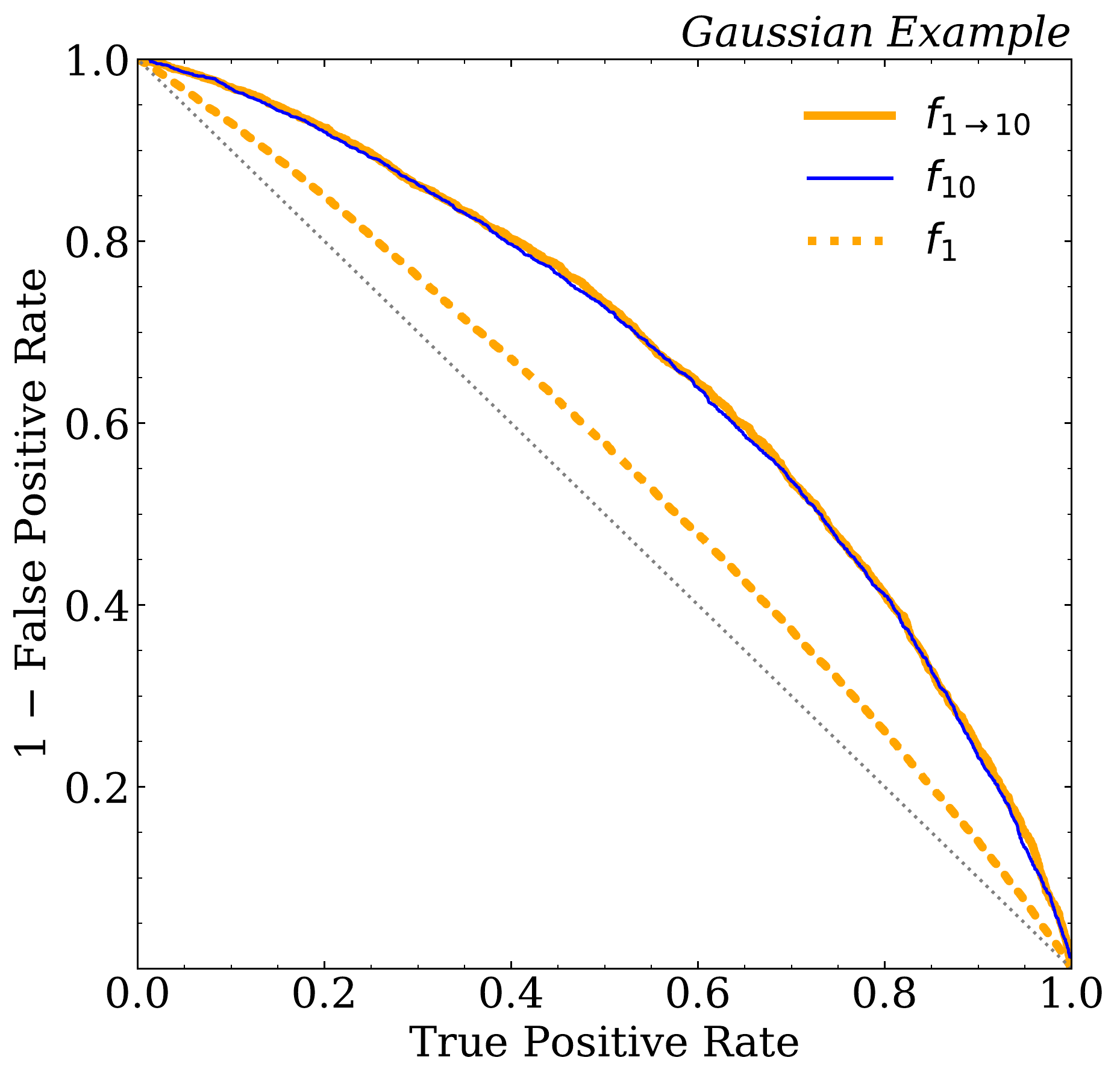}  \label{fig:Gaussian-b}}
 \caption{Classification in the two Gaussian example.
  (a) A histogram of the Gaussian random variable $X$, for the ``signal'' ($x_0 = 0.1$) and background ($x_0 = -0.1$).
  (b) ROC curves for various binary classifiers.  From the single-event classifier $f_1$, we can construct a multi-event classifier $f_{1 \to 10}$ that matches the performance of a classifier trained on 10 events simultaneously ($f_{10}$).
  }
 \label{fig:Gaussian}
 \end{figure*}

Our first case study involves one-dimensional Gaussian random variables.
As shown in \Fig{Gaussian-a}, we consider two Gaussian distributions $X\sim\mathcal{N}(\pm\epsilon,1)$, with slightly different means ($x_0 = \pm \epsilon$) but the same variance ($\sigma = 1$).
Here, the ``signal'' has positive mean while the ``background'' has negative mean, and we take $\epsilon=0.1$ for concreteness.

Both the per-instance ($f_1$) and per-ensemble ($f_N$) classifiers are parametrized by neural networks and implemented using \textsc{Keras}~\cite{keras} with the \textsc{Tensorflow} backend~\cite{tensorflow} and optimized with \textsc{Adam}~\cite{adam}.
We use the binary cross entropy loss function so \Eq{c_to_p} is needed to convert the classifier output to a likelihood ratio.
Each classifier consists of two hidden layers with 128 nodes per layer.
Rectified Linear Unit (ReLU) activation functions are used for the intermediate layers while sigmoid activation is used for the last layer.
The only difference between the per-instance and per-ensemble networks is that the input layer has one input for $f_1$ but $N$ inputs for $f_N$.

We train each network with 50,000 events to minimize the binary cross entropy loss function, and we test the performance with an additional 50,000 events.
For each network, we train for up to 1000 epochs with a batch size of 10\%, which means that the number of batches per epoch is the same, as is the number of events considered per batch.  
The training is stopped if the validation loss does not decrease for 20 consecutive epochs (early stopping).
For the ensemble network, we take $N = 10$.
We did not do any detailed hyperparameter optimization for these studies.

In \Fig{Gaussian-b}, we show the performance of the resulting classifiers $f_1$ and $f_{10}$.
We checked that the $f_1$ classifier parametrized by a neural network has essentially the same performance as an analytic function derived by taking the ratio of Gaussian probability densities, which means that the neural network $f_1$ is nearly optimal.
As expected, the per-instance classifier $f_1$ has a worse receiver operating characteristic (ROC) curve than the per-ensemble classifier $f_{10}$.
This is not a relevant comparison, however, because the two are solving different classification tasks (i.e.~classifying individual events as coming from signal or background versus classifying an ensemble of $N = 10$ events as all coming from signal or background).
With \Eq{f_1_to_N}, we can use $f_1$ to build a $10$-instance classifier $f_{1\rightarrow 10}$, whose ROC curve is nearly identical to $f_{10}$, if not even slightly better.
Thus, as expected from \Eq{iid_lr_relation}, all of the information in the 10-instance classifier is contained in the per-instance classifier.

\subsubsection{Dijet Resonance Search}
\label{sec:manyfromone_bsm}

 \begin{figure*}[t]
 \centering
\begin{minipage}[t]{0.95\columnwidth}
 \subfloat[]{\includegraphics[width=0.91\columnwidth]{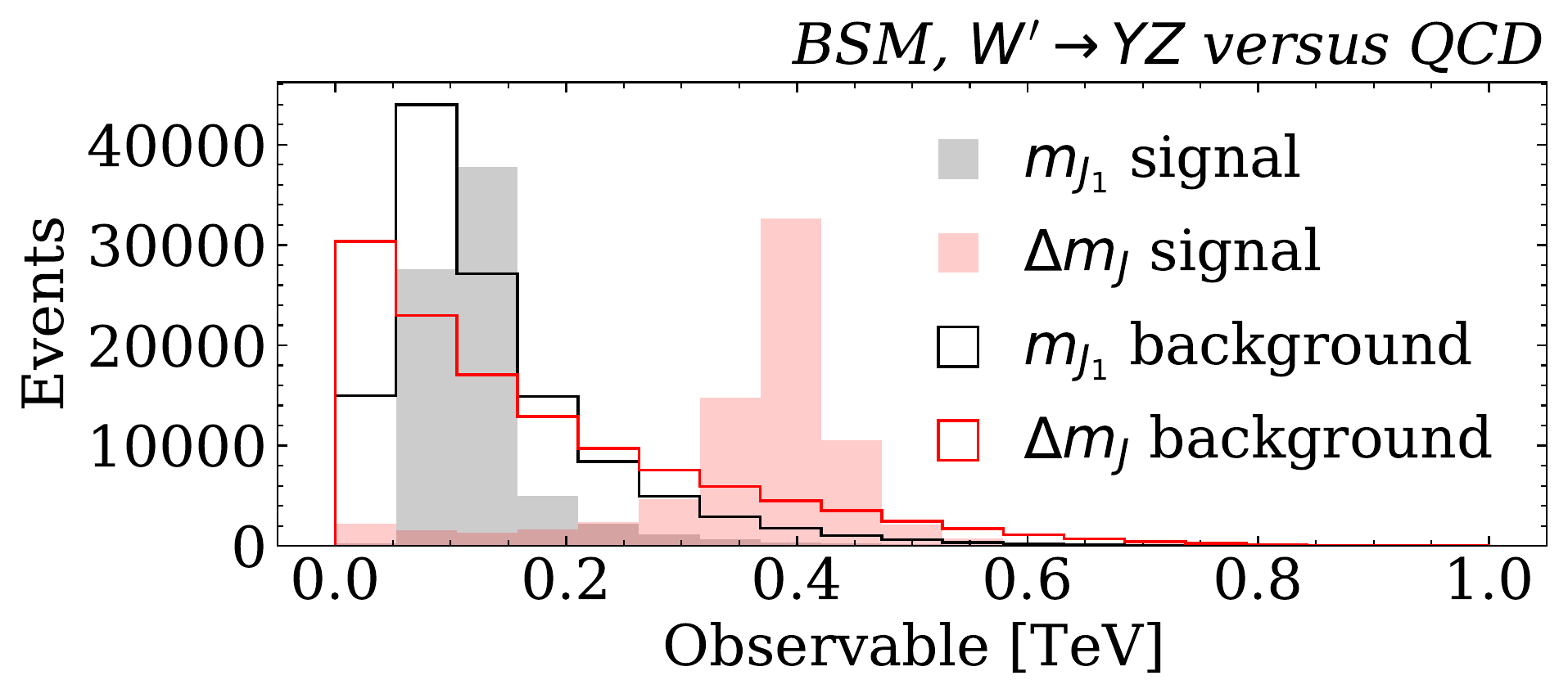}  \label{fig:BSM-a}}\\
  \subfloat[]{\includegraphics[width=0.91\columnwidth]{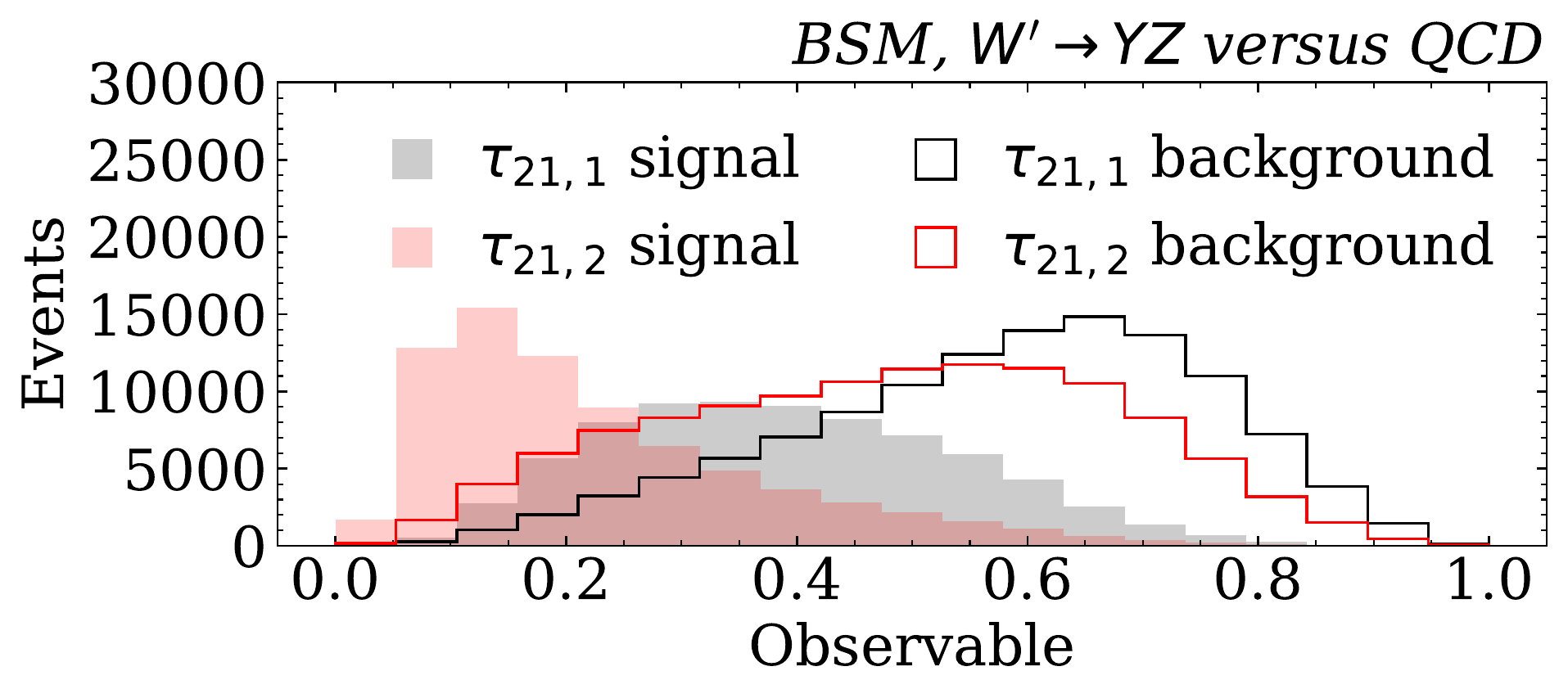}  \label{fig:BSM-b}}
  \end{minipage}
  $\qquad$
  \begin{minipage}[t]{0.95\columnwidth}
 \subfloat[]{\includegraphics[width=0.91\columnwidth]{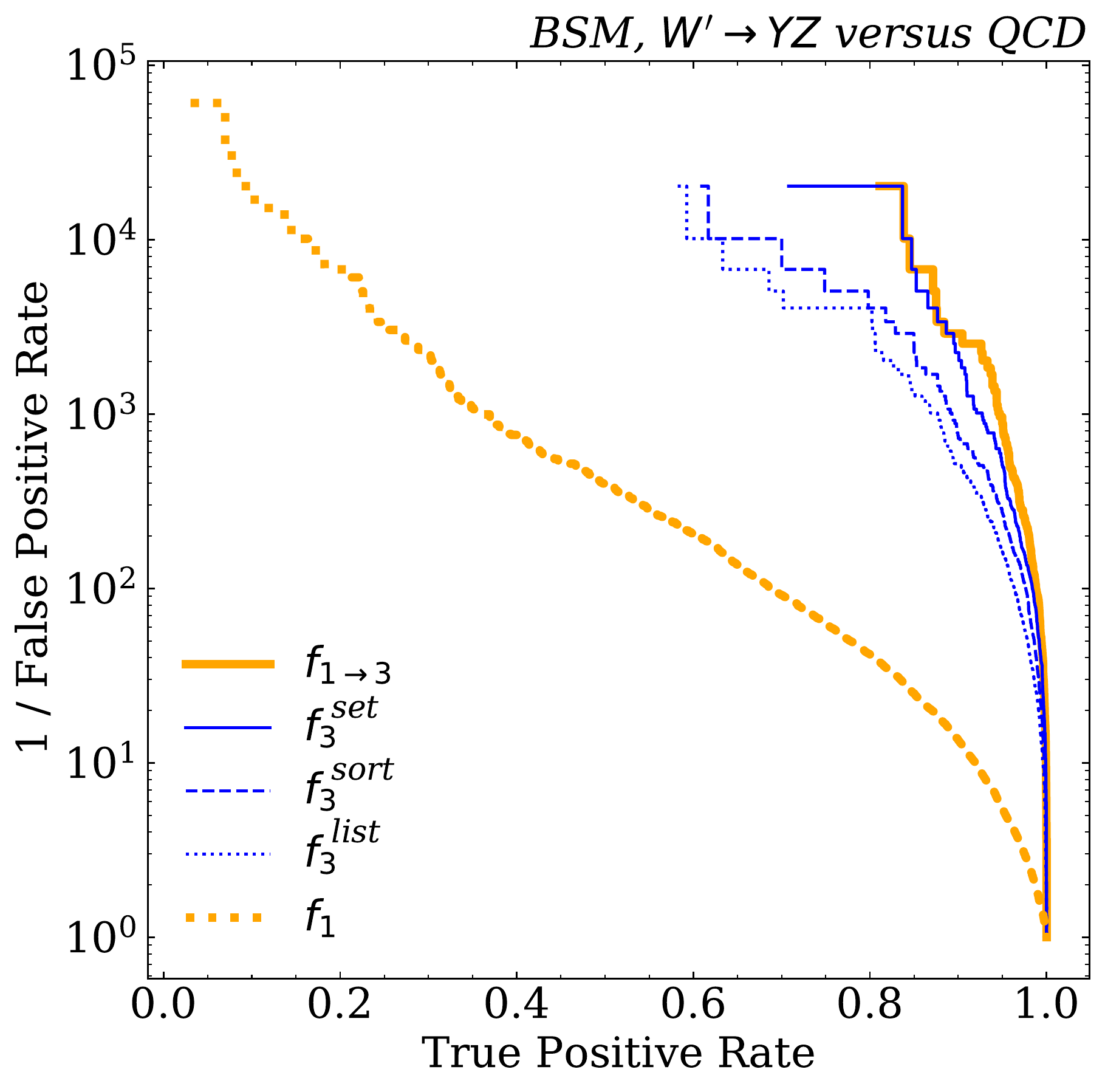}  \label{fig:BSM-c}}
 \end{minipage}
 \caption{Classification in the dijet resonance search example.  (a,b)  Histograms of the four jet features for the signal ($W' \to X Y$) and background (QCD dijet) processes.  
 (c) ROC curves for various binary classifiers.  The multi-event classifier $f_{1 \to 3}$ (built from $f_1$) outperforms three classifiers trained on triplets of events:  $f_3^\text{list}$ with randomly ordered inputs, $f_3^\text{sort}$ with sorted inputs, and $f_3^\text{set}$ based on the deep sets/PFN strategy in \Eq{pfn} with built-in permutation invariance.}
 \label{fig:BSM}
 \end{figure*}

We now consider an example from collider physics, motivated by a search for new beyond-the-Standard-Model (BSM) particles in a dijet final state.
The simulations used for this study were produced for the LHC Olympics 2020 community challenge~\cite{gregor_kasieczka_2019_2629073}.
The background process involves generic quantum chromodynamics (QCD) dijet events with a requirement of at least one such jet with transverse momentum $p_T>1.3$ TeV.
The signal process involves the production of a hypothetical new resonance $W'$ with mass $m_{W'}=3.5$ TeV, which decays via $W'\rightarrow XY$ to two hypothetical particles $X$ and $Y$ of masses 500 GeV and 100 GeV, respectively.
Each of the $X$ and $Y$ particles decays promptly into pairs of quarks.
Due to the mass hierarchy between the $W'$ boson and its decay products, the final state is characterized by two large-radius jets with two-prong substructure.
The background and signal are generated using \textsc{Pythia}~8.219~\cite{Sjostrand:2006za,Sjostrand:2007gs}.
A detector simulation is performed with \textsc{Delphes}~3.4.1~\cite{deFavereau:2013fsa,Mertens:2015kba,Selvaggi:2014mya} using the default CMS detector card.
Particle flow objects are used as inputs to jet clustering, implemented with \textsc{FastJet}~3.2.1~\cite{Cacciari:2011ma,Cacciari:2005hq} and the anti-$k_t$ algorithm~\cite{Cacciari:2008gp} using $R=1.0$ for the radius parameter.
Events are required to have a reconstructed dijet mass within the range $m_{JJ} < [3.3,3.7]\,\text{GeV}$.

Four features are used to train our classifiers: the invariant mass of the lighter jet, the mass difference of the leading two jets, and the $N$-subjettiess ratios $\tau_{21}$~\cite{Thaler:2011gf,Thaler:2010tr} of the leading two jets.
The observable $\tau_{21}$ quantifies the degree to which a jet is characterized by two subjets or one subjet, with smaller values indicating two-prong substructure.
The mass features are recorded in units of TeV so that they are numerically $\mathcal{O}(1)$.
Histograms of the four features for signal and background are shown in \Figs{BSM-a}{BSM-b}.
The signal jet masses are localized at the $X$ and $Y$ masses and the $\tau_{21}$ observables are shifted towards lower values, indicating that the jets have two-prong substructure.

We train a per-instance classifier ($f_1$) and a per-ensemble classifier ($f_3$) using the same tools as for the Gaussian example above, again using binary cross entropy for the loss function.
Because signal and background are so well separated in this example, we restrict our attention to $N = 3$ to avoid saturating the performance.
Note that this is an artificially constructed classification problem, since in a more realistic context one would be trying to estimate the signal fraction in an event ensemble, not classify triplets of events as all coming from signal or background.

For $f_1$, the neural network architecture is the same as \Ref{2010.03569} with four hidden layers, each with 64 nodes and ReLU activation, and an output layer with sigmoid activation. 
For $f_3$, the neural network involves $4 \times 3 = 12$ inputs, and the penultimate hidden layer is adjusted to have 128 nodes, yielding a marginal performance gain. 
In both cases, about 100,000 events are used for testing and training, with roughly balanced classes.
All of the networks are trained for up to 1000 epochs with the same early stopping condition as in the Gaussian case and with a batch size of 10\%.
Following \Eq{f_1_to_N}, we construct a tri-event classifier $f_{1\rightarrow 3}$ from $f_1$.

The ROC curves for $f_3$ and $f_{1 \to 3}$ are shown in \Fig{BSM-c}, with $f_1$ also shown for completeness.
Interestingly, the $f_{1\rightarrow 3}$ classifier trained on single events significantly outperforms $f_3$ trained on multiple events.
There are a variety of reasons for this, but one important deficiency of the $f_3$ classifier is that it does not respect the permutation symmetry of its inputs.
Because events are IID distributed, there is no natural ordering of the events, but the fully connected architecture we are using imposes an artificial ordering.
Inspired by \Ref{Flesher:2020kuy}, we can break the permutation symmetry of the inputs by imposing a particular order on the events.
Specifically, we train a network $f_3^\text{sort}$ where the triplet of events is sorted by their leading jet mass.
Using $f_3^\text{sort}$ yields a small gain in performance seen in \Fig{BSM}, but not enough to close the gap with $f_{1\rightarrow 3}$. 

 \begin{figure*}[t]
 \centering
 \subfloat[]{\includegraphics[width=0.91\columnwidth]{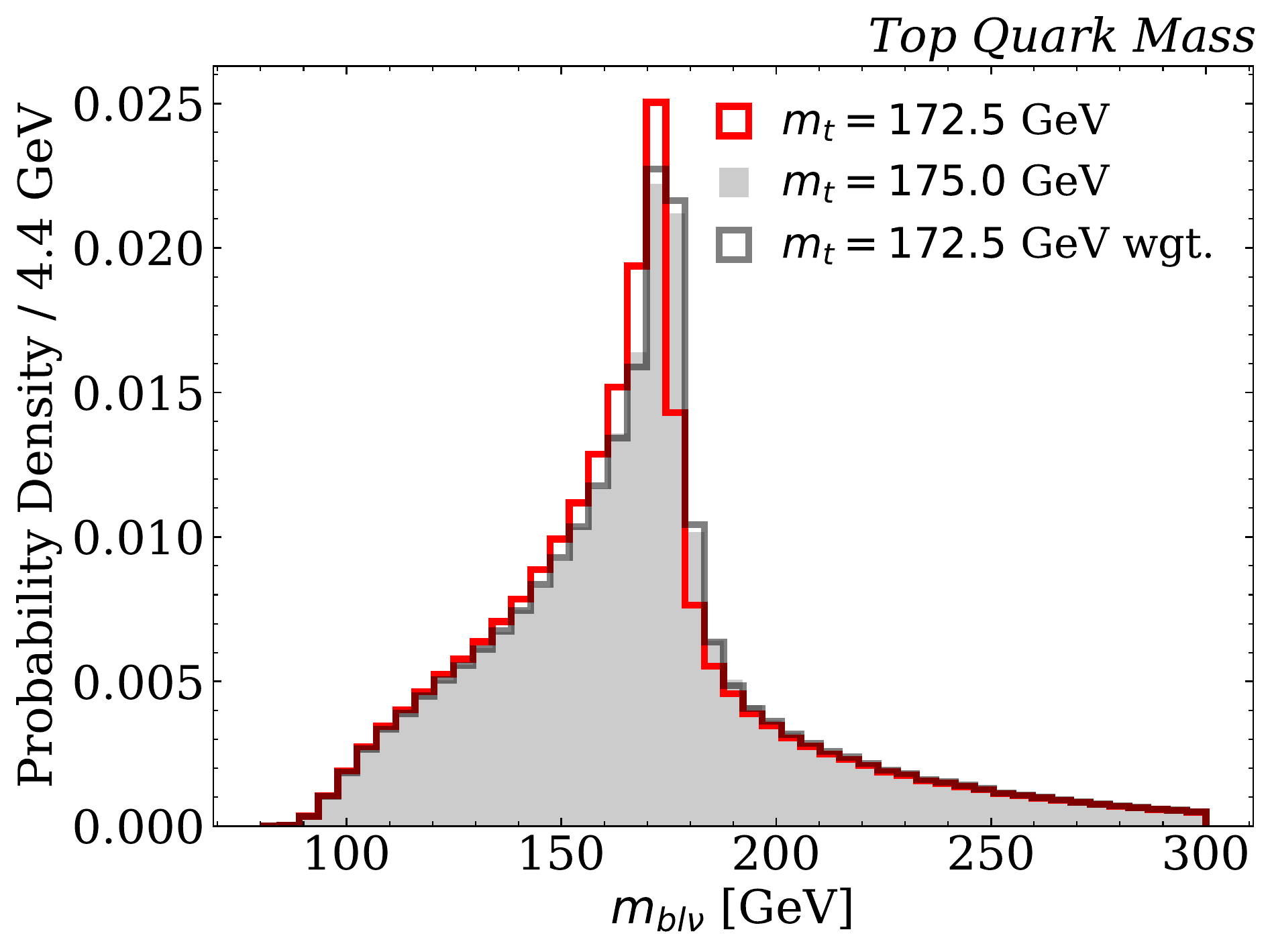}  \label{fig:top-a}}
 $\qquad$
 \subfloat[]{\includegraphics[width=0.91\columnwidth]{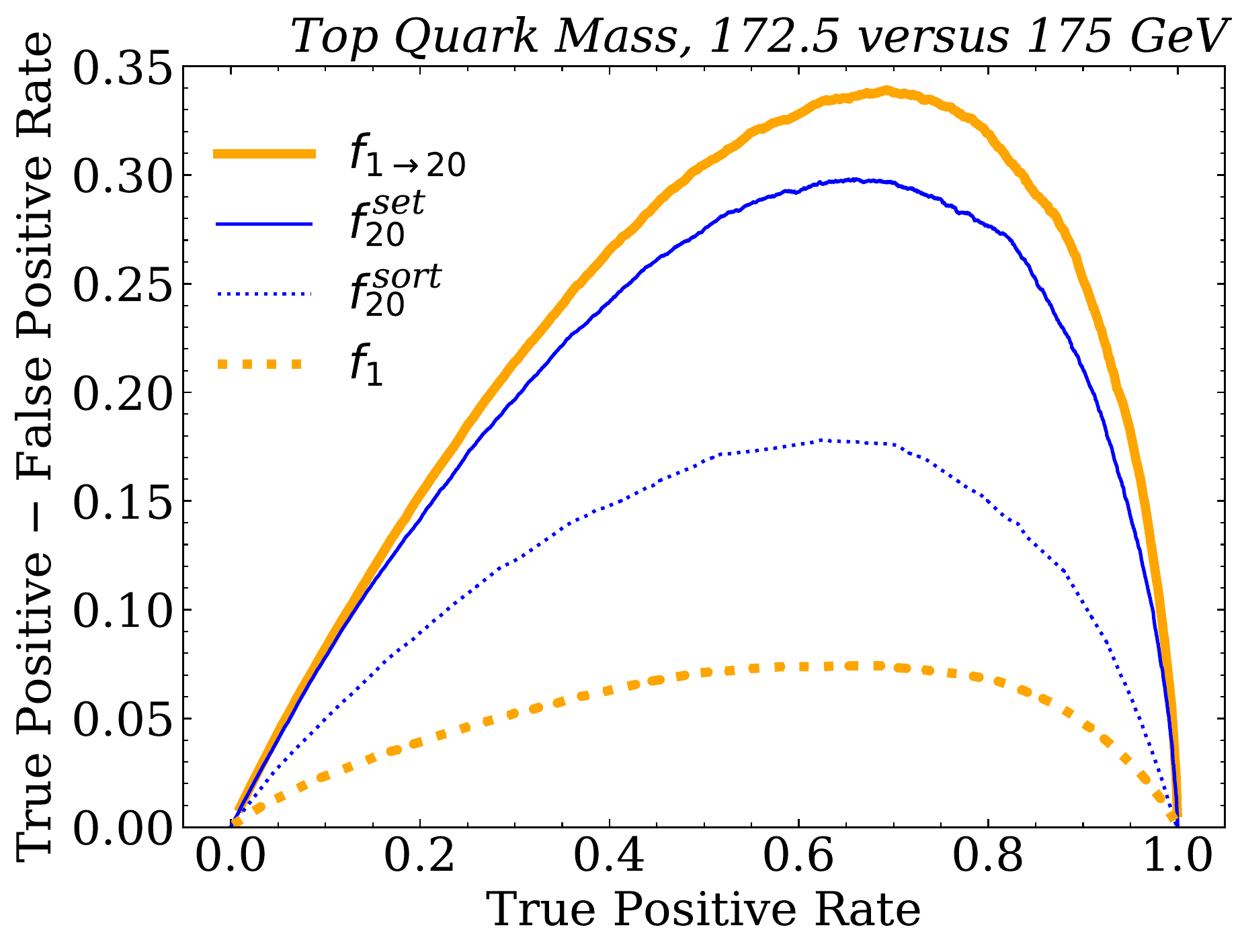}  \label{fig:top-b}}
 \caption{Classification in the top quark mass example.
 (a)  A histogram of $m_{b_1\mu\nu}$ for top quark masses of 172.5 GeV and 175 GeV.  The ``wgt.'' curve is explained later in \Sec{top_mass_regression}, where we test the performance of a likelihood reweighting.
(b) The difference in efficiency for the 172.5 GeV top quark mass sample (true positive) and the 175 GeV top quark mass sample (false positive) as a function of the true positive rate for various binary classifiers.  Once again, a multi-event classifier ($f_{1 \to 20}$) built from the single-event classifier ($f_{1}$) has the best performance.  For the classifiers trained to process 20 events simultaneously, the deep sets/PFN approach ($f_{20}^\text{set}$) does better than sorting the inputs ($f_{20}^\text{sort}$).}
 \label{fig:top}
 \end{figure*}

A more powerful way to account for the permutation symmetry among events is to explicitly build a permutation-invariant neural network architecture.
For this purpose, we use the deep sets approach~\cite{10.5555/3294996.3295098}.
In the particle physics context, deep sets were first used to construct particle flow networks (PFNs)~\cite{Komiske:2018cqr}, where the inputs involve sets of particles.
Here, we are interested in sets of events, though we will still use the PFN code from the \url{https://energyflow.network/} package.
Following \Refs{10.5555/3294996.3295098,Komiske:2018cqr}, we decompose our set-based classifier as:
\begin{align}
\label{eq:pfn}
    f^\text{set}_N(\vec{x}) = F\left(\sum_{i=1}^N\Phi(x_i)\right),
\end{align}
where $F:\mathbb{R}^L\rightarrow [0,1]$ and $\Phi:\mathbb{E}\rightarrow\mathbb{R}^L$ are neural networks that are simultaneously optimized.
The network $\Phi$ embeds single events $x_i$ into a $L$-dimensional latent space.
The sum operator in \Eq{pfn} guarantees that $f^\text{set}_N$ is invariant under permutations $x_{\sigma(i)}$ for $\sigma\in S_N$, the permutation group acting on $N$ elements.
We use the default parameters from the PFN code, with $L=128$, $\Phi$ having two hidden layers with 100 nodes each, and $F$ having three hidden nodes with 100 nodes each.
The same learning strategy (up to 1000 epochs, early stopping, 10\% batch size) as the other networks is used for the PFN.

The performance of $f^\text{set}_3$ is shown in \Fig{BSM}, which gets much closer to matching the performance of $f_{1\rightarrow 3}$.
Part of this improvement is due to enforcing the permutation symmetry, though there is also a potential gain from the fact the PFN we used for $f^\text{set}_3$ has more trainable weights than the fully connected network for $f^\text{sort}_3$.
All of the $f_3$ variants were considerably more difficult to train than $f_{1 \to 3}$, likely for the reason discussed in \Sec{gradients}.
Thus, we have empirical evidence for the superiority of single-event training for multi-event classification.

 \subsubsection{Top Quark Mass Measurement}
 \label{sec:manyfromone_top}

Our third and final example is motivated by the top quark mass measurement, as recently studied in \Refs{2010.03569,Flesher:2020kuy}.
Extracting the top quark mass is really a regression problem, which we investigate in \Sec{regression_study}.
Here, we consider a related classification task to distinguish two event samples generated with different top quark masses (172.5 GeV and 175 GeV).
This is a realistic hypothesis testing task that requires full event ensemble information, though only per-instance training as we will see.

We use the same dataset as \Ref{2010.03569}.
Top quark pair production is generated using \textsc{Pythia}~8.230~\cite{Sjostrand:2006za,Sjostrand:2007gs}
 and detector effects are modeled with \textsc{Delphes} 3.4.1~
\cite{deFavereau:2013fsa,Mertens:2015kba,Selvaggi:2014mya} using the default CMS run card.
After the production and decay steps $t\bar{t} \to b W^+ \bar{b} W^-$, one of the $W$ bosons is forced to decay to $\mu^+\nu$ while the other $W$ boson decays hadronically.
Each event is recorded as a variable-length set of objects, consisting of jets, muons, and neutrinos.
At simulation-level, the neutrino is replaced with the missing transverse momentum.
Generator-level and simulation-level jets are clustered with the anti-$k_t$ algorithm using $R=0.4$ and the simulation-level jet is labeled as $b$-tagged if the highest energy parton inside the nearest generator-level jet ($\Delta R < 0.5$) is a $b$ quark.
Jets are required to have $p_T>20$~GeV and they can only be $b$-tagged if $|\eta|<2.5$.
Furthermore, jets overlapping with the muon are removed.

Events are only saved if they have at least two $b$-tagged jets and at least two additional non $b$-tagged jets.
The $b$-jet closest to the muon in rapidity-azimuth is labeled $b_1$.
Of the remaining $b$-tagged jets, the highest $p_T$ one is labeled $b_2$.
The two highest $p_T$ non-$b$-tagged jets are labeled $j_1$ and $j_2$, and typically come from the $W$ boson.
(Imposing the $W$ mass constraint on $j_1$ and $j_2$ would yield lower efficiency, though without significantly impacting the results.)
The four-momentum of the detector-level neutrino ($\nu$) is determined by solving the quadratic equation for the $W$ boson mass; if there is no solution, the mass is set to zero, while if there are two real solutions, the one with the smaller $|p_z|$ is selected.
Four observables are formed for performing the top quark mass extraction, given by the following invariant masses: $m_{b_1\mu\nu}$, $m_{b_2\mu\nu}$, $m_{b_1j_1j_2}$, and $m_{b_2j_1j_2}$.
A histogram of $m_{b_1\mu\nu}$ is shown for illustration in \Fig{top-a}.

We use the same neural network architectures and training procedure as in the BSM example above, with 1.5 million events per fixed-mass sample.
The only difference is that the batch size is set to 0.1\% in order to keep the number of examples to be $\mathcal{O}(1000)$.
For the per-ensemble classifier, we take $N = 20$, though of course for a realistic hypothesis testing situation, $N$ would be as large as the number of top quark events recorded in data.
To capture the permutation invariance of the inputs, we construct $f_{20}^{\text{set}}$ using the deep sets approach in \Eq{pfn}.
We also build a classifier $f_{1 \to 20}$ from the per-instance classifier $f_1$ using \Eq{f_1_to_N}.

In \Fig{top-b}, we see that $f_{1 \to 20}$ and $f_{20}^{\text{set}}$ have comparable performance, though $f_{1 \to 20}$ is noticeably better.
Some of this improvement may be due to differences in the network architecture, but we suspect that most of the gain is due to the more efficient training in the per-instance case.
We checked that very poor performance is obtained for a classifier $f_{20}$ lacking permutation invariance, with a ROC curve that was not that much better than $f_1$ alone.
Explicitly breaking the invariance by sorting the inputs based on $m_{b_1\mu\nu}$ does help a little, as indicated by the $f_{20}^{\text{sort}}$ curve in \Fig{top-b}, but does not reach the set-based approach.

 \begin{figure}[t]
 \centering
 \includegraphics[width=0.91\columnwidth]{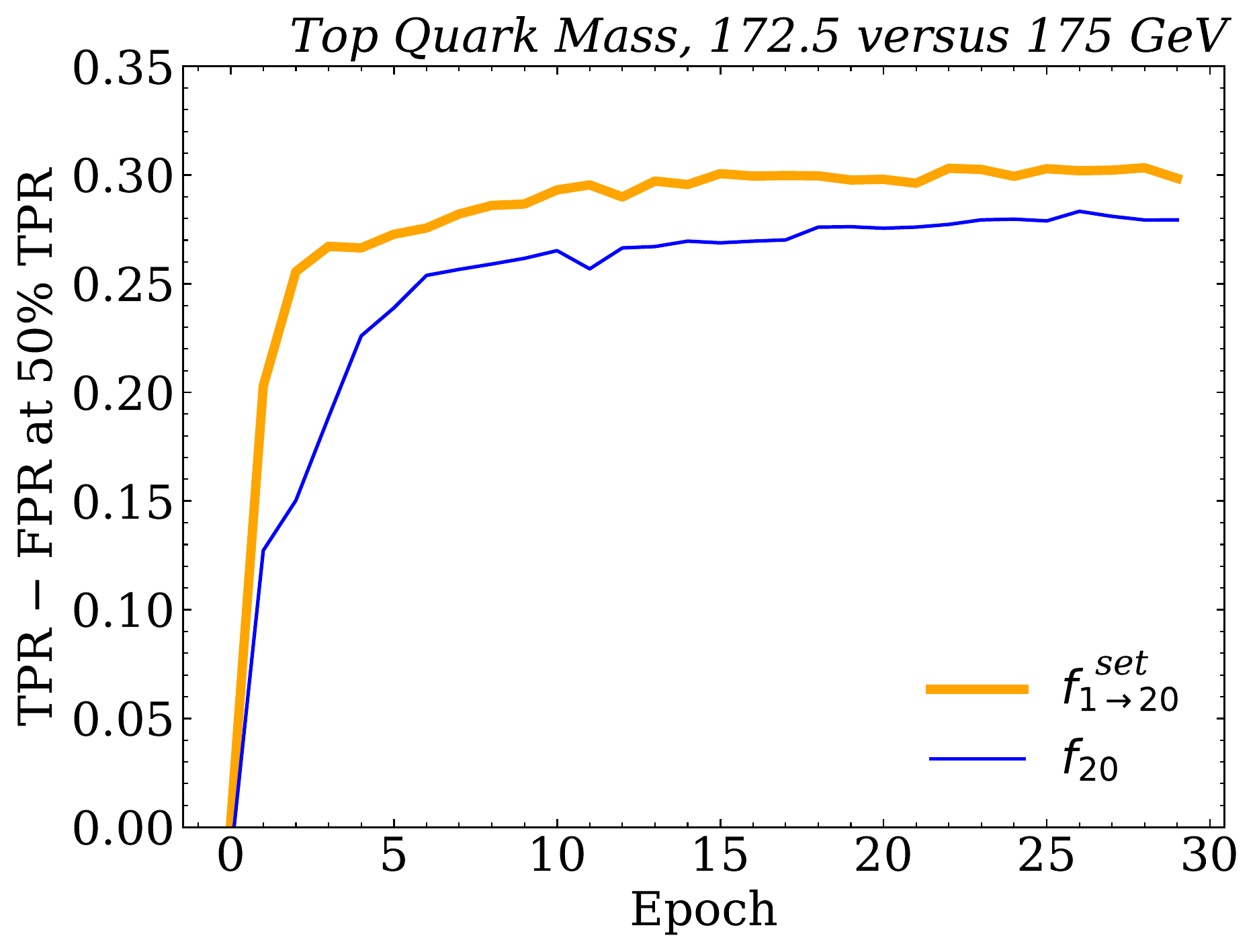}
 \caption{Computational performance of single-event versus multi-event training.
 Shown is the efficiency for the 175 GeV sample (false positive) for a fixed 50\% efficiency for the 172.5 GeV sample (true positive), plotted as a function of training epoch.
 Single-event training ($f_{1 \to 20}$) outperforms multi-event training ($f_{20}^\text{set}$), where both methods go through the full data set per epoch.
 }
 \label{fig:top2}
 \end{figure}

 Given the similar performance of $f_{1\rightarrow 20}$ and $f_{20}^\text{set}$, it is interesting to examine which learning strategy is more computationally efficient.
In \Fig{top2}, we compare the performance as a function of the training epoch, using the difference of the true and false positive rates at a fixed 50\% signal efficiency. 
 In each epoch, both $f_{1\rightarrow 20}$ and $f_{20}^\text{set}$ see the full ensemble of events, so this is an apples-to-apples comparison as far as data usage is concerned.
  In particular, we plot this information per epoch instead of per compute time to avoid differences due to the structure of the neural networks.
  (There is not an easy way to control for possible differences in the training time due to the differences in the network structures, since the underlying tasks are different.) 
 The $f_{1\rightarrow 20}$ classifier trains much faster, in agreement with the analysis in \Sec{gradients}, even though the ultimate asymptotic performance is similar for both classifiers.
Once again, we see better empirical behavior from $f_{1 \to 20}$ trained on one event at a time version $f_{20}^{\text{set}}$ trained on multiple events simultaneously.%
\footnote{Away from the asymptotic limit, one could try to improve the empirical per-ensemble performance through data augmentation.  Data augmentation is a generic strategy to help neural networks learn symmetries, and the IID structure can be reinforced by showing the network new ensembles built from sampling instances from the existing ensembles.}

\subsection{Classifiers: Single-Event from Multi-Event}
\label{sec:onefrommany}

In general, one cannot take a multi-event classifier $f_N$ and extract a single-event classifier $f_1$.
It is, however, possible to construct a special $\tilde{f}_N$ network such that one can interpret a subnetwork as a per-event classifier, as discussed in \Sec{per_ensemble_classification}.
When using the MLC loss function, we can use the functional form in \Eq{tildef_N}, where $\tilde{f}_N$ is a product of $f_{N \to 1}$ terms.
Training $\tilde{f}_N$, where the only trainable weights are contained in $f_{N \to 1}$, we can learn a single-event classifier $f_{N \to 1}$ from multi-event samples.

For the binary cross entropy loss used in our case studies, where \Eq{c_to_p} is needed to convert the classifier to a likelihood ratio, we have to introduce a slightly different structure than \Eq{tildef_N}.
Let $f_N^\text{set}$ be a permutation-invariant classifier, as defined in \Eq{pfn} using the deep sets/PFN strategy.
Taking the latent space dimension to be $L=1$, the $\Phi$ network can be interpreted as a single-event classifier.
Because the $\Phi$ network outputs are pooled via summation, we can build an optimal multi-event classifier if $\Phi$ learns the \emph{logarithm} of the likelihood ratio; cf.\ \Eq{iid_lr_relation}.
With this insight, we can fix the $F$ function to achieve the same asymptotic performance as a trainable $F$ by setting:
\begin{align}
\label{eq:specialF}
    F(\vec{x}) = \frac{\exp\big(\sum_{i = 1}^N \Phi(x_i)\big)}{1+\exp\big(\sum_{i = 1}^N \Phi(x_i)\big)}\,.
\end{align}
Using \Eq{c_to_p}, one can check that this $F$ is monotonically related to the ensemble likelihood ratio.
Similarly, $\Phi$ will be monotonically related to the optimal $f_1$, which we call $f_{N \to 1}$ for the remainder of this discussion.

 \begin{figure}[t]
 \centering
 \includegraphics[width=0.91\columnwidth]{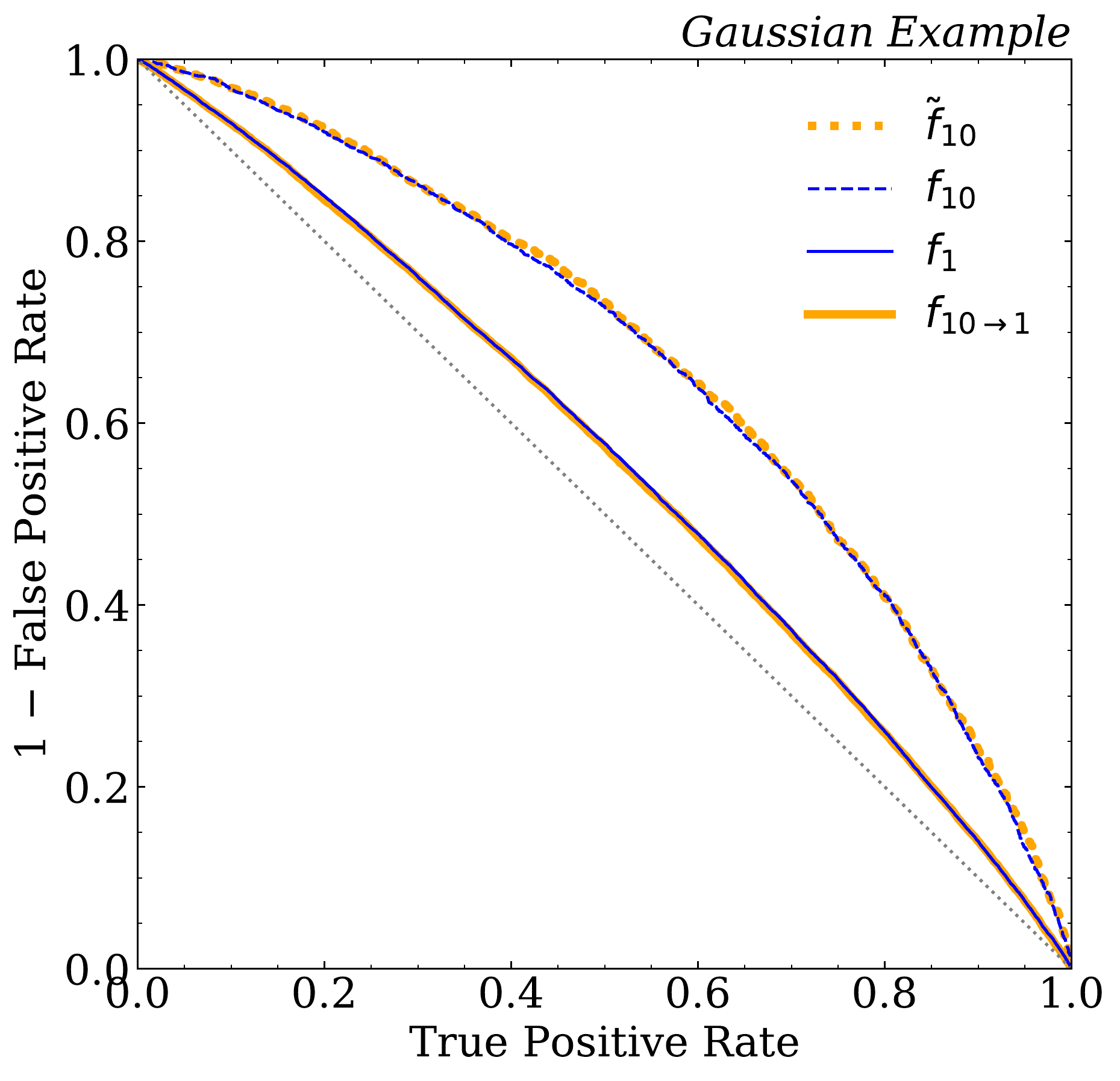}
 \caption{Revisiting the ROC curves for the two Gaussian example from \Fig{Gaussian-b}.
 The multi-event classifier $\tilde{f}_{10}$ with the restricted functional form in \Eq{specialF} has the same performance as $f_{10}$ with no restrictions.
 Using $\tilde{f}_{10}$, we can construct a single-event classifier $\tilde{f}_{10 \to 1}$ with the same performance as $f_1$ trained directly.}
 \label{fig:Gaussian2}
 \end{figure}

This construction is demonstrated in \Fig{Gaussian2} for the Gaussian example.
We see that the deep sets architecture with the fixed form of \Eq{specialF} ($\tilde{f}^{\rm set}_{10}$) has the same or better performance as the 10-instance fully-connected classifier with more network capacity ($f_{10}$).
Similarly, the $\Phi$ function used as a single-event classifier ($f_{10\rightarrow 1}$)  has nearly the same performance as an independently trained single-event classifier ($f_1$).

 \begin{figure}[t]
 \centering
 \includegraphics[width=0.91\columnwidth]{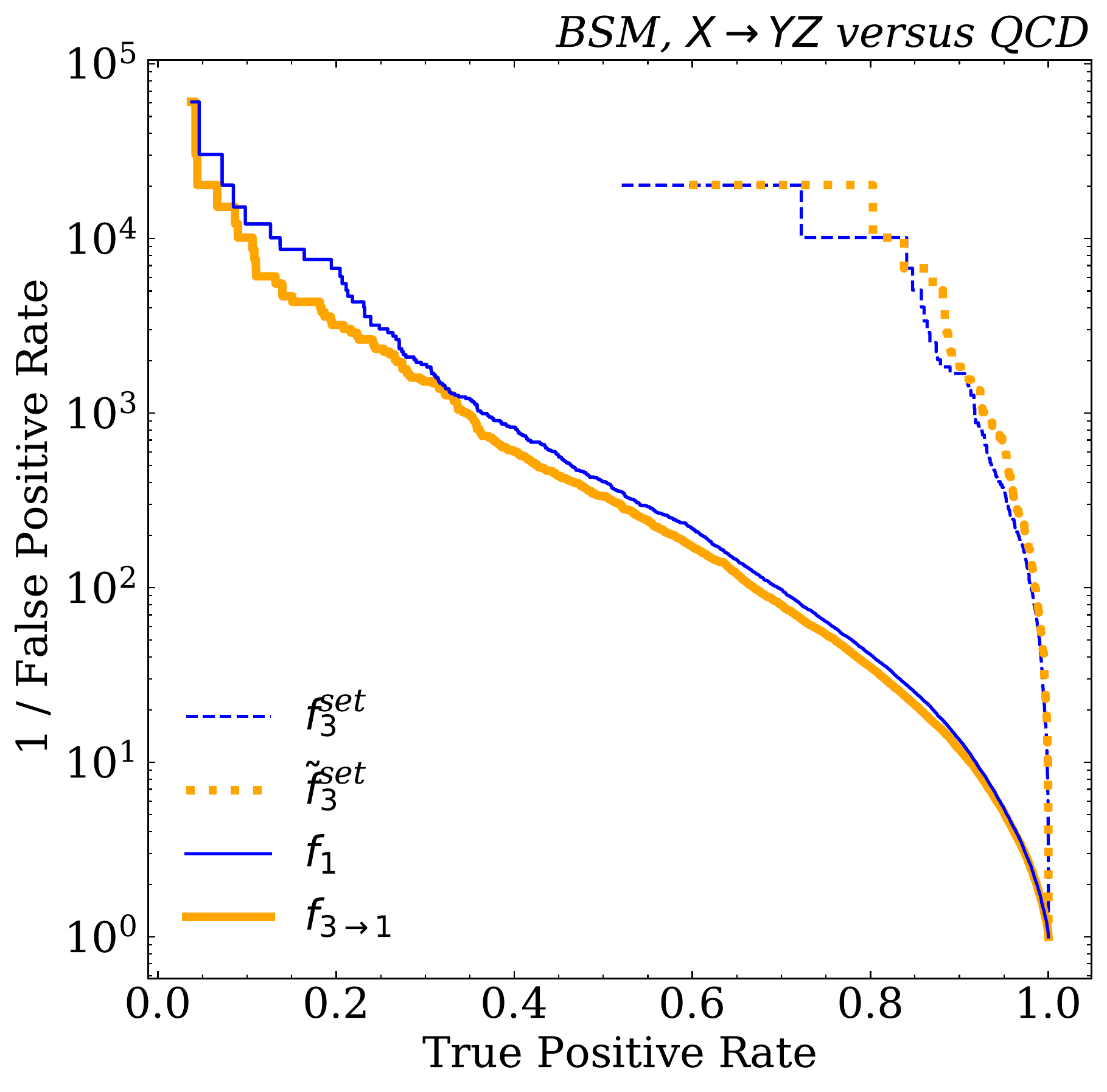}
 \caption{Revisiting the ROC curves for the dijet resonance search example in \Fig{BSM-c}.
The set-based multi-event classifiers $\tilde{f}_3^{\rm set}$ and $f_3^{\rm set}$ have similar performance, but we can use the former to construct a single-event classifier $f_{3 \to 1}$.
This construction is not as effective as performing single-event training directly ($f_1$).
}
 \label{fig:BSM2}
 \end{figure}

The same conclusion holds for the BSM classification task, shown in \Fig{BSM2}.
The only difference between the set-based architectures $\tilde{f}_3^\text{set}$ and $f_3^\text{set}$ is that the former uses the fixed functional form in \Eq{specialF}.
The fact that they achieve nearly the same performance is ensured by the IID relation in \Eq{iid_lr_relation}.
The per-instance $f_{3\rightarrow 1}$ network extracted from $\tilde{f}_3^\text{set}$ is not quite as powerful as the $f_{1}$ network trained independently on single events, as expected from the gradient issue discussed in \Sec{gradients}.
While we found no benefit to extracting a single-event classifier from a multi-event classifier, it is satisfying to see these IID-derived theoretical predictions  borne out in these empirical examples.

\subsection{Comparison of Regression Strategies}
\label{sec:regression_study}

We now consider the regression methods introduced in \Sec{regression}.
For classification, the mapping between per-instance and per-ensemble information is relatively straightforward.
For regression, though, per-ensemble regression is structurally dissimilar from per-instance regression because of the need to integrate over priors on the regression parameters.
Nevertheless, we can perform per-ensemble regression by first mapping the problem to per-instance parametrized classification.

We compare three different regression strategies for our empirical studies.
The first method is a maximum-likelihood analysis, using the form in \Eq{theta_star} based on the single-event parametrized classifier in \Eq{paramed}.
The second method is per-instance direct regression, using the construction in \Eqs{directregression}{conditionalprobabilitytoclassifier} based on the same classifier as above.
The third method is per-ensemble direct regression, based on minimizing the mean squared error loss in \Eq{directregressionMSE}.

\subsubsection{Gaussian Mean Example}

Our first regression study is based on the same one-dimensional Gaussian distributions as \Sec{manyfromone_gaussian}.
The prior distribution for the Gaussian means is taken to be uniform with $\mu \in [-0.5,0.5]$, while the variance is fixed at $\sigma = 1$.
A training dataset is created from 100 examples each from 10,000 values of the Gaussian mean, for a total of one million training data points.
For the reference sample $p(x|\theta_0)$ needed to build the single-event parametrized classifier $f(x,\mu)$ in \Eq{paramed}, we create a second dataset with one million examples drawn from a standard normal distribution (i.e.\ $\mu  = 0$).
To implement the $p(\theta)$ term in the second line of \Eq{paramclass}, each example $x_i$ from the reference dataset is assigned a random mean value picked from the variable-mean dataset.

We train a parametrized neural network to distinguish the variable-mean datasets from the reference dataset.
This network takes as input two features: one component of $\vec{x}$ and the random mean value $\mu$.
The architecture consists of three hidden layers with $(64, 128, 64)$ nodes per layer and ReLU activation.
The output layer has a single node and sigmoid activation.
Binary cross entropy is used to train the classifier and \Eq{c_to_p} is used to convert it to the likelihood ratio form $f(x,\mu)$.
The model is trained for 1000 epochs with early stopping and a batch size of 10\% of the training statistics.

The same learned function $f(x,\mu)$ is used for both the maximum likelihood analysis and per-instance direct regression.
For the maximum-likelihood analysis, the optimization in \Eq{theta_star} is performed over a fixed grid with 20 evenly spaced values in $\mu \in [-0.5,0.5]$.
For per-instance direct regression, the function $f_N(\vec{x},\mu)$ in \Eq{conditionalprobabilitytoclassifier} is constructed by taking a product of $f(x,\mu)$ outputs over all 100 examples in a given ensemble data point $\vec{x}$.
The integrals in \Eqs{directregression}{conditionalprobabilitytoclassifier} are approximated by evaluating $f_N(\vec{x},\mu)$ at 20 evenly spaced $\mu$ values between $-0.5$ and $0.5$ and then adding their values; this is possible because the prior is uniform.

The per-ensemble direct regression approach uses a neural network $g_N$ that takes as input 100 values (i.e.\ all of $\vec{x}$) and predicts a single mean value.
This network has the same architecture as $f(x,\mu)$, except it directly takes as input $\vec{x}$ and has linear (instead of a sigmoid) activation for the output layer, since the predicted mean can be both positive or negative.  
It is trained to minimize the mean squared error loss in \Eq{directregressionMSE}.

 \begin{figure}[t]
 \centering
 \includegraphics[width=0.91\columnwidth]{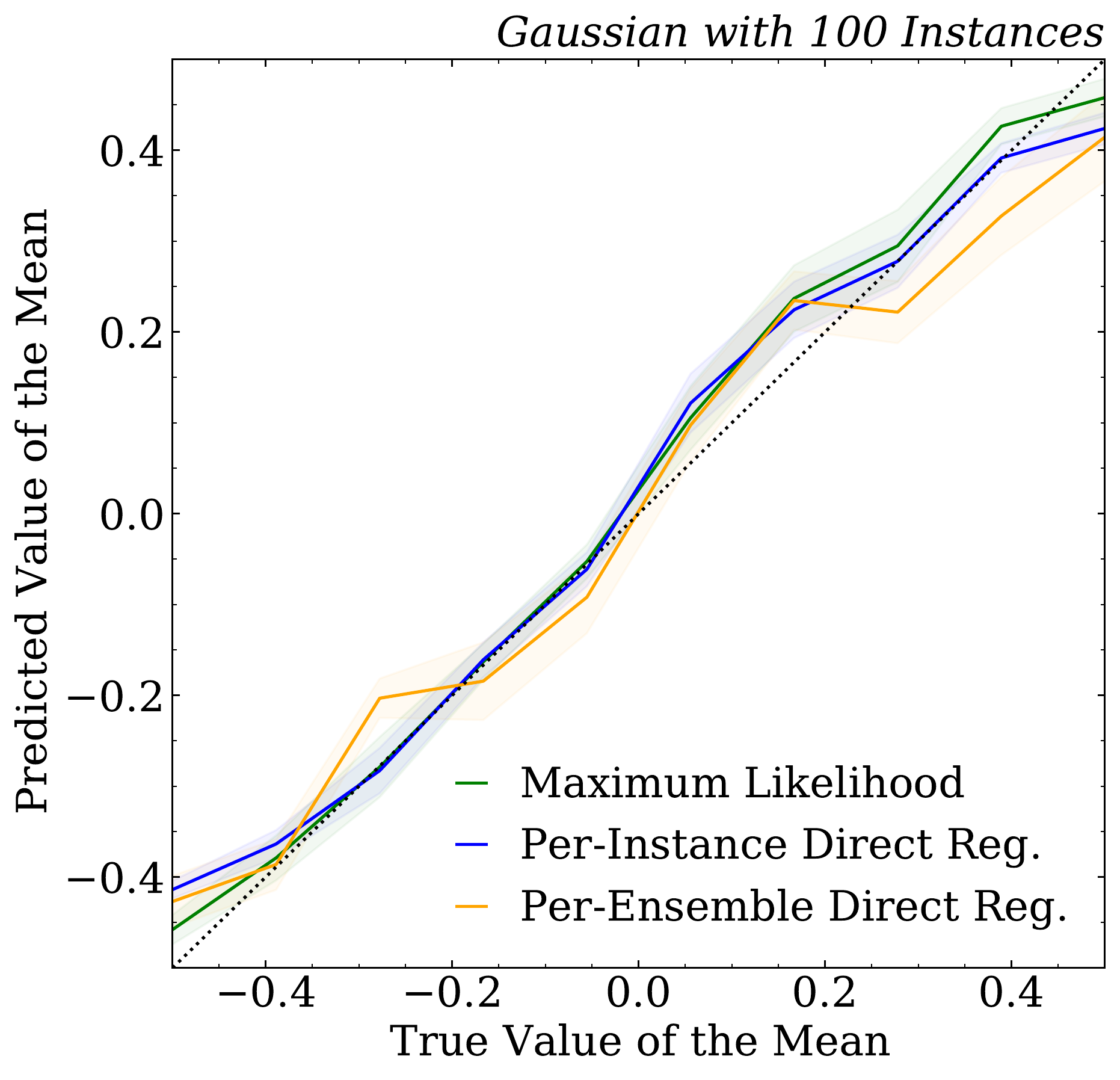}
 \caption{
 Comparison of regression methods with the Gaussian example, with the predicted value of the mean plotted against the true value of the mean.
 The regression involves analyzing 100 instances drawn from the same Gaussian distribution.
 Bands are the standard deviation of the predictions over 10,000 generated samples.
 The per-instance direct regression uses single-event training, yet achieves comparable performance to per-ensemble direct regression that processes 100 events simultaneously.}
 \label{fig:Gaussianregression}
 \end{figure}

In \Fig{Gaussianregression}, we see that all three approaches give nearly the same results in terms of bias and variance.
Strictly speaking, maximum likelihood and direct regression are different tasks so their behavior could be different.
For per-instance and per-ensemble direct regression, they are constructed to yield the same asymptotic behavior, but there will be differences due to, e.g., the finite approximations to the integrals.
Note that maximum likelihood and per-instance direct regression only use neural networks that process per-instance inputs; information about the rest of the events is used only through the training procedure.
Thus, we have empirical evidence that per-ensemble regression can be accomplished via per-instance training.

\subsubsection{Top Quark Mass Measurement}
\label{sec:top_mass_regression}

As a physics example of regression, we consider extracting the top quark mass.
Here, the top quark mass is the regression target and the setup is similar to the Gaussian example above.
We use the same event generation as \Sec{manyfromone_top}, but now with top quark mass parameters sampled uniformly at random in $m_t \in [170,180]~\text{GeV}$.
As with the Gaussian example, a variable-mass dataset is created.  In this case, we have 100 events for each of 100,000 sampled top quark mass values. 
The reference sample uses a top quark mass of 172.5 GeV.
Due to event selection effects, the actual number of events for each top quark mass value varies from set-to-set, with a mean of about 40 events.
Because this event selection has a slight top quark mass dependence, this yields an effective non-uniform prior on $m_t$, which we account for when assigning dummy mass values to the reference sample.

The parametrized classifier now takes five inputs: the four mass features from \Sec{manyfromone_top} ($m_{b_1\mu\nu}$, $m_{b_2\mu\nu}$, $m_{b_1j_1j_2}$, and $m_{b_2j_1j_2}$) plus the top quark mass used for event generation.
The neural network has three hidden layers with 50 nodes per layer and ReLU activation, and a single node output layer with sigmoid activation.
We train 100 models and take the median as the classifier output, using \Eq{c_to_p} to convert it to the likelihood ratio $f(x,m_t)$.
Each model is trained for 1000 epochs with early stopping with a patience of 20 epochs and a batch size of 0.1\%.
To test the fidelity of the training, we extract the estimated likelihood ratio of $m_t = 175~\text{GeV}$ over $m_t = 172.5~\text{GeV}$ and use it to reweight the $172.5~\text{GeV}$ sample.
From \Fig{top-a}, we see that we achieve good reweighting performance despite the relatively limited training data.

 \begin{figure*}[t]
 \centering
  \subfloat[]{\includegraphics[width=0.91\columnwidth]{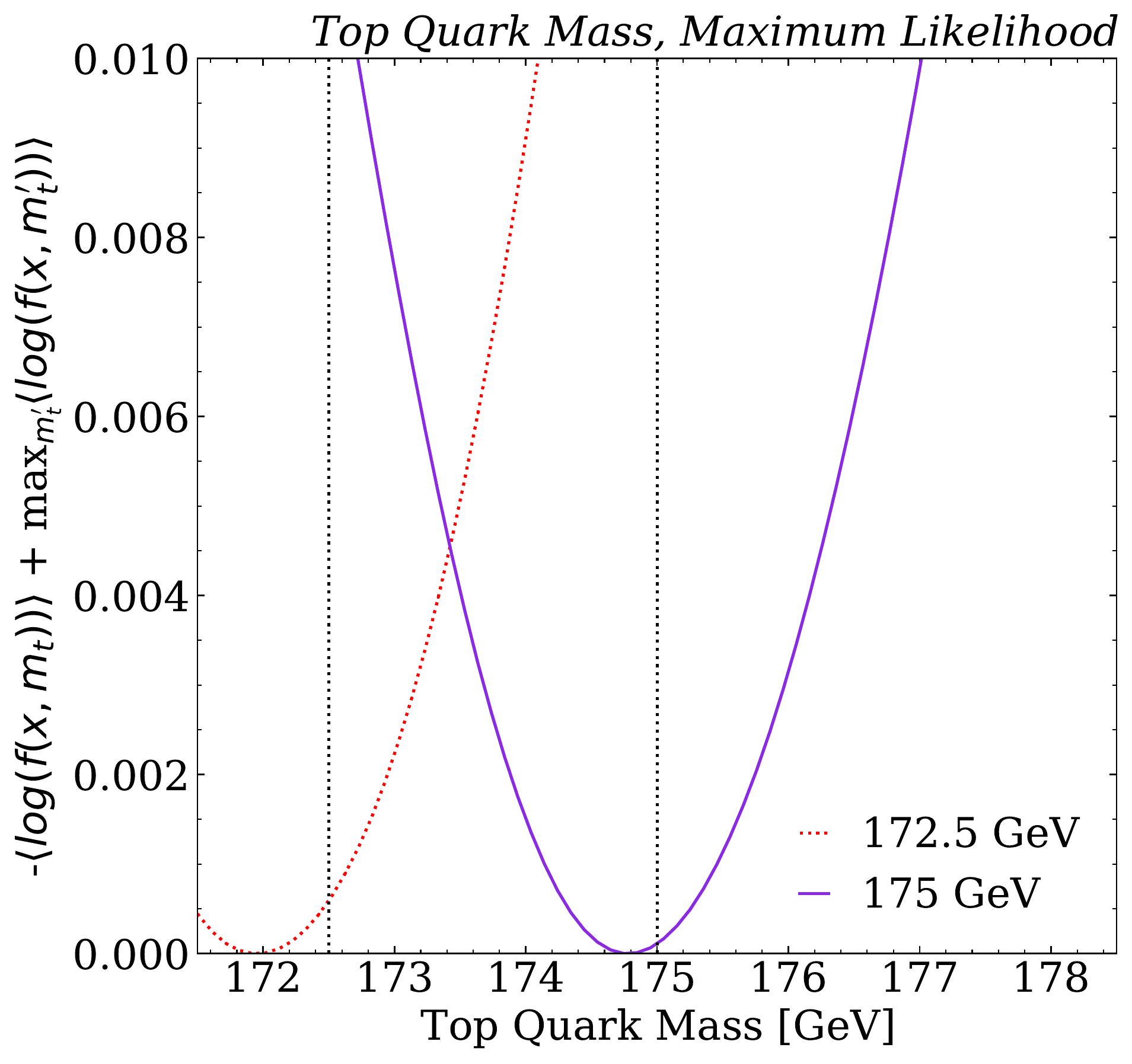}  \label{fig:Topregression-a}}
  $\qquad$
 \subfloat[]{\includegraphics[width=0.87\columnwidth]{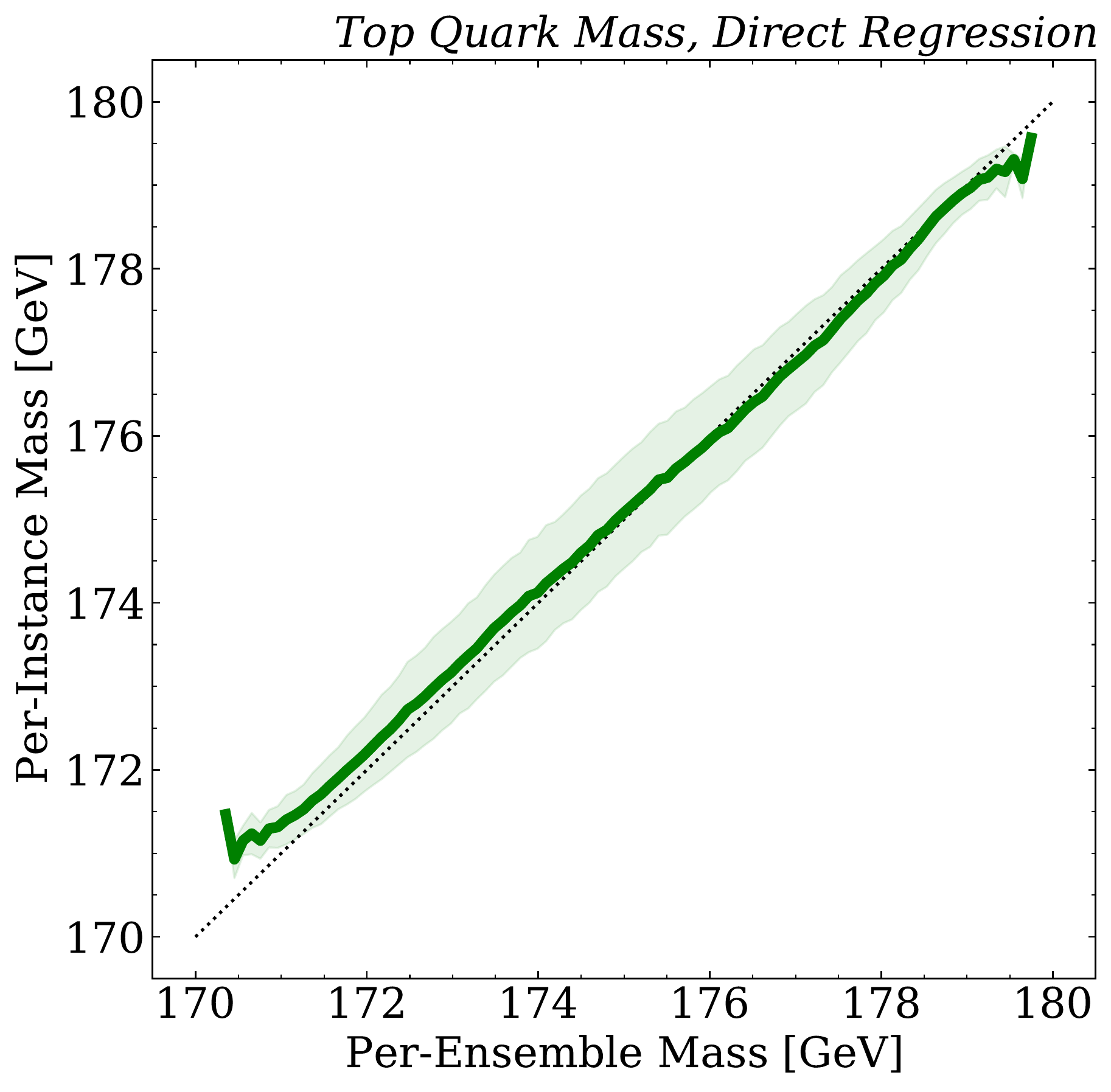}  \label{fig:Topregression-b}}
 \caption{Regression in the top quark mass example.
 (a) An estimate of the log likelihood for samples generated with 172.5 and 175 GeV top quark masses.
 The vertical axis has been shifted such that the minimum value is at zero.
 Note that the axis represents the average log likelihood which is a factor of $N_\text{events}$ different from the total log likelihood.
 (b) Correlation between the per-instance predicted mass and the per-ensemble predicted mass in the context of direct regression.  The per-ensemble mass values are put in bins of 0.1 GeV width, and the bands represent the standard deviation of the per-instance mass values in each bin.}
 \label{fig:Topregression}
 \end{figure*} 

The maximum likelihood analysis is performed by scanning the learned log likelihood estimate over a fixed grid with 100 uniformly spaced steps in $m_t \in [170,180]~\text{GeV}$.
In \Fig{Topregression-a}, we show this scan where the target data comes from the high statistics 172.5 GeV and 175 GeV samples from \Sec{manyfromone_top}.
As desired, the minimum of the parabolic shapes are near the input top quark masses.

For the per-instance direct regression, we follow the same strategy as in the Gaussian case to convert $f(x,m_t)$ into an estimate of $\mathbb{E}[m_t|\vec{x}]$.
The integrals in \Eqs{directregression}{conditionalprobabilitytoclassifier} are approximated by sampling 50 random top quark masses per set of 100 following the probability density from the training dataset.  
Because 40 events are insufficient to make a precision measurement of the top quark mass, we find a noticeable bias between the estimated and true top mass values, which is exacerbated by edge effects at the ends of the training range.
For this reason, we do not show a direct analog to \Fig{Gaussianregression}, though this bias could be overcome with much larger training datasets with many more than 100 examples per mass value.

For the per-ensemble direct regression, we use the deep sets approach in \Eq{pfn} to handle the permutation-invariance of the inputs.
This approach is also well suited to handle the large variation in the number of events in each set due to the event selection effect.
We again use PFNs for our practical implementation.
We use the default PFN hyperparameters from the \url{https://energyflow.network/} package, except we use linear activation in the output layer and the mean squared error loss function.
We found that it was important for the model accuracy to standardize both the inputs and outputs of the network.
Note that this is a different per-ensemble direct regression setup than used in \Ref{Flesher:2020kuy}, which found excellent performance using linear regression on sorted inputs.

In \Fig{Topregression-b}, we compare the output of per-ensemble direct regression to the output of per-instance direct regression.
We find a very strong correlation between these two very different approaches to computing the same quantity $\mathbb{E}[m_t|\vec{x}]$.  
The band in \Fig{Topregression-b} is the standard deviation over data sets with a true mass in the same one of the 100 bins that are evenly spaced between 170 and 180 GeV.
A key advantage of the per-instance approach is that it does not need to be retrained if more events are acquired.
By contrast, the per-ensemble approach is only valid for event samples that have the same sizes as were used during training.

\subsection{Beyond Regression Example}
\label{sec:beyondregression}

As remarked in \Sec{beyond}, the ideas discussed above apply to learning tasks beyond just standard classification and regression.
As one simple example to illustrate this, we consider the Gaussian classification task from \Sec{manyfromone_gaussian} and compute the mutual information between the Gaussian feature and the label.
This quantifies how much information is available in the feature for classification and can be directly compared with other features and other classification tasks.

 \begin{figure}[t]
 \centering
 \includegraphics[width=0.91\columnwidth]{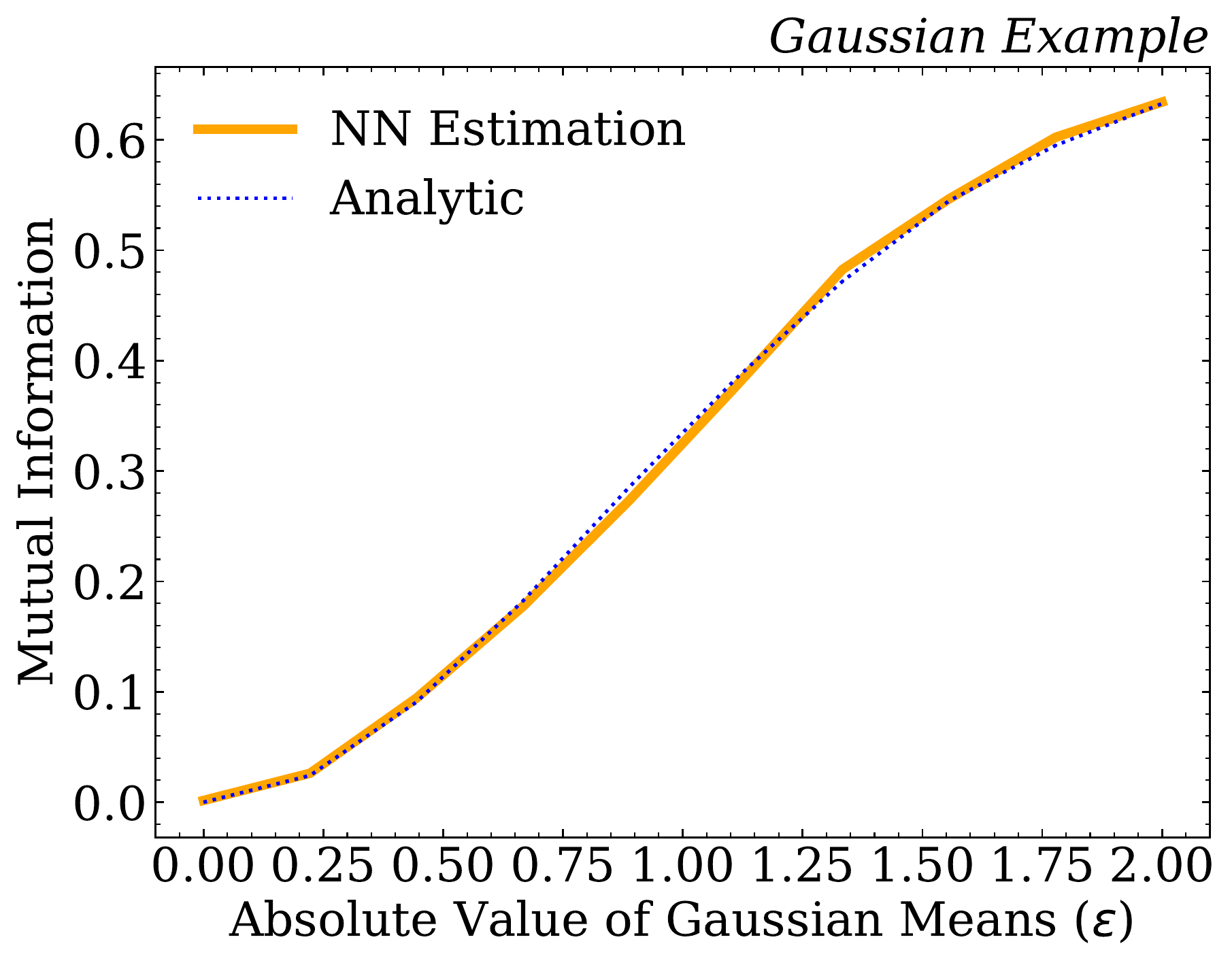}
 \caption{
 Mutual information between a Gaussian feature and a label, where the `` signal'' ($x_0 = \epsilon$) and `` background'' ($x_0 = - \epsilon$) have opposite means.
 The estimate using the MLC loss approach shows good agreement with the exact analytic expression.
 }
 \label{fig:GaussianMI}
 \end{figure}

For this illustration, $10^5$ events are generated each from two Gaussian distributions with means $\pm |\epsilon|$ for fixed $\epsilon$.
The mutual information is estimated using a per-instance classifier as described in \Sec{beyond} and also computed analytically via \Eq{MI}.
For the per-instance classifier, we use a neural network that processes two inputs (label and feature), has two hidden layers with ReLU activation, and has a single node sigmoid output.
The classification task is to distinguish the nominal dataset from one where the labels are assigned uniformly at random to the features.
The value of the MLC loss yields an estimate of the mutual information.

The mutual information results are presented in \Fig{GaussianMI}, as a function of $\epsilon$.
As expected, the neural network strategy yields an excellent approximation to the analytic calculation.
Note that this strategy does require any binning and naturally extends to high-dimensional data, since the core component is a neural network classifier.
We leave an investigation of this approach in the particle physics context to future work.

%\FloatBarrier

%%%%%%%%%%%%%%%%%%%%%%%%%%%%%%%%%%%%%%%%%%%%%%%%%%%%%%%
\section{Conclusions}
\label{sec:conclusions}

We have demonstrated a connection between classifiers trained on single events and those that process multiple events at the same time.
One can take a generic single-event classifier and build an $N$-event classifier using simple arithmetic operations.
Such classifiers tend to out-perform generic $N$-event classifiers, since we can enforce the IID assumptions into the learning task.
This performance gap can be mostly recovered by deploying a classifier that respects the permutation invariance of the set of $N$ events.
We used the deep sets/PFN architecture~\cite{10.5555/3294996.3295098,Komiske:2018cqr} for this purpose, but other set-based architectures such as graph neural networks~\cite{10.1109/TNN.2008.2005605,Shlomi:2020gdn} would also be appropriate.

An amusing feature of the deep sets approach is that we can use it to reverse-engineer a single-event classifier from a multi-event classifier by restricting the latent space to be one-dimensional and fixing a static output function.
Even after enforcing these additional structures, though, we found both theoretically and empirically that the loss function gradients are better behaved for single-event classifiers than multi-event classifiers.
Going beyond classification, we explained how various regression tasks can be phrased in terms of per-instance parametrized classification, yielding similar performance to per-ensemble direct regression.
We also mentioned how to compute distances and divergences between probability densities without requiring explicit density estimation.
These results hold for any data sample satisfying the IID property.

Ultimately, we did not find any formal or practical advantage for training a multi-event classifier instead of a single-event classifier, as least for the cases we studied.
With a carefully selected multi-event architecture, one can achieve similar performance to a scaled-up per-event classifier, but the latter will typically train faster.
For direct regression, the per-ensemble strategy might be conceptually simpler than the per-instance method, though the per-instance methods allow for a simpler treatment of variably-sized data sets.
Note that there may be situations where a simplifying assumption (e.g.~the linear regression model in~\Ref{Flesher:2020kuy}) could yield better per-ensemble behavior than indicated by our case studies.
At minimum, we hope this paper has demystified aspects of per-ensemble learning and highlighted some interesting features of the MLC loss function.

Going beyond the IID assumption, the duality between per-instance classifiers and per-ensemble classifiers could have applications to problems with approximate independence.
For example, flavor tagging algorithms have traditionally exploited the approximate independence of individual track features within a jet~\cite{Aaboud:2018xwy,Chatrchyan:2012jua}.
Similarly, emissions in the Lund jet plane~\cite{Andersson:1988gp,Dreyer:2018nbf} are approximately independent, with exact independence in the strongly ordered limit of QCD.
In both contexts, the instances are particles (or particle-like features) and the ensemble is the jet.
A potentially powerful training procedure for these situations might be to first train a per-particle classifier, then build a per-jet classifier using the constructions described in this paper, and finally let the network train further to learn interdependencies between the particles.

\section*{Code and Data}

The code for this paper can be found at \url{https://github.com/bnachman/EnsembleLearning}. The physics datasets are hosted on Zenodo at~\Ref{anders_andreassen_2020_4067673} for the top quark dataset and~\Ref{kasieczka_gregor_2019_4287846} for the BSM dataset. 

\begin{acknowledgments}

We thank Anders Andreassen, Patrick Komiske, and Eric Metodiev for discussions about the MLC loss.
We thank Rikab Gambhir and Ian Convy for discussions about mutual information.
We thank Adi Suresh for discussions about the regression task with the classifier loss.
We thank Katherine Fraiser, Yue Lai, Duff Neill, Bryan Ostdiek, Mateusz Ploskon, Felix Ringer, and  Matthew Schwartz for useful comments on our manuscript.
BN is supported by the U.S. Department of Energy (DOE), Office of Science under contract DE-AC02-05CH11231.
JT is supported by the National Science Foundation under Cooperative Agreement PHY-2019786 (The NSF AI Institute for Artificial Intelligence and Fundamental Interactions, \url{http://iaifi.org/}), and by the U.S. DOE Office of High Energy Physics under grant number DE-SC0012567.
BN would also like to thank NVIDIA for providing Volta GPUs for neural network training.

\end{acknowledgments}

\appendix

\section{Deriving Maximum Likelihood Classifier Loss}
\label{app:maxlikeloss}

Beyond just the practical value of learning the likelihood ratio, the MLC loss in \Eq{maxlikeloss} has a nice interpretation in terms of learning probability distributions.

Consider trying to learn a function $f(x)$ that is a normalized probability distribution, up to a Jacobian factor $j(x)$:
\begin{equation}
\label{eq:jf_norm}
    \int dx \, j(x) f(x) = 1.
\end{equation}
We are given samples from a probability distribution $q(x)$, and we want to learn $f(x)$ such that
\begin{equation}
f(x) \to \frac{q(x)}{j(x)}.
\end{equation}
In other words, we want to learn a function $f(x)$ that reproduces the sampled distribution $q(x)$ after including the Jacobian factor.
This problem was studied in \Ref{Andreassen:2018apy}, albeit in a context where $f(x)$ had a restricted functional form such that \Eq{jf_norm} was automatically enforced.

 \begin{figure}[t]
 \centering
 \includegraphics[width=0.91\columnwidth]{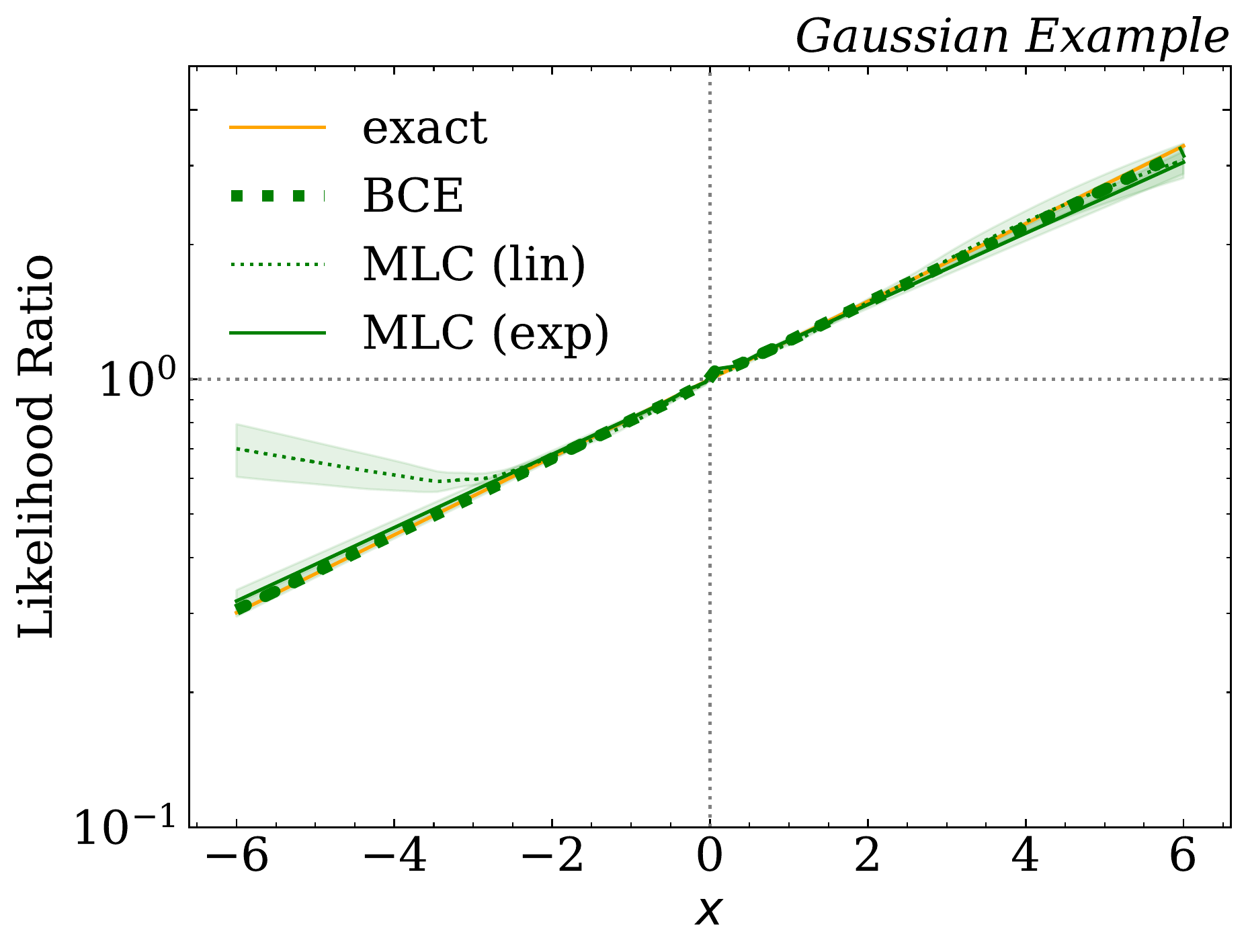}
 \caption{ 
 A demonstration of the MLC loss for learning the likelihood ratio directly, using the Gaussian example from \Fig{Gaussian-a}.
 The linear (lin) and exponential (exp) parametrizations perform similarly. 
 Shown for comparison is likelihood ratio
 computed using the binary cross entropy (BCE) loss that requires the manipulation in \Eq{c_to_p}.}
 \label{fig:MLC}
 \end{figure}

One strategy to accomplish this is to minimize the cross entropy of $f(x)$ with respect to $q(x)$, since the smallest cross entropy is obtained when $f(x)$ has the same information content as $q(x)$.
The associated loss functional is:
\begin{equation}
\label{eq:loss_q_j_story_with_Lagrange}
L[f] = - \int dx \, q(x) \log f(x) - \lambda \left(1 - \int dx \, j(x) f(x) \right),
\end{equation}
where the first term is the cross entropy and $\lambda$ is a Lagrange multiplier to enforce the normalization condition in \Eq{jf_norm}.
Taking the functional derivative of \Eq{loss_q_j_story_with_Lagrange} with respect to $f(x)$ and setting it equal to zero, we find the extremum condition:
\begin{equation}
\label{eq:loss_q_j_story_with_Lagrange_derivative}
- \frac{q(x)}{f(x)} + \lambda \, j(x) = 0.
\end{equation}
Multiplying both sides of this equation by $f(x)$ and integrating over $x$ to set the Lagrange multiplier, we find that \Eq{loss_q_j_story_with_Lagrange_derivative} is solved for
\begin{equation}
\lambda = 1, \qquad f(x) = \frac{q(x)}{j(x)},
\end{equation}
so $f(x)$ learns the $q(x)/j(x)$ ratio as desired.

In the special case that $j(x)$ is itself a normalized probability distribution, we can substitute for the Lagrange multiplier and rewrite \Eq{loss_q_j_story_with_Lagrange} in the following form:
\begin{equation}
\label{eq:loss_q_j_story_without_Lagrange}
L[f] = - \int dx \, \Big( q(x) \log f(x) + j(x) (1 -f(x)) \Big).
\end{equation}
Identifying $q(x) = p(x|\thetaup)$ and $j(x) = p(x|\thetadown)$, this is precisely the MLC loss in \Eq{maxlikeloss}.
Therefore, we have an intuitive understanding of the MLC loss as trying to maximize the (log) likelihood of $f(x)$ with respect to $p(x|\thetaup)$, subject to the constraint that $f(x) \, p(x|\thetadown)$ is a proper probability distribution.

In \Fig{MLC}, we plot the learned likelihood ratio between the two Gaussian samples from \Fig{Gaussian-a}, comparing the performance of MLC against binary cross entropy and the exact analytic expression.  In all cases, a network is trained with 100 epochs and early stopping with a patience of 10 epochs.
We also compare the MLC loss against the $C(f) = \exp f$ variant discussed in footnote~\ref{footnote:modMLC}.
We see that both the linear (i.e.~$C(f) = f$) and exponential parametrizations perform similarly in the region with ample data.
That said, the exponential parametrization has a more robust extrapolation towards the edges, yielding similar behavior to binary cross entropy.  
Note that the exponential parametrization of the MLC loss was used in \Ref{DAgnolo:2018cun}.

\bibliography{myrefs}

%apsrev4-2.bst 2019-01-14 (MD) hand-edited version of apsrev4-1.bst
%Control: key (0)
%Control: author (8) initials jnrlst
%Control: editor formatted (1) identically to author
%Control: production of article title (0) allowed
%Control: page (0) single
%Control: year (1) truncated
%Control: production of eprint (0) enabled
\begin{thebibliography}{71}%
\makeatletter
\providecommand \@ifxundefined [1]{%
 \@ifx{#1\undefined}
}%
\providecommand \@ifnum [1]{%
 \ifnum #1\expandafter \@firstoftwo
 \else \expandafter \@secondoftwo
 \fi
}%
\providecommand \@ifx [1]{%
 \ifx #1\expandafter \@firstoftwo
 \else \expandafter \@secondoftwo
 \fi
}%
\providecommand \natexlab [1]{#1}%
\providecommand \enquote  [1]{``#1''}%
\providecommand \bibnamefont  [1]{#1}%
\providecommand \bibfnamefont [1]{#1}%
\providecommand \citenamefont [1]{#1}%
\providecommand \href@noop [0]{\@secondoftwo}%
\providecommand \href [0]{\begingroup \@sanitize@url \@href}%
\providecommand \@href[1]{\@@startlink{#1}\@@href}%
\providecommand \@@href[1]{\endgroup#1\@@endlink}%
\providecommand \@sanitize@url [0]{\catcode `\\12\catcode `\$12\catcode
  `\&12\catcode `\#12\catcode `\^12\catcode `\_12\catcode `\%12\relax}%
\providecommand \@@startlink[1]{}%
\providecommand \@@endlink[0]{}%
\providecommand \url  [0]{\begingroup\@sanitize@url \@url }%
\providecommand \@url [1]{\endgroup\@href {#1}{\urlprefix }}%
\providecommand \urlprefix  [0]{URL }%
\providecommand \Eprint [0]{\href }%
\providecommand \doibase [0]{https://doi.org/}%
\providecommand \selectlanguage [0]{\@gobble}%
\providecommand \bibinfo  [0]{\@secondoftwo}%
\providecommand \bibfield  [0]{\@secondoftwo}%
\providecommand \translation [1]{[#1]}%
\providecommand \BibitemOpen [0]{}%
\providecommand \bibitemStop [0]{}%
\providecommand \bibitemNoStop [0]{.\EOS\space}%
\providecommand \EOS [0]{\spacefactor3000\relax}%
\providecommand \BibitemShut  [1]{\csname bibitem#1\endcsname}%
\let\auto@bib@innerbib\@empty
%</preamble>
\bibitem [{\citenamefont {Larkoski}\ \emph {et~al.}(2020)\citenamefont
  {Larkoski}, \citenamefont {Moult},\ and\ \citenamefont
  {Nachman}}]{Larkoski:2017jix}%
  \BibitemOpen
  \bibfield  {author} {\bibinfo {author} {\bibfnamefont {A.~J.}\ \bibnamefont
  {Larkoski}}, \bibinfo {author} {\bibfnamefont {I.}~\bibnamefont {Moult}},\
  and\ \bibinfo {author} {\bibfnamefont {B.}~\bibnamefont {Nachman}},\
  }\bibfield  {title} {\bibinfo {title} {{Jet Substructure at the Large Hadron
  Collider: A Review of Recent Advances in Theory and Machine Learning}},\
  }\href {https://doi.org/10.1016/j.physrep.2019.11.001} {\bibfield  {journal}
  {\bibinfo  {journal} {Phys. Rept.}\ }\textbf {\bibinfo {volume} {841}},\
  \bibinfo {pages} {1} (\bibinfo {year} {2020})},\ \Eprint
  {https://arxiv.org/abs/1709.04464} {arXiv:1709.04464 [hep-ph]} \BibitemShut
  {NoStop}%
\bibitem [{\citenamefont {Guest}\ \emph {et~al.}(2018)\citenamefont {Guest},
  \citenamefont {Cranmer},\ and\ \citenamefont {Whiteson}}]{Guest:2018yhq}%
  \BibitemOpen
  \bibfield  {author} {\bibinfo {author} {\bibfnamefont {D.}~\bibnamefont
  {Guest}}, \bibinfo {author} {\bibfnamefont {K.}~\bibnamefont {Cranmer}},\
  and\ \bibinfo {author} {\bibfnamefont {D.}~\bibnamefont {Whiteson}},\
  }\bibfield  {title} {\bibinfo {title} {{Deep Learning and its Application to
  LHC Physics}},\ }\href {https://doi.org/10.1146/annurev-nucl-101917-021019}
  {\bibfield  {journal} {\bibinfo  {journal} {Ann. Rev. Nucl. Part. Sci.}\
  }\textbf {\bibinfo {volume} {68}},\ \bibinfo {pages} {161} (\bibinfo {year}
  {2018})},\ \Eprint {https://arxiv.org/abs/1806.11484} {arXiv:1806.11484
  [hep-ex]} \BibitemShut {NoStop}%
\bibitem [{\citenamefont {Albertsson}\ \emph {et~al.}(2018)\citenamefont
  {Albertsson} \emph {et~al.}}]{Albertsson:2018maf}%
  \BibitemOpen
  \bibfield  {author} {\bibinfo {author} {\bibfnamefont {K.}~\bibnamefont
  {Albertsson}} \emph {et~al.},\ }\bibfield  {title} {\bibinfo {title}
  {{Machine Learning in High Energy Physics Community White Paper}},\
  }\href@noop {} {\  (\bibinfo {year} {2018})},\ \Eprint
  {https://arxiv.org/abs/1807.02876} {arXiv:1807.02876 [physics.comp-ph]}
  \BibitemShut {NoStop}%
%%CITATION = ARXIV:1807.02876;%%
\bibitem [{\citenamefont {Radovic}\ \emph {et~al.}(2018)\citenamefont {Radovic}
  \emph {et~al.}}]{Radovic:2018dip}%
  \BibitemOpen
  \bibfield  {author} {\bibinfo {author} {\bibfnamefont {A.}~\bibnamefont
  {Radovic}} \emph {et~al.},\ }\bibfield  {title} {\bibinfo {title} {{Machine
  learning at the energy and intensity frontiers of particle physics}},\ }\href
  {https://doi.org/10.1038/s41586-018-0361-2} {\bibfield  {journal} {\bibinfo
  {journal} {Nature}\ }\textbf {\bibinfo {volume} {560}},\ \bibinfo {pages}
  {41} (\bibinfo {year} {2018})}\BibitemShut {NoStop}%
\bibitem [{\citenamefont {Bourilkov}(2020)}]{Bourilkov:2019yoi}%
  \BibitemOpen
  \bibfield  {author} {\bibinfo {author} {\bibfnamefont {D.}~\bibnamefont
  {Bourilkov}},\ }\bibfield  {title} {\bibinfo {title} {{Machine and Deep
  Learning Applications in Particle Physics}},\ }\href
  {https://doi.org/10.1142/S0217751X19300199} {\bibfield  {journal} {\bibinfo
  {journal} {Int. J. Mod. Phys. A}\ }\textbf {\bibinfo {volume} {34}},\
  \bibinfo {pages} {1930019} (\bibinfo {year} {2020})},\ \Eprint
  {https://arxiv.org/abs/1912.08245} {arXiv:1912.08245 [physics.data-an]}
  \BibitemShut {NoStop}%
\bibitem [{\citenamefont {{HEP ML Community}}()}]{hepmllivingreview}%
  \BibitemOpen
  \bibfield  {author} {\bibinfo {author} {\bibnamefont {{HEP ML Community}}},\
  }\href {https://iml-wg.github.io/HEPML-LivingReview/} {\bibinfo {title} {{A
  Living Review of Machine Learning for Particle Physics}}}\BibitemShut
  {NoStop}%
\bibitem [{\citenamefont {Lai}(2018)}]{Lai:2018ixk}%
  \BibitemOpen
  \bibfield  {author} {\bibinfo {author} {\bibfnamefont {Y.~S.}\ \bibnamefont
  {Lai}},\ }\bibfield  {title} {\bibinfo {title} {{Automated Discovery of Jet
  Substructure Analyses}},\ }\href@noop {} {\  (\bibinfo {year} {2018})},\
  \Eprint {https://arxiv.org/abs/1810.00835} {arXiv:1810.00835 [nucl-th]}
  \BibitemShut {NoStop}%
\bibitem [{\citenamefont {Khosa}\ \emph {et~al.}(2019)\citenamefont {Khosa},
  \citenamefont {Sanz},\ and\ \citenamefont {Soughton}}]{Khosa:2019kxd}%
  \BibitemOpen
  \bibfield  {author} {\bibinfo {author} {\bibfnamefont {C.~K.}\ \bibnamefont
  {Khosa}}, \bibinfo {author} {\bibfnamefont {V.}~\bibnamefont {Sanz}},\ and\
  \bibinfo {author} {\bibfnamefont {M.}~\bibnamefont {Soughton}},\ }\bibfield
  {title} {\bibinfo {title} {{Using Machine Learning to disentangle LHC
  signatures of Dark Matter candidates}},\ }\href@noop {} {\  (\bibinfo {year}
  {2019})},\ \Eprint {https://arxiv.org/abs/1910.06058} {arXiv:1910.06058
  [hep-ph]} \BibitemShut {NoStop}%
\bibitem [{\citenamefont {Du}\ \emph {et~al.}(2020)\citenamefont {Du},
  \citenamefont {Zhou}, \citenamefont {Steinheimer}, \citenamefont {Pang},
  \citenamefont {Motornenko}, \citenamefont {Zong}, \citenamefont {Wang},\ and\
  \citenamefont {St\"ocker}}]{Du:2019civ}%
  \BibitemOpen
  \bibfield  {author} {\bibinfo {author} {\bibfnamefont {Y.-L.}\ \bibnamefont
  {Du}}, \bibinfo {author} {\bibfnamefont {K.}~\bibnamefont {Zhou}}, \bibinfo
  {author} {\bibfnamefont {J.}~\bibnamefont {Steinheimer}}, \bibinfo {author}
  {\bibfnamefont {L.-G.}\ \bibnamefont {Pang}}, \bibinfo {author}
  {\bibfnamefont {A.}~\bibnamefont {Motornenko}}, \bibinfo {author}
  {\bibfnamefont {H.-S.}\ \bibnamefont {Zong}}, \bibinfo {author}
  {\bibfnamefont {X.-N.}\ \bibnamefont {Wang}},\ and\ \bibinfo {author}
  {\bibfnamefont {H.}~\bibnamefont {St\"ocker}},\ }\bibfield  {title} {\bibinfo
  {title} {{Identifying the nature of the QCD transition in relativistic
  collision of heavy nuclei with deep learning}},\ }\href
  {https://doi.org/10.1140/epjc/s10052-020-8030-7} {\bibfield  {journal}
  {\bibinfo  {journal} {Eur. Phys. J. C}\ }\textbf {\bibinfo {volume} {80}},\
  \bibinfo {pages} {516} (\bibinfo {year} {2020})},\ \Eprint
  {https://arxiv.org/abs/1910.11530} {arXiv:1910.11530 [hep-ph]} \BibitemShut
  {NoStop}%
\bibitem [{\citenamefont {Mullin}\ \emph {et~al.}(2019)\citenamefont {Mullin},
  \citenamefont {Pacey}, \citenamefont {Parker}, \citenamefont {White},\ and\
  \citenamefont {Williams}}]{Mullin:2019mmh}%
  \BibitemOpen
  \bibfield  {author} {\bibinfo {author} {\bibfnamefont {A.}~\bibnamefont
  {Mullin}}, \bibinfo {author} {\bibfnamefont {H.}~\bibnamefont {Pacey}},
  \bibinfo {author} {\bibfnamefont {M.}~\bibnamefont {Parker}}, \bibinfo
  {author} {\bibfnamefont {M.}~\bibnamefont {White}},\ and\ \bibinfo {author}
  {\bibfnamefont {S.}~\bibnamefont {Williams}},\ }\bibfield  {title} {\bibinfo
  {title} {{Does SUSY have friends? A new approach for LHC event analysis}},\
  }\href@noop {} {\  (\bibinfo {year} {2019})},\ \Eprint
  {https://arxiv.org/abs/1912.10625} {arXiv:1912.10625 [hep-ph]} \BibitemShut
  {NoStop}%
\bibitem [{\citenamefont {Chang}\ \emph {et~al.}(2020)\citenamefont {Chang},
  \citenamefont {Chen},\ and\ \citenamefont {Chiang}}]{Chang:2020rtc}%
  \BibitemOpen
  \bibfield  {author} {\bibinfo {author} {\bibfnamefont {S.}~\bibnamefont
  {Chang}}, \bibinfo {author} {\bibfnamefont {T.-K.}\ \bibnamefont {Chen}},\
  and\ \bibinfo {author} {\bibfnamefont {C.-W.}\ \bibnamefont {Chiang}},\
  }\bibfield  {title} {\bibinfo {title} {{Distinguishing $W'$ Signals at Hadron
  Colliders Using Neural Networks}},\ }\href@noop {} {\  (\bibinfo {year}
  {2020})},\ \Eprint {https://arxiv.org/abs/2007.14586} {arXiv:2007.14586
  [hep-ph]} \BibitemShut {NoStop}%
\bibitem [{\citenamefont {Flesher}\ \emph {et~al.}(2020)\citenamefont
  {Flesher}, \citenamefont {Fraser}, \citenamefont {Hutchison}, \citenamefont
  {Ostdiek},\ and\ \citenamefont {Schwartz}}]{Flesher:2020kuy}%
  \BibitemOpen
  \bibfield  {author} {\bibinfo {author} {\bibfnamefont {F.}~\bibnamefont
  {Flesher}}, \bibinfo {author} {\bibfnamefont {K.}~\bibnamefont {Fraser}},
  \bibinfo {author} {\bibfnamefont {C.}~\bibnamefont {Hutchison}}, \bibinfo
  {author} {\bibfnamefont {B.}~\bibnamefont {Ostdiek}},\ and\ \bibinfo {author}
  {\bibfnamefont {M.~D.}\ \bibnamefont {Schwartz}},\ }\bibfield  {title}
  {\bibinfo {title} {{Parameter Inference from Event Ensembles and the
  Top-Quark Mass}},\ }\href@noop {} {\  (\bibinfo {year} {2020})},\ \Eprint
  {https://arxiv.org/abs/2011.04666} {arXiv:2011.04666 [hep-ph]} \BibitemShut
  {NoStop}%
\bibitem [{\citenamefont {Lazzarin}\ \emph {et~al.}(2020)\citenamefont
  {Lazzarin}, \citenamefont {Alioli},\ and\ \citenamefont
  {Carrazza}}]{Lazzarin:2020uvv}%
  \BibitemOpen
  \bibfield  {author} {\bibinfo {author} {\bibfnamefont {M.}~\bibnamefont
  {Lazzarin}}, \bibinfo {author} {\bibfnamefont {S.}~\bibnamefont {Alioli}},\
  and\ \bibinfo {author} {\bibfnamefont {S.}~\bibnamefont {Carrazza}},\
  }\bibfield  {title} {\bibinfo {title} {{MCNNTUNES: tuning Shower Monte Carlo
  generators with machine learning}},\ }\href@noop {} {\  (\bibinfo {year}
  {2020})},\ \Eprint {https://arxiv.org/abs/2010.02213} {arXiv:2010.02213
  [physics.comp-ph]} \BibitemShut {NoStop}%
\bibitem [{\citenamefont {Lai}\ \emph {et~al.}(2020)\citenamefont {Lai},
  \citenamefont {Neill}, \citenamefont {P\l{}osko\'n},\ and\ \citenamefont
  {Ringer}}]{Lai:2020byl}%
  \BibitemOpen
  \bibfield  {author} {\bibinfo {author} {\bibfnamefont {Y.~S.}\ \bibnamefont
  {Lai}}, \bibinfo {author} {\bibfnamefont {D.}~\bibnamefont {Neill}}, \bibinfo
  {author} {\bibfnamefont {M.}~\bibnamefont {P\l{}osko\'n}},\ and\ \bibinfo
  {author} {\bibfnamefont {F.}~\bibnamefont {Ringer}},\ }\bibfield  {title}
  {\bibinfo {title} {{Explainable machine learning of the underlying physics of
  high-energy particle collisions}},\ }\href@noop {} {\  (\bibinfo {year}
  {2020})},\ \Eprint {https://arxiv.org/abs/2012.06582} {arXiv:2012.06582
  [hep-ph]} \BibitemShut {NoStop}%
\bibitem [{\citenamefont {Neyman}\ and\ \citenamefont
  {Pearson}(1933)}]{neyman1933ix}%
  \BibitemOpen
  \bibfield  {author} {\bibinfo {author} {\bibfnamefont {J.}~\bibnamefont
  {Neyman}}\ and\ \bibinfo {author} {\bibfnamefont {E.~S.}\ \bibnamefont
  {Pearson}},\ }\bibfield  {title} {\bibinfo {title} {On the problem of the
  most efficient tests of statistical hypotheses},\ }\href@noop {} {\bibfield
  {journal} {\bibinfo  {journal} {Phil. Trans. R. Soc. Lond. A}\ }\textbf
  {\bibinfo {volume} {231}},\ \bibinfo {pages} {289} (\bibinfo {year}
  {1933})}\BibitemShut {NoStop}%
\bibitem [{\citenamefont {Hastie}\ \emph {et~al.}(2001)\citenamefont {Hastie},
  \citenamefont {Tibshirani},\ and\ \citenamefont
  {Friedman}}]{hastie01statisticallearning}%
  \BibitemOpen
  \bibfield  {author} {\bibinfo {author} {\bibfnamefont {T.}~\bibnamefont
  {Hastie}}, \bibinfo {author} {\bibfnamefont {R.}~\bibnamefont {Tibshirani}},\
  and\ \bibinfo {author} {\bibfnamefont {J.}~\bibnamefont {Friedman}},\
  }\href@noop {} {\emph {\bibinfo {title} {The Elements of Statistical
  Learning}}},\ Springer Series in Statistics\ (\bibinfo  {publisher} {Springer
  New York Inc.},\ \bibinfo {address} {New York, NY, USA},\ \bibinfo {year}
  {2001})\BibitemShut {NoStop}%
\bibitem [{\citenamefont {Sugiyama}\ \emph {et~al.}(2012)\citenamefont
  {Sugiyama}, \citenamefont {Suzuki},\ and\ \citenamefont
  {Kanamori}}]{sugiyama_suzuki_kanamori_2012}%
  \BibitemOpen
  \bibfield  {author} {\bibinfo {author} {\bibfnamefont {M.}~\bibnamefont
  {Sugiyama}}, \bibinfo {author} {\bibfnamefont {T.}~\bibnamefont {Suzuki}},\
  and\ \bibinfo {author} {\bibfnamefont {T.}~\bibnamefont {Kanamori}},\ }\href
  {https://doi.org/10.1017/CBO9781139035613} {\emph {\bibinfo {title} {Density
  Ratio Estimation in Machine Learning}}}\ (\bibinfo  {publisher} {Cambridge
  University Press},\ \bibinfo {year} {2012})\BibitemShut {NoStop}%
\bibitem [{\citenamefont {{A. Andreassen, S. Hsu, B. Nachman, N. Suaysom, A.
  Suresh}}(2020)}]{2010.03569}%
  \BibitemOpen
  \bibfield  {author} {\bibinfo {author} {\bibnamefont {{A. Andreassen, S. Hsu,
  B. Nachman, N. Suaysom, A. Suresh}}},\ }\bibfield  {title} {\bibinfo {title}
  {{Parameter Estimation using Neural Networks in the Presence of Detector
  Effects}},\ }\href@noop {} {\  (\bibinfo {year} {2020})},\ \Eprint
  {https://arxiv.org/abs/2010.03569} {arXiv:2010.03569 [hep-ph]} \BibitemShut
  {NoStop}%
\bibitem [{\citenamefont {Andreassen}\ and\ \citenamefont
  {Nachman}(2020)}]{1907.08209}%
  \BibitemOpen
  \bibfield  {author} {\bibinfo {author} {\bibfnamefont {A.}~\bibnamefont
  {Andreassen}}\ and\ \bibinfo {author} {\bibfnamefont {B.}~\bibnamefont
  {Nachman}},\ }\bibfield  {title} {\bibinfo {title} {{Neural Networks for Full
  Phase-space Reweighting and Parameter Tuning}},\ }\href
  {https://doi.org/10.1103/PhysRevD.101.091901} {\bibfield  {journal} {\bibinfo
   {journal} {Phys. Rev. D}\ }\textbf {\bibinfo {volume} {101}},\ \bibinfo
  {pages} {091901(R)} (\bibinfo {year} {2020})},\ \Eprint
  {https://arxiv.org/abs/1907.08209} {arXiv:1907.08209 [hep-ph]} \BibitemShut
  {NoStop}%
\bibitem [{\citenamefont {Stoye}\ \emph {et~al.}(2018)\citenamefont {Stoye},
  \citenamefont {Brehmer}, \citenamefont {Louppe}, \citenamefont {Pavez},\ and\
  \citenamefont {Cranmer}}]{Stoye:2018ovl}%
  \BibitemOpen
  \bibfield  {author} {\bibinfo {author} {\bibfnamefont {M.}~\bibnamefont
  {Stoye}}, \bibinfo {author} {\bibfnamefont {J.}~\bibnamefont {Brehmer}},
  \bibinfo {author} {\bibfnamefont {G.}~\bibnamefont {Louppe}}, \bibinfo
  {author} {\bibfnamefont {J.}~\bibnamefont {Pavez}},\ and\ \bibinfo {author}
  {\bibfnamefont {K.}~\bibnamefont {Cranmer}},\ }\bibfield  {title} {\bibinfo
  {title} {{Likelihood-free inference with an improved cross-entropy
  estimator}},\ }\href@noop {} {\  (\bibinfo {year} {2018})},\ \Eprint
  {https://arxiv.org/abs/1808.00973} {arXiv:1808.00973 [stat.ML]} \BibitemShut
  {NoStop}%
%%CITATION = ARXIV:1808.00973;%%
\bibitem [{\citenamefont {Hollingsworth}\ and\ \citenamefont
  {Whiteson}(2020)}]{Hollingsworth:2020kjg}%
  \BibitemOpen
  \bibfield  {author} {\bibinfo {author} {\bibfnamefont {J.}~\bibnamefont
  {Hollingsworth}}\ and\ \bibinfo {author} {\bibfnamefont {D.}~\bibnamefont
  {Whiteson}},\ }\bibfield  {title} {\bibinfo {title} {{Resonance Searches with
  Machine Learned Likelihood Ratios}},\ }\href@noop {} {\  (\bibinfo {year}
  {2020})},\ \Eprint {https://arxiv.org/abs/2002.04699} {arXiv:2002.04699
  [hep-ph]} \BibitemShut {NoStop}%
%%CITATION = ARXIV:2002.04699;%%
\bibitem [{\citenamefont {Brehmer}\ \emph
  {et~al.}(2018{\natexlab{a}})\citenamefont {Brehmer}, \citenamefont {Cranmer},
  \citenamefont {Louppe},\ and\ \citenamefont {Pavez}}]{Brehmer:2018kdj}%
  \BibitemOpen
  \bibfield  {author} {\bibinfo {author} {\bibfnamefont {J.}~\bibnamefont
  {Brehmer}}, \bibinfo {author} {\bibfnamefont {K.}~\bibnamefont {Cranmer}},
  \bibinfo {author} {\bibfnamefont {G.}~\bibnamefont {Louppe}},\ and\ \bibinfo
  {author} {\bibfnamefont {J.}~\bibnamefont {Pavez}},\ }\bibfield  {title}
  {\bibinfo {title} {{Constraining Effective Field Theories with Machine
  Learning}},\ }\href {https://doi.org/10.1103/PhysRevLett.121.111801}
  {\bibfield  {journal} {\bibinfo  {journal} {Phys. Rev. Lett.}\ }\textbf
  {\bibinfo {volume} {121}},\ \bibinfo {pages} {111801} (\bibinfo {year}
  {2018}{\natexlab{a}})},\ \Eprint {https://arxiv.org/abs/1805.00013}
  {arXiv:1805.00013 [hep-ph]} \BibitemShut {NoStop}%
\bibitem [{\citenamefont {Brehmer}\ \emph
  {et~al.}(2018{\natexlab{b}})\citenamefont {Brehmer}, \citenamefont {Cranmer},
  \citenamefont {Louppe},\ and\ \citenamefont {Pavez}}]{Brehmer:2018eca}%
  \BibitemOpen
  \bibfield  {author} {\bibinfo {author} {\bibfnamefont {J.}~\bibnamefont
  {Brehmer}}, \bibinfo {author} {\bibfnamefont {K.}~\bibnamefont {Cranmer}},
  \bibinfo {author} {\bibfnamefont {G.}~\bibnamefont {Louppe}},\ and\ \bibinfo
  {author} {\bibfnamefont {J.}~\bibnamefont {Pavez}},\ }\bibfield  {title}
  {\bibinfo {title} {{A Guide to Constraining Effective Field Theories with
  Machine Learning}},\ }\href {https://doi.org/10.1103/PhysRevD.98.052004}
  {\bibfield  {journal} {\bibinfo  {journal} {Phys. Rev. D}\ }\textbf {\bibinfo
  {volume} {98}},\ \bibinfo {pages} {052004} (\bibinfo {year}
  {2018}{\natexlab{b}})},\ \Eprint {https://arxiv.org/abs/1805.00020}
  {arXiv:1805.00020 [hep-ph]} \BibitemShut {NoStop}%
\bibitem [{\citenamefont {Brehmer}\ \emph
  {et~al.}(2020{\natexlab{a}})\citenamefont {Brehmer}, \citenamefont {Kling},
  \citenamefont {Espejo},\ and\ \citenamefont {Cranmer}}]{Brehmer:2019xox}%
  \BibitemOpen
  \bibfield  {author} {\bibinfo {author} {\bibfnamefont {J.}~\bibnamefont
  {Brehmer}}, \bibinfo {author} {\bibfnamefont {F.}~\bibnamefont {Kling}},
  \bibinfo {author} {\bibfnamefont {I.}~\bibnamefont {Espejo}},\ and\ \bibinfo
  {author} {\bibfnamefont {K.}~\bibnamefont {Cranmer}},\ }\bibfield  {title}
  {\bibinfo {title} {{MadMiner: Machine learning-based inference for particle
  physics}},\ }\href {https://doi.org/10.1007/s41781-020-0035-2} {\bibfield
  {journal} {\bibinfo  {journal} {Comput. Softw. Big Sci.}\ }\textbf {\bibinfo
  {volume} {4}},\ \bibinfo {pages} {3} (\bibinfo {year}
  {2020}{\natexlab{a}})},\ \Eprint {https://arxiv.org/abs/1907.10621}
  {arXiv:1907.10621 [hep-ph]} \BibitemShut {NoStop}%
\bibitem [{\citenamefont {Brehmer}\ \emph
  {et~al.}(2020{\natexlab{b}})\citenamefont {Brehmer}, \citenamefont {Louppe},
  \citenamefont {Pavez},\ and\ \citenamefont {Cranmer}}]{Brehmer:2018hga}%
  \BibitemOpen
  \bibfield  {author} {\bibinfo {author} {\bibfnamefont {J.}~\bibnamefont
  {Brehmer}}, \bibinfo {author} {\bibfnamefont {G.}~\bibnamefont {Louppe}},
  \bibinfo {author} {\bibfnamefont {J.}~\bibnamefont {Pavez}},\ and\ \bibinfo
  {author} {\bibfnamefont {K.}~\bibnamefont {Cranmer}},\ }\bibfield  {title}
  {\bibinfo {title} {{Mining gold from implicit models to improve
  likelihood-free inference}},\ }\href
  {https://doi.org/10.1073/pnas.1915980117} {\bibfield  {journal} {\bibinfo
  {journal} {Proc. Nat. Acad. Sci.}\ ,\ \bibinfo {pages} {201915980}} (\bibinfo
  {year} {2020}{\natexlab{b}})},\ \Eprint {https://arxiv.org/abs/1805.12244}
  {arXiv:1805.12244 [stat.ML]} \BibitemShut {NoStop}%
\bibitem [{\citenamefont {Cranmer}\ \emph {et~al.}(2015)\citenamefont
  {Cranmer}, \citenamefont {Pavez},\ and\ \citenamefont
  {Louppe}}]{Cranmer:2015bka}%
  \BibitemOpen
  \bibfield  {author} {\bibinfo {author} {\bibfnamefont {K.}~\bibnamefont
  {Cranmer}}, \bibinfo {author} {\bibfnamefont {J.}~\bibnamefont {Pavez}},\
  and\ \bibinfo {author} {\bibfnamefont {G.}~\bibnamefont {Louppe}},\
  }\bibfield  {title} {\bibinfo {title} {{Approximating Likelihood Ratios with
  Calibrated Discriminative Classifiers}},\ }\href@noop {} {\  (\bibinfo {year}
  {2015})},\ \Eprint {https://arxiv.org/abs/1506.02169} {arXiv:1506.02169
  [stat.AP]} \BibitemShut {NoStop}%
\bibitem [{\citenamefont {Badiali}\ \emph {et~al.}(2020)\citenamefont
  {Badiali}, \citenamefont {Di~Bello}, \citenamefont {Frattari}, \citenamefont
  {Gross}, \citenamefont {Ippolito}, \citenamefont {Kado},\ and\ \citenamefont
  {Shlomi}}]{Badiali:2020wal}%
  \BibitemOpen
  \bibfield  {author} {\bibinfo {author} {\bibfnamefont {C.}~\bibnamefont
  {Badiali}}, \bibinfo {author} {\bibfnamefont {F.}~\bibnamefont {Di~Bello}},
  \bibinfo {author} {\bibfnamefont {G.}~\bibnamefont {Frattari}}, \bibinfo
  {author} {\bibfnamefont {E.}~\bibnamefont {Gross}}, \bibinfo {author}
  {\bibfnamefont {V.}~\bibnamefont {Ippolito}}, \bibinfo {author}
  {\bibfnamefont {M.}~\bibnamefont {Kado}},\ and\ \bibinfo {author}
  {\bibfnamefont {J.}~\bibnamefont {Shlomi}},\ }\bibfield  {title} {\bibinfo
  {title} {{Efficiency Parameterization with Neural Networks}},\ }\href@noop {}
  {\  (\bibinfo {year} {2020})},\ \Eprint {https://arxiv.org/abs/2004.02665}
  {arXiv:2004.02665 [hep-ex]} \BibitemShut {NoStop}%
\bibitem [{\citenamefont {Andreassen}\ \emph
  {et~al.}(2020{\natexlab{a}})\citenamefont {Andreassen}, \citenamefont
  {Nachman},\ and\ \citenamefont {Shih}}]{Andreassen:2020nkr}%
  \BibitemOpen
  \bibfield  {author} {\bibinfo {author} {\bibfnamefont {A.}~\bibnamefont
  {Andreassen}}, \bibinfo {author} {\bibfnamefont {B.}~\bibnamefont
  {Nachman}},\ and\ \bibinfo {author} {\bibfnamefont {D.}~\bibnamefont
  {Shih}},\ }\bibfield  {title} {\bibinfo {title} {{Simulation Assisted
  Likelihood-free Anomaly Detection}},\ }\href
  {https://doi.org/10.1103/PhysRevD.101.095004} {\bibfield  {journal} {\bibinfo
   {journal} {Phys. Rev. D}\ }\textbf {\bibinfo {volume} {101}},\ \bibinfo
  {pages} {095004} (\bibinfo {year} {2020}{\natexlab{a}})},\ \Eprint
  {https://arxiv.org/abs/2001.05001} {arXiv:2001.05001 [hep-ph]} \BibitemShut
  {NoStop}%
\bibitem [{\citenamefont {Andreassen}\ \emph
  {et~al.}(2020{\natexlab{b}})\citenamefont {Andreassen}, \citenamefont
  {Komiske}, \citenamefont {Metodiev}, \citenamefont {Nachman},\ and\
  \citenamefont {Thaler}}]{Andreassen:2019cjw}%
  \BibitemOpen
  \bibfield  {author} {\bibinfo {author} {\bibfnamefont {A.}~\bibnamefont
  {Andreassen}}, \bibinfo {author} {\bibfnamefont {P.~T.}\ \bibnamefont
  {Komiske}}, \bibinfo {author} {\bibfnamefont {E.~M.}\ \bibnamefont
  {Metodiev}}, \bibinfo {author} {\bibfnamefont {B.}~\bibnamefont {Nachman}},\
  and\ \bibinfo {author} {\bibfnamefont {J.}~\bibnamefont {Thaler}},\
  }\bibfield  {title} {\bibinfo {title} {{OmniFold: A Method to Simultaneously
  Unfold All Observables}},\ }\href
  {https://doi.org/10.1103/PhysRevLett.124.182001} {\bibfield  {journal}
  {\bibinfo  {journal} {Phys. Rev. Lett.}\ }\textbf {\bibinfo {volume} {124}},\
  \bibinfo {pages} {182001} (\bibinfo {year} {2020}{\natexlab{b}})},\ \Eprint
  {https://arxiv.org/abs/1911.09107} {arXiv:1911.09107 [hep-ph]} \BibitemShut
  {NoStop}%
\bibitem [{\citenamefont {Erdmann}\ \emph {et~al.}(2019)\citenamefont
  {Erdmann}, \citenamefont {Fischer}, \citenamefont {Noll}, \citenamefont
  {Rath}, \citenamefont {Rieger},\ and\ \citenamefont
  {Schmidt}}]{Fischer-ACAT2019}%
  \BibitemOpen
  \bibfield  {author} {\bibinfo {author} {\bibfnamefont {M.}~\bibnamefont
  {Erdmann}}, \bibinfo {author} {\bibfnamefont {B.}~\bibnamefont {Fischer}},
  \bibinfo {author} {\bibfnamefont {D.}~\bibnamefont {Noll}}, \bibinfo {author}
  {\bibfnamefont {Y.}~\bibnamefont {Rath}}, \bibinfo {author} {\bibfnamefont
  {M.}~\bibnamefont {Rieger}},\ and\ \bibinfo {author} {\bibfnamefont
  {D.}~\bibnamefont {Schmidt}},\ }\bibfield  {title} {\bibinfo {title}
  {{Adversarial Neural Network-based data-simulation corrections for
  jet-tagging at CMS}},\ }in\ \href@noop {} {\emph {\bibinfo {booktitle}
  {{Proc. 19th Int. Workshop on Adv. Comp., Anal. Techn. in Phys. Research,
  ACAT2019}}}}\ (\bibinfo {year} {2019})\BibitemShut {NoStop}%
\bibitem [{\citenamefont {{Nguyen}}\ \emph {et~al.}(2005)\citenamefont
  {{Nguyen}}, \citenamefont {{Wainwright}},\ and\ \citenamefont
  {{Jordan}}}]{2005math.....10521N}%
  \BibitemOpen
  \bibfield  {author} {\bibinfo {author} {\bibfnamefont {X.}~\bibnamefont
  {{Nguyen}}}, \bibinfo {author} {\bibfnamefont {M.~J.}\ \bibnamefont
  {{Wainwright}}},\ and\ \bibinfo {author} {\bibfnamefont {M.~I.}\ \bibnamefont
  {{Jordan}}},\ }\bibfield  {title} {\bibinfo {title} {{On surrogate loss
  functions and $f$-divergences}},\ }\href@noop {} {\bibfield  {journal}
  {\bibinfo  {journal} {arXiv Mathematics e-prints}\ ,\ \bibinfo {eid}
  {math/0510521}} (\bibinfo {year} {2005})},\ \Eprint
  {https://arxiv.org/abs/math/0510521} {arXiv:math/0510521 [math.ST]}
  \BibitemShut {NoStop}%
\bibitem [{\citenamefont {D'Agnolo}\ and\ \citenamefont
  {Wulzer}(2019)}]{DAgnolo:2018cun}%
  \BibitemOpen
  \bibfield  {author} {\bibinfo {author} {\bibfnamefont {R.~T.}\ \bibnamefont
  {D'Agnolo}}\ and\ \bibinfo {author} {\bibfnamefont {A.}~\bibnamefont
  {Wulzer}},\ }\bibfield  {title} {\bibinfo {title} {{Learning New Physics from
  a Machine}},\ }\href {https://doi.org/10.1103/PhysRevD.99.015014} {\bibfield
  {journal} {\bibinfo  {journal} {Phys. Rev. D}\ }\textbf {\bibinfo {volume}
  {99}},\ \bibinfo {pages} {015014} (\bibinfo {year} {2019})},\ \Eprint
  {https://arxiv.org/abs/1806.02350} {arXiv:1806.02350 [hep-ph]} \BibitemShut
  {NoStop}%
\bibitem [{\citenamefont {D'Agnolo}\ \emph {et~al.}(2019)\citenamefont
  {D'Agnolo}, \citenamefont {Grosso}, \citenamefont {Pierini}, \citenamefont
  {Wulzer},\ and\ \citenamefont {Zanetti}}]{DAgnolo:2019vbw}%
  \BibitemOpen
  \bibfield  {author} {\bibinfo {author} {\bibfnamefont {R.~T.}\ \bibnamefont
  {D'Agnolo}}, \bibinfo {author} {\bibfnamefont {G.}~\bibnamefont {Grosso}},
  \bibinfo {author} {\bibfnamefont {M.}~\bibnamefont {Pierini}}, \bibinfo
  {author} {\bibfnamefont {A.}~\bibnamefont {Wulzer}},\ and\ \bibinfo {author}
  {\bibfnamefont {M.}~\bibnamefont {Zanetti}},\ }\bibfield  {title} {\bibinfo
  {title} {{Learning Multivariate New Physics}},\ }\href@noop {} {\  (\bibinfo
  {year} {2019})},\ \Eprint {https://arxiv.org/abs/1912.12155}
  {arXiv:1912.12155 [hep-ph]} \BibitemShut {NoStop}%
\bibitem [{\citenamefont {Andreassen}\ \emph {et~al.}(2019)\citenamefont
  {Andreassen}, \citenamefont {Feige}, \citenamefont {Frye},\ and\
  \citenamefont {Schwartz}}]{Andreassen:2018apy}%
  \BibitemOpen
  \bibfield  {author} {\bibinfo {author} {\bibfnamefont {A.}~\bibnamefont
  {Andreassen}}, \bibinfo {author} {\bibfnamefont {I.}~\bibnamefont {Feige}},
  \bibinfo {author} {\bibfnamefont {C.}~\bibnamefont {Frye}},\ and\ \bibinfo
  {author} {\bibfnamefont {M.~D.}\ \bibnamefont {Schwartz}},\ }\bibfield
  {title} {\bibinfo {title} {{JUNIPR: a Framework for Unsupervised Machine
  Learning in Particle Physics}},\ }\href
  {https://doi.org/10.1140/epjc/s10052-019-6607-9} {\bibfield  {journal}
  {\bibinfo  {journal} {Eur. Phys. J. C}\ }\textbf {\bibinfo {volume} {79}},\
  \bibinfo {pages} {102} (\bibinfo {year} {2019})},\ \Eprint
  {https://arxiv.org/abs/1804.09720} {arXiv:1804.09720 [hep-ph]} \BibitemShut
  {NoStop}%
\bibitem [{\citenamefont {Brehmer}\ and\ \citenamefont
  {Cranmer}(2020)}]{brehmer2020flows}%
  \BibitemOpen
  \bibfield  {author} {\bibinfo {author} {\bibfnamefont {J.}~\bibnamefont
  {Brehmer}}\ and\ \bibinfo {author} {\bibfnamefont {K.}~\bibnamefont
  {Cranmer}},\ }\href@noop {} {\bibinfo {title} {Flows for simultaneous
  manifold learning and density estimation}} (\bibinfo {year} {2020}),\ \Eprint
  {https://arxiv.org/abs/2003.13913} {arXiv:2003.13913 [stat.ML]} \BibitemShut
  {NoStop}%
\bibitem [{\citenamefont {Nachman}\ and\ \citenamefont
  {Shih}(2020)}]{Nachman:2020lpy}%
  \BibitemOpen
  \bibfield  {author} {\bibinfo {author} {\bibfnamefont {B.}~\bibnamefont
  {Nachman}}\ and\ \bibinfo {author} {\bibfnamefont {D.}~\bibnamefont {Shih}},\
  }\bibfield  {title} {\bibinfo {title} {{Anomaly Detection with Density
  Estimation}},\ }\href {https://doi.org/10.1103/PhysRevD.101.075042}
  {\bibfield  {journal} {\bibinfo  {journal} {Phys. Rev. D}\ }\textbf {\bibinfo
  {volume} {101}},\ \bibinfo {pages} {075042} (\bibinfo {year} {2020})},\
  \Eprint {https://arxiv.org/abs/2001.04990} {arXiv:2001.04990 [hep-ph]}
  \BibitemShut {NoStop}%
\bibitem [{\citenamefont {Metodiev}\ \emph {et~al.}(2017)\citenamefont
  {Metodiev}, \citenamefont {Nachman},\ and\ \citenamefont
  {Thaler}}]{Metodiev:2017vrx}%
  \BibitemOpen
  \bibfield  {author} {\bibinfo {author} {\bibfnamefont {E.~M.}\ \bibnamefont
  {Metodiev}}, \bibinfo {author} {\bibfnamefont {B.}~\bibnamefont {Nachman}},\
  and\ \bibinfo {author} {\bibfnamefont {J.}~\bibnamefont {Thaler}},\
  }\bibfield  {title} {\bibinfo {title} {{Classification without labels:
  Learning from mixed samples in high energy physics}},\ }\href
  {https://doi.org/10.1007/JHEP10(2017)174} {\bibfield  {journal} {\bibinfo
  {journal} {JHEP}\ }\textbf {\bibinfo {volume} {10}},\ \bibinfo {pages}
  {174}},\ \Eprint {https://arxiv.org/abs/1708.02949} {arXiv:1708.02949
  [hep-ph]} \BibitemShut {NoStop}%
\bibitem [{\citenamefont {Baldi}\ \emph {et~al.}(2016)\citenamefont {Baldi},
  \citenamefont {Cranmer}, \citenamefont {Faucett}, \citenamefont {Sadowski},\
  and\ \citenamefont {Whiteson}}]{Baldi:2016fzo}%
  \BibitemOpen
  \bibfield  {author} {\bibinfo {author} {\bibfnamefont {P.}~\bibnamefont
  {Baldi}}, \bibinfo {author} {\bibfnamefont {K.}~\bibnamefont {Cranmer}},
  \bibinfo {author} {\bibfnamefont {T.}~\bibnamefont {Faucett}}, \bibinfo
  {author} {\bibfnamefont {P.}~\bibnamefont {Sadowski}},\ and\ \bibinfo
  {author} {\bibfnamefont {D.}~\bibnamefont {Whiteson}},\ }\bibfield  {title}
  {\bibinfo {title} {{Parameterized neural networks for high-energy physics}},\
  }\href {https://doi.org/10.1140/epjc/s10052-016-4099-4} {\bibfield  {journal}
  {\bibinfo  {journal} {Eur. Phys. J. C}\ }\textbf {\bibinfo {volume} {76}},\
  \bibinfo {pages} {235} (\bibinfo {year} {2016})},\ \Eprint
  {https://arxiv.org/abs/1601.07913} {arXiv:1601.07913 [hep-ex]} \BibitemShut
  {NoStop}%
\bibitem [{\citenamefont {{S. Cheong, A. Cukierman, B. Nachman, M. Safdari, A.
  Schwartzman}}(2020)}]{1910.03773}%
  \BibitemOpen
  \bibfield  {author} {\bibinfo {author} {\bibnamefont {{S. Cheong, A.
  Cukierman, B. Nachman, M. Safdari, A. Schwartzman}}},\ }\bibfield  {title}
  {\bibinfo {title} {{Parametrizing the Detector Response with Neural
  Networks}},\ }\href {https://doi.org/10.1088/1748-0221/15/01/P01030}
  {\bibfield  {journal} {\bibinfo  {journal} {JINST}\ }\textbf {\bibinfo
  {volume} {15}},\ \bibinfo {pages} {P01030}},\ \Eprint
  {https://arxiv.org/abs/1910.03773} {arXiv:1910.03773 [physics.data-an]}
  \BibitemShut {NoStop}%
\bibitem [{\citenamefont {Fix}\ and\ \citenamefont {Hodges~Jr.}(1951)}]{knn1}%
  \BibitemOpen
  \bibfield  {author} {\bibinfo {author} {\bibfnamefont {E.}~\bibnamefont
  {Fix}}\ and\ \bibinfo {author} {\bibfnamefont {J.~L.}\ \bibnamefont
  {Hodges~Jr.}},\ }\bibfield  {title} {\bibinfo {title} {{Discriminatory
  analysis-nonparametric discrimination: consistency properties}},\ }\href@noop
  {} {\bibfield  {journal} {\bibinfo  {journal} {USAF School of Aviation
  Medicine, Project Number 21-49-004, Report Number 4}\ } (\bibinfo {year}
  {1951})}\BibitemShut {NoStop}%
\bibitem [{\citenamefont {Cover~M.}\ and\ \citenamefont {Hart}(1967)}]{knn2}%
  \BibitemOpen
  \bibfield  {author} {\bibinfo {author} {\bibfnamefont {T.}~\bibnamefont
  {Cover~M.}}\ and\ \bibinfo {author} {\bibfnamefont {P.~E.}\ \bibnamefont
  {Hart}},\ }\bibfield  {title} {\bibinfo {title} {{Nearest neighbor pattern
  classification}},\ }\href@noop {} {\bibfield  {journal} {\bibinfo  {journal}
  {IEEE Transactions on Information Theory}\ }\textbf {\bibinfo {volume}
  {13}},\ \bibinfo {pages} {21} (\bibinfo {year} {1967})}\BibitemShut {NoStop}%
\bibitem [{\citenamefont {Goodfellow}\ \emph {et~al.}(2014)\citenamefont
  {Goodfellow}, \citenamefont {Pouget-Abadie}, \citenamefont {Mirza},
  \citenamefont {Xu}, \citenamefont {Warde-Farley}, \citenamefont {Ozair},
  \citenamefont {Courville},\ and\ \citenamefont {Bengio}}]{NIPS2014_5ca3e9b1}%
  \BibitemOpen
  \bibfield  {author} {\bibinfo {author} {\bibfnamefont {I.}~\bibnamefont
  {Goodfellow}}, \bibinfo {author} {\bibfnamefont {J.}~\bibnamefont
  {Pouget-Abadie}}, \bibinfo {author} {\bibfnamefont {M.}~\bibnamefont
  {Mirza}}, \bibinfo {author} {\bibfnamefont {B.}~\bibnamefont {Xu}}, \bibinfo
  {author} {\bibfnamefont {D.}~\bibnamefont {Warde-Farley}}, \bibinfo {author}
  {\bibfnamefont {S.}~\bibnamefont {Ozair}}, \bibinfo {author} {\bibfnamefont
  {A.}~\bibnamefont {Courville}},\ and\ \bibinfo {author} {\bibfnamefont
  {Y.}~\bibnamefont {Bengio}},\ }\bibfield  {title} {\bibinfo {title}
  {Generative adversarial nets},\ }in\ \href
  {https://proceedings.neurips.cc/paper/2014/file/5ca3e9b122f61f8f06494c97b1afccf3-Paper.pdf}
  {\emph {\bibinfo {booktitle} {Advances in Neural Information Processing
  Systems}}},\ Vol.~\bibinfo {volume} {27},\ \bibinfo {editor} {edited by\
  \bibinfo {editor} {\bibfnamefont {Z.}~\bibnamefont {Ghahramani}}, \bibinfo
  {editor} {\bibfnamefont {M.}~\bibnamefont {Welling}}, \bibinfo {editor}
  {\bibfnamefont {C.}~\bibnamefont {Cortes}}, \bibinfo {editor} {\bibfnamefont
  {N.}~\bibnamefont {Lawrence}},\ and\ \bibinfo {editor} {\bibfnamefont
  {K.~Q.}\ \bibnamefont {Weinberger}}}\ (\bibinfo  {publisher} {Curran
  Associates, Inc.},\ \bibinfo {year} {2014})\ pp.\ \bibinfo {pages}
  {2672--2680}\BibitemShut {NoStop}%
\bibitem [{\citenamefont {{Salimans}}\ \emph {et~al.}(2016)\citenamefont
  {{Salimans}}, \citenamefont {{Goodfellow}}, \citenamefont {{Zaremba}},
  \citenamefont {{Cheung}}, \citenamefont {{Radford}},\ and\ \citenamefont
  {{Chen}}}]{2016arXiv160603498S}%
  \BibitemOpen
  \bibfield  {author} {\bibinfo {author} {\bibfnamefont {T.}~\bibnamefont
  {{Salimans}}}, \bibinfo {author} {\bibfnamefont {I.}~\bibnamefont
  {{Goodfellow}}}, \bibinfo {author} {\bibfnamefont {W.}~\bibnamefont
  {{Zaremba}}}, \bibinfo {author} {\bibfnamefont {V.}~\bibnamefont {{Cheung}}},
  \bibinfo {author} {\bibfnamefont {A.}~\bibnamefont {{Radford}}},\ and\
  \bibinfo {author} {\bibfnamefont {X.}~\bibnamefont {{Chen}}},\ }\bibfield
  {title} {\bibinfo {title} {{Improved Techniques for Training GANs}},\
  }\href@noop {} {\bibfield  {journal} {\bibinfo  {journal} {arXiv e-prints}\
  ,\ \bibinfo {eid} {arXiv:1606.03498}} (\bibinfo {year} {2016})},\ \Eprint
  {https://arxiv.org/abs/1606.03498} {arXiv:1606.03498 [cs.LG]} \BibitemShut
  {NoStop}%
\bibitem [{\citenamefont {Kingma}\ and\ \citenamefont
  {Welling}(2014)}]{kingma2014autoencoding}%
  \BibitemOpen
  \bibfield  {author} {\bibinfo {author} {\bibfnamefont {D.~P.}\ \bibnamefont
  {Kingma}}\ and\ \bibinfo {author} {\bibfnamefont {M.}~\bibnamefont
  {Welling}},\ }\bibfield  {title} {\bibinfo {title} {Auto-encoding variational
  bayes.},\ }in\ \href
  {http://dblp.uni-trier.de/db/conf/iclr/iclr2014.html#KingmaW13} {\emph
  {\bibinfo {booktitle} {ICLR}}},\ \bibinfo {editor} {edited by\ \bibinfo
  {editor} {\bibfnamefont {Y.}~\bibnamefont {Bengio}}\ and\ \bibinfo {editor}
  {\bibfnamefont {Y.}~\bibnamefont {LeCun}}}\ (\bibinfo {year}
  {2014})\BibitemShut {NoStop}%
\bibitem [{\citenamefont {Rezende}\ and\ \citenamefont
  {Mohamed}(2015)}]{pmlr-v37-rezende15}%
  \BibitemOpen
  \bibfield  {author} {\bibinfo {author} {\bibfnamefont {D.}~\bibnamefont
  {Rezende}}\ and\ \bibinfo {author} {\bibfnamefont {S.}~\bibnamefont
  {Mohamed}},\ }\bibfield  {title} {\bibinfo {title} {Variational inference
  with normalizing flows},\ }in\ \href
  {http://proceedings.mlr.press/v37/rezende15.html} {\emph {\bibinfo
  {booktitle} {Proceedings of the 32nd International Conference on Machine
  Learning}}},\ \bibinfo {series} {Proceedings of Machine Learning Research},
  Vol.~\bibinfo {volume} {37},\ \bibinfo {editor} {edited by\ \bibinfo {editor}
  {\bibfnamefont {F.}~\bibnamefont {Bach}}\ and\ \bibinfo {editor}
  {\bibfnamefont {D.}~\bibnamefont {Blei}}}\ (\bibinfo  {publisher} {PMLR},\
  \bibinfo {address} {Lille, France},\ \bibinfo {year} {2015})\ pp.\ \bibinfo
  {pages} {1530--1538}\BibitemShut {NoStop}%
\bibitem [{\citenamefont {Larkoski}\ \emph {et~al.}(2014)\citenamefont
  {Larkoski}, \citenamefont {Thaler},\ and\ \citenamefont
  {Waalewijn}}]{Larkoski:2014pca}%
  \BibitemOpen
  \bibfield  {author} {\bibinfo {author} {\bibfnamefont {A.~J.}\ \bibnamefont
  {Larkoski}}, \bibinfo {author} {\bibfnamefont {J.}~\bibnamefont {Thaler}},\
  and\ \bibinfo {author} {\bibfnamefont {W.~J.}\ \bibnamefont {Waalewijn}},\
  }\bibfield  {title} {\bibinfo {title} {{Gaining (Mutual) Information about
  Quark/Gluon Discrimination}},\ }\href
  {https://doi.org/10.1007/JHEP11(2014)129} {\bibfield  {journal} {\bibinfo
  {journal} {JHEP}\ }\textbf {\bibinfo {volume} {11}},\ \bibinfo {pages}
  {129}},\ \Eprint {https://arxiv.org/abs/1408.3122} {arXiv:1408.3122 [hep-ph]}
  \BibitemShut {NoStop}%
\bibitem [{\citenamefont {Carrara}\ and\ \citenamefont
  {Ernst}(2019)}]{carrara2019estimation}%
  \BibitemOpen
  \bibfield  {author} {\bibinfo {author} {\bibfnamefont {N.}~\bibnamefont
  {Carrara}}\ and\ \bibinfo {author} {\bibfnamefont {J.}~\bibnamefont
  {Ernst}},\ }\href@noop {} {\bibinfo {title} {On the estimation of mutual
  information}} (\bibinfo {year} {2019}),\ \Eprint
  {https://arxiv.org/abs/1910.00365} {arXiv:1910.00365 [physics.data-an]}
  \BibitemShut {NoStop}%
\bibitem [{\citenamefont {Chollet}(2017)}]{keras}%
  \BibitemOpen
  \bibfield  {author} {\bibinfo {author} {\bibfnamefont {F.}~\bibnamefont
  {Chollet}},\ }\href@noop {} {\bibinfo {title} {Keras}},\ \bibinfo
  {howpublished} {\url{https://github.com/fchollet/keras}} (\bibinfo {year}
  {2017})\BibitemShut {NoStop}%
\bibitem [{\citenamefont {Abadi}\ \emph {et~al.}(2016)\citenamefont {Abadi},
  \citenamefont {Barham}, \citenamefont {Chen}, \citenamefont {Chen},
  \citenamefont {Davis}, \citenamefont {Dean}, \citenamefont {Devin},
  \citenamefont {Ghemawat}, \citenamefont {Irving}, \citenamefont {Isard} \emph
  {et~al.}}]{tensorflow}%
  \BibitemOpen
  \bibfield  {author} {\bibinfo {author} {\bibfnamefont {M.}~\bibnamefont
  {Abadi}}, \bibinfo {author} {\bibfnamefont {P.}~\bibnamefont {Barham}},
  \bibinfo {author} {\bibfnamefont {J.}~\bibnamefont {Chen}}, \bibinfo {author}
  {\bibfnamefont {Z.}~\bibnamefont {Chen}}, \bibinfo {author} {\bibfnamefont
  {A.}~\bibnamefont {Davis}}, \bibinfo {author} {\bibfnamefont
  {J.}~\bibnamefont {Dean}}, \bibinfo {author} {\bibfnamefont {M.}~\bibnamefont
  {Devin}}, \bibinfo {author} {\bibfnamefont {S.}~\bibnamefont {Ghemawat}},
  \bibinfo {author} {\bibfnamefont {G.}~\bibnamefont {Irving}}, \bibinfo
  {author} {\bibfnamefont {M.}~\bibnamefont {Isard}}, \emph {et~al.},\
  }\bibfield  {title} {\bibinfo {title} {Tensorflow: A system for large-scale
  machine learning.},\ }in\ \href@noop {} {\emph {\bibinfo {booktitle}
  {OSDI}}},\ Vol.~\bibinfo {volume} {16}\ (\bibinfo {year} {2016})\ pp.\
  \bibinfo {pages} {265--283}\BibitemShut {NoStop}%
\bibitem [{\citenamefont {Kingma}\ and\ \citenamefont {Ba}(2014)}]{adam}%
  \BibitemOpen
  \bibfield  {author} {\bibinfo {author} {\bibfnamefont {D.}~\bibnamefont
  {Kingma}}\ and\ \bibinfo {author} {\bibfnamefont {J.}~\bibnamefont {Ba}},\
  }\bibfield  {title} {\bibinfo {title} {Adam: A method for stochastic
  optimization},\ }\href@noop {} {\  (\bibinfo {year} {2014})},\ \Eprint
  {https://arxiv.org/abs/1412.6980} {arXiv:1412.6980 [cs]} \BibitemShut
  {NoStop}%
\bibitem [{\citenamefont {Kasieczka}\ \emph
  {et~al.}(2019{\natexlab{a}})\citenamefont {Kasieczka}, \citenamefont
  {Nachman},\ and\ \citenamefont {Shih}}]{gregor_kasieczka_2019_2629073}%
  \BibitemOpen
  \bibfield  {author} {\bibinfo {author} {\bibfnamefont {G.}~\bibnamefont
  {Kasieczka}}, \bibinfo {author} {\bibfnamefont {B.}~\bibnamefont {Nachman}},\
  and\ \bibinfo {author} {\bibfnamefont {D.}~\bibnamefont {Shih}},\ }\bibfield
  {title} {\bibinfo {title} {{R\&D Dataset for LHC Olympics 2020 Anomaly
  Detection Challenge}},\ }\href {https://doi.org/10.5281/zenodo.2629073}
  {10.5281/zenodo.2629073} (\bibinfo {year} {2019}{\natexlab{a}}),\ \bibinfo
  {note} {https://doi.org/10.5281/zenodo.2629073}\BibitemShut {NoStop}%
\bibitem [{\citenamefont {Sj{\"o}strand}\ \emph {et~al.}(2006)\citenamefont
  {Sj{\"o}strand}, \citenamefont {Mrenna},\ and\ \citenamefont
  {Skands}}]{Sjostrand:2006za}%
  \BibitemOpen
  \bibfield  {author} {\bibinfo {author} {\bibfnamefont {T.}~\bibnamefont
  {Sj{\"o}strand}}, \bibinfo {author} {\bibfnamefont {S.}~\bibnamefont
  {Mrenna}},\ and\ \bibinfo {author} {\bibfnamefont {P.~Z.}\ \bibnamefont
  {Skands}},\ }\bibfield  {title} {\bibinfo {title} {{PYTHIA 6.4 Physics and
  Manual}},\ }\href {https://doi.org/10.1088/1126-6708/2006/05/026} {\bibfield
  {journal} {\bibinfo  {journal} {JHEP}\ }\textbf {\bibinfo {volume} {05}},\
  \bibinfo {pages} {026}},\ \Eprint {https://arxiv.org/abs/hep-ph/0603175}
  {arXiv:hep-ph/0603175 [hep-ph]} \BibitemShut {NoStop}%
%%CITATION = HEP-PH/0603175;%%
\bibitem [{\citenamefont {Sj{\"o}strand}\ \emph {et~al.}(2008)\citenamefont
  {Sj{\"o}strand}, \citenamefont {Mrenna},\ and\ \citenamefont
  {Skands}}]{Sjostrand:2007gs}%
  \BibitemOpen
  \bibfield  {author} {\bibinfo {author} {\bibfnamefont {T.}~\bibnamefont
  {Sj{\"o}strand}}, \bibinfo {author} {\bibfnamefont {S.}~\bibnamefont
  {Mrenna}},\ and\ \bibinfo {author} {\bibfnamefont {P.~Z.}\ \bibnamefont
  {Skands}},\ }\bibfield  {title} {\bibinfo {title} {{A Brief Introduction to
  PYTHIA 8.1}},\ }\href {https://doi.org/10.1016/j.cpc.2008.01.036} {\bibfield
  {journal} {\bibinfo  {journal} {Comput. Phys. Commun.}\ }\textbf {\bibinfo
  {volume} {178}},\ \bibinfo {pages} {852} (\bibinfo {year} {2008})},\ \Eprint
  {https://arxiv.org/abs/0710.3820} {arXiv:0710.3820 [hep-ph]} \BibitemShut
  {NoStop}%
%%CITATION = ARXIV:0710.3820;%%
\bibitem [{\citenamefont {de~Favereau}\ \emph {et~al.}(2014)\citenamefont
  {de~Favereau}, \citenamefont {Delaere}, \citenamefont {Demin}, \citenamefont
  {Giammanco}, \citenamefont {Lema{\^\i}tre}, \citenamefont {Mertens},\ and\
  \citenamefont {Selvaggi}}]{deFavereau:2013fsa}%
  \BibitemOpen
  \bibfield  {author} {\bibinfo {author} {\bibfnamefont {J.}~\bibnamefont
  {de~Favereau}}, \bibinfo {author} {\bibfnamefont {C.}~\bibnamefont
  {Delaere}}, \bibinfo {author} {\bibfnamefont {P.}~\bibnamefont {Demin}},
  \bibinfo {author} {\bibfnamefont {A.}~\bibnamefont {Giammanco}}, \bibinfo
  {author} {\bibfnamefont {V.}~\bibnamefont {Lema{\^\i}tre}}, \bibinfo {author}
  {\bibfnamefont {A.}~\bibnamefont {Mertens}},\ and\ \bibinfo {author}
  {\bibfnamefont {M.}~\bibnamefont {Selvaggi}} (\bibinfo {collaboration}
  {DELPHES 3}),\ }\bibfield  {title} {\bibinfo {title} {{DELPHES 3, A modular
  framework for fast simulation of a generic collider experiment}},\ }\href
  {https://doi.org/10.1007/JHEP02(2014)057} {\bibfield  {journal} {\bibinfo
  {journal} {JHEP}\ }\textbf {\bibinfo {volume} {02}},\ \bibinfo {pages}
  {057}},\ \Eprint {https://arxiv.org/abs/1307.6346} {arXiv:1307.6346 [hep-ex]}
  \BibitemShut {NoStop}%
%%CITATION = ARXIV:1307.6346;%%
\bibitem [{\citenamefont {Mertens}(2015)}]{Mertens:2015kba}%
  \BibitemOpen
  \bibfield  {author} {\bibinfo {author} {\bibfnamefont {A.}~\bibnamefont
  {Mertens}},\ }\bibfield  {title} {\bibinfo {title} {{New features in Delphes
  3}},\ }\bibfield  {booktitle} {\emph {\bibinfo {booktitle} {{Proceedings,
  16th International workshop on Advanced Computing and Analysis Techniques in
  physics (ACAT 14): Prague, Czech Republic, September 1-5, 2014}}},\ }\href
  {https://doi.org/10.1088/1742-6596/608/1/012045} {\bibfield  {journal}
  {\bibinfo  {journal} {J. Phys. Conf. Ser.}\ }\textbf {\bibinfo {volume}
  {608}},\ \bibinfo {pages} {012045} (\bibinfo {year} {2015})}\BibitemShut
  {NoStop}%
%%CITATION = 00462,608,012045;%%
\bibitem [{\citenamefont {Selvaggi}(2014)}]{Selvaggi:2014mya}%
  \BibitemOpen
  \bibfield  {author} {\bibinfo {author} {\bibfnamefont {M.}~\bibnamefont
  {Selvaggi}},\ }\bibfield  {title} {\bibinfo {title} {{DELPHES 3: A modular
  framework for fast-simulation of generic collider experiments}},\ }\bibfield
  {booktitle} {\emph {\bibinfo {booktitle} {{Proceedings, 15th International
  Workshop on Advanced Computing and Analysis Techniques in Physics Research
  (ACAT 2013): Beijing, China, May 16-21, 2013}}},\ }\href
  {https://doi.org/10.1088/1742-6596/523/1/012033} {\bibfield  {journal}
  {\bibinfo  {journal} {J. Phys. Conf. Ser.}\ }\textbf {\bibinfo {volume}
  {523}},\ \bibinfo {pages} {012033} (\bibinfo {year} {2014})}\BibitemShut
  {NoStop}%
%%CITATION = 00462,523,012033;%%
\bibitem [{\citenamefont {Cacciari}\ \emph {et~al.}(2012)\citenamefont
  {Cacciari}, \citenamefont {Salam},\ and\ \citenamefont
  {Soyez}}]{Cacciari:2011ma}%
  \BibitemOpen
  \bibfield  {author} {\bibinfo {author} {\bibfnamefont {M.}~\bibnamefont
  {Cacciari}}, \bibinfo {author} {\bibfnamefont {G.~P.}\ \bibnamefont
  {Salam}},\ and\ \bibinfo {author} {\bibfnamefont {G.}~\bibnamefont {Soyez}},\
  }\bibfield  {title} {\bibinfo {title} {{FastJet User Manual}},\ }\href
  {https://doi.org/10.1140/epjc/s10052-012-1896-2} {\bibfield  {journal}
  {\bibinfo  {journal} {Eur. Phys. J.}\ }\textbf {\bibinfo {volume} {C72}},\
  \bibinfo {pages} {1896} (\bibinfo {year} {2012})},\ \Eprint
  {https://arxiv.org/abs/1111.6097} {arXiv:1111.6097 [hep-ph]} \BibitemShut
  {NoStop}%
%%CITATION = ARXIV:1111.6097;%%
\bibitem [{\citenamefont {Cacciari}\ and\ \citenamefont
  {Salam}(2006)}]{Cacciari:2005hq}%
  \BibitemOpen
  \bibfield  {author} {\bibinfo {author} {\bibfnamefont {M.}~\bibnamefont
  {Cacciari}}\ and\ \bibinfo {author} {\bibfnamefont {G.~P.}\ \bibnamefont
  {Salam}},\ }\bibfield  {title} {\bibinfo {title} {{Dispelling the $N^{3}$
  myth for the $k_t$ jet-finder}},\ }\href
  {https://doi.org/10.1016/j.physletb.2006.08.037} {\bibfield  {journal}
  {\bibinfo  {journal} {Phys. Lett.}\ }\textbf {\bibinfo {volume} {B641}},\
  \bibinfo {pages} {57} (\bibinfo {year} {2006})},\ \Eprint
  {https://arxiv.org/abs/hep-ph/0512210} {arXiv:hep-ph/0512210 [hep-ph]}
  \BibitemShut {NoStop}%
%%CITATION = HEP-PH/0512210;%%
\bibitem [{\citenamefont {Cacciari}\ \emph {et~al.}(2008)\citenamefont
  {Cacciari}, \citenamefont {Salam},\ and\ \citenamefont
  {Soyez}}]{Cacciari:2008gp}%
  \BibitemOpen
  \bibfield  {author} {\bibinfo {author} {\bibfnamefont {M.}~\bibnamefont
  {Cacciari}}, \bibinfo {author} {\bibfnamefont {G.~P.}\ \bibnamefont
  {Salam}},\ and\ \bibinfo {author} {\bibfnamefont {G.}~\bibnamefont {Soyez}},\
  }\bibfield  {title} {\bibinfo {title} {{The anti-$k_t$ jet clustering
  algorithm}},\ }\href {https://doi.org/10.1088/1126-6708/2008/04/063}
  {\bibfield  {journal} {\bibinfo  {journal} {JHEP}\ }\textbf {\bibinfo
  {volume} {04}},\ \bibinfo {pages} {063}},\ \Eprint
  {https://arxiv.org/abs/0802.1189} {arXiv:0802.1189 [hep-ph]} \BibitemShut
  {NoStop}%
%%CITATION = ARXIV:0802.1189;%%
\bibitem [{\citenamefont {Thaler}\ and\ \citenamefont
  {Van~Tilburg}(2012)}]{Thaler:2011gf}%
  \BibitemOpen
  \bibfield  {author} {\bibinfo {author} {\bibfnamefont {J.}~\bibnamefont
  {Thaler}}\ and\ \bibinfo {author} {\bibfnamefont {K.}~\bibnamefont
  {Van~Tilburg}},\ }\bibfield  {title} {\bibinfo {title} {{Maximizing Boosted
  Top Identification by Minimizing N-subjettiness}},\ }\href
  {https://doi.org/10.1007/JHEP02(2012)093} {\bibfield  {journal} {\bibinfo
  {journal} {JHEP}\ }\textbf {\bibinfo {volume} {02}},\ \bibinfo {pages}
  {093}},\ \Eprint {https://arxiv.org/abs/1108.2701} {arXiv:1108.2701 [hep-ph]}
  \BibitemShut {NoStop}%
%%CITATION = ARXIV:1108.2701;%%
\bibitem [{\citenamefont {Thaler}\ and\ \citenamefont
  {Van~Tilburg}(2011)}]{Thaler:2010tr}%
  \BibitemOpen
  \bibfield  {author} {\bibinfo {author} {\bibfnamefont {J.}~\bibnamefont
  {Thaler}}\ and\ \bibinfo {author} {\bibfnamefont {K.}~\bibnamefont
  {Van~Tilburg}},\ }\bibfield  {title} {\bibinfo {title} {{Identifying Boosted
  Objects with N-subjettiness}},\ }\href
  {https://doi.org/10.1007/JHEP03(2011)015} {\bibfield  {journal} {\bibinfo
  {journal} {JHEP}\ }\textbf {\bibinfo {volume} {03}},\ \bibinfo {pages}
  {015}},\ \Eprint {https://arxiv.org/abs/1011.2268} {arXiv:1011.2268 [hep-ph]}
  \BibitemShut {NoStop}%
%%CITATION = ARXIV:1011.2268;%%
\bibitem [{\citenamefont {Zaheer}\ \emph {et~al.}(2017)\citenamefont {Zaheer},
  \citenamefont {Kottur}, \citenamefont {Ravanbhakhsh}, \citenamefont
  {P\'{o}czos}, \citenamefont {Salakhutdinov},\ and\ \citenamefont
  {Smola}}]{10.5555/3294996.3295098}%
  \BibitemOpen
  \bibfield  {author} {\bibinfo {author} {\bibfnamefont {M.}~\bibnamefont
  {Zaheer}}, \bibinfo {author} {\bibfnamefont {S.}~\bibnamefont {Kottur}},
  \bibinfo {author} {\bibfnamefont {S.}~\bibnamefont {Ravanbhakhsh}}, \bibinfo
  {author} {\bibfnamefont {B.}~\bibnamefont {P\'{o}czos}}, \bibinfo {author}
  {\bibfnamefont {R.}~\bibnamefont {Salakhutdinov}},\ and\ \bibinfo {author}
  {\bibfnamefont {A.~J.}\ \bibnamefont {Smola}},\ }\bibfield  {title} {\bibinfo
  {title} {Deep sets},\ }in\ \href@noop {} {\emph {\bibinfo {booktitle}
  {Proceedings of the 31st International Conference on Neural Information
  Processing Systems}}},\ \bibinfo {series and number} {NIPS'17}\ (\bibinfo
  {publisher} {Curran Associates Inc.},\ \bibinfo {address} {Red Hook, NY,
  USA},\ \bibinfo {year} {2017})\ p.\ \bibinfo {pages}
  {3394–3404}\BibitemShut {NoStop}%
\bibitem [{\citenamefont {Komiske}\ \emph {et~al.}(2019)\citenamefont
  {Komiske}, \citenamefont {Metodiev},\ and\ \citenamefont
  {Thaler}}]{Komiske:2018cqr}%
  \BibitemOpen
  \bibfield  {author} {\bibinfo {author} {\bibfnamefont {P.~T.}\ \bibnamefont
  {Komiske}}, \bibinfo {author} {\bibfnamefont {E.~M.}\ \bibnamefont
  {Metodiev}},\ and\ \bibinfo {author} {\bibfnamefont {J.}~\bibnamefont
  {Thaler}},\ }\bibfield  {title} {\bibinfo {title} {{Energy Flow Networks:
  Deep Sets for Particle Jets}},\ }\href
  {https://doi.org/10.1007/JHEP01(2019)121} {\bibfield  {journal} {\bibinfo
  {journal} {JHEP}\ }\textbf {\bibinfo {volume} {01}},\ \bibinfo {pages}
  {121}},\ \Eprint {https://arxiv.org/abs/1810.05165} {arXiv:1810.05165
  [hep-ph]} \BibitemShut {NoStop}%
\bibitem [{\citenamefont {Scarselli}\ \emph {et~al.}(2009)\citenamefont
  {Scarselli}, \citenamefont {Gori}, \citenamefont {Tsoi}, \citenamefont
  {Hagenbuchner},\ and\ \citenamefont {Monfardini}}]{10.1109/TNN.2008.2005605}%
  \BibitemOpen
  \bibfield  {author} {\bibinfo {author} {\bibfnamefont {F.}~\bibnamefont
  {Scarselli}}, \bibinfo {author} {\bibfnamefont {M.}~\bibnamefont {Gori}},
  \bibinfo {author} {\bibfnamefont {A.~C.}\ \bibnamefont {Tsoi}}, \bibinfo
  {author} {\bibfnamefont {M.}~\bibnamefont {Hagenbuchner}},\ and\ \bibinfo
  {author} {\bibfnamefont {G.}~\bibnamefont {Monfardini}},\ }\bibfield  {title}
  {\bibinfo {title} {The graph neural network model},\ }\href
  {https://doi.org/10.1109/TNN.2008.2005605} {\bibfield  {journal} {\bibinfo
  {journal} {Trans. Neur. Netw.}\ }\textbf {\bibinfo {volume} {20}},\ \bibinfo
  {pages} {61–80} (\bibinfo {year} {2009})}\BibitemShut {NoStop}%
\bibitem [{\citenamefont {Shlomi}\ \emph {et~al.}(2020)\citenamefont {Shlomi},
  \citenamefont {Battaglia},\ and\ \citenamefont {Vlimant}}]{Shlomi:2020gdn}%
  \BibitemOpen
  \bibfield  {author} {\bibinfo {author} {\bibfnamefont {J.}~\bibnamefont
  {Shlomi}}, \bibinfo {author} {\bibfnamefont {P.}~\bibnamefont {Battaglia}},\
  and\ \bibinfo {author} {\bibfnamefont {J.-R.}\ \bibnamefont {Vlimant}},\
  }\bibfield  {title} {\bibinfo {title} {{Graph Neural Networks in Particle
  Physics}}\ }\href {https://doi.org/10.1088/2632-2153/abbf9a}
  {10.1088/2632-2153/abbf9a} (\bibinfo {year} {2020}),\ \Eprint
  {https://arxiv.org/abs/2007.13681} {arXiv:2007.13681 [hep-ex]} \BibitemShut
  {NoStop}%
\bibitem [{\citenamefont {Aaboud}\ \emph {et~al.}(2018)\citenamefont {Aaboud}
  \emph {et~al.}}]{Aaboud:2018xwy}%
  \BibitemOpen
  \bibfield  {author} {\bibinfo {author} {\bibfnamefont {M.}~\bibnamefont
  {Aaboud}} \emph {et~al.} (\bibinfo {collaboration} {ATLAS}),\ }\bibfield
  {title} {\bibinfo {title} {{Measurements of b-jet tagging efficiency with the
  ATLAS detector using $ t\overline{t} $ events at $ \sqrt{s}=13 $ TeV}},\
  }\href {https://doi.org/10.1007/JHEP08(2018)089} {\bibfield  {journal}
  {\bibinfo  {journal} {JHEP}\ }\textbf {\bibinfo {volume} {08}},\ \bibinfo
  {pages} {089}},\ \Eprint {https://arxiv.org/abs/1805.01845} {arXiv:1805.01845
  [hep-ex]} \BibitemShut {NoStop}%
\bibitem [{\citenamefont {Chatrchyan}\ \emph {et~al.}(2013)\citenamefont
  {Chatrchyan} \emph {et~al.}}]{Chatrchyan:2012jua}%
  \BibitemOpen
  \bibfield  {author} {\bibinfo {author} {\bibfnamefont {S.}~\bibnamefont
  {Chatrchyan}} \emph {et~al.} (\bibinfo {collaboration} {CMS}),\ }\bibfield
  {title} {\bibinfo {title} {{Identification of b-Quark Jets with the CMS
  Experiment}},\ }\href {https://doi.org/10.1088/1748-0221/8/04/P04013}
  {\bibfield  {journal} {\bibinfo  {journal} {JINST}\ }\textbf {\bibinfo
  {volume} {8}},\ \bibinfo {pages} {P04013}},\ \Eprint
  {https://arxiv.org/abs/1211.4462} {arXiv:1211.4462 [hep-ex]} \BibitemShut
  {NoStop}%
\bibitem [{\citenamefont {Andersson}\ \emph {et~al.}(1989)\citenamefont
  {Andersson}, \citenamefont {Gustafson}, \citenamefont {Lonnblad},\ and\
  \citenamefont {Pettersson}}]{Andersson:1988gp}%
  \BibitemOpen
  \bibfield  {author} {\bibinfo {author} {\bibfnamefont {B.}~\bibnamefont
  {Andersson}}, \bibinfo {author} {\bibfnamefont {G.}~\bibnamefont
  {Gustafson}}, \bibinfo {author} {\bibfnamefont {L.}~\bibnamefont
  {Lonnblad}},\ and\ \bibinfo {author} {\bibfnamefont {U.}~\bibnamefont
  {Pettersson}},\ }\bibfield  {title} {\bibinfo {title} {{Coherence Effects in
  Deep Inelastic Scattering}},\ }\href {https://doi.org/10.1007/BF01550942}
  {\bibfield  {journal} {\bibinfo  {journal} {Z. Phys. C}\ }\textbf {\bibinfo
  {volume} {43}},\ \bibinfo {pages} {625} (\bibinfo {year} {1989})}\BibitemShut
  {NoStop}%
\bibitem [{\citenamefont {Dreyer}\ \emph {et~al.}(2018)\citenamefont {Dreyer},
  \citenamefont {Salam},\ and\ \citenamefont {Soyez}}]{Dreyer:2018nbf}%
  \BibitemOpen
  \bibfield  {author} {\bibinfo {author} {\bibfnamefont {F.~A.}\ \bibnamefont
  {Dreyer}}, \bibinfo {author} {\bibfnamefont {G.~P.}\ \bibnamefont {Salam}},\
  and\ \bibinfo {author} {\bibfnamefont {G.}~\bibnamefont {Soyez}},\ }\bibfield
   {title} {\bibinfo {title} {{The Lund Jet Plane}},\ }\href
  {https://doi.org/10.1007/JHEP12(2018)064} {\bibfield  {journal} {\bibinfo
  {journal} {JHEP}\ }\textbf {\bibinfo {volume} {12}},\ \bibinfo {pages}
  {064}},\ \Eprint {https://arxiv.org/abs/1807.04758} {arXiv:1807.04758
  [hep-ph]} \BibitemShut {NoStop}%
\bibitem [{\citenamefont {Andreassen}\ \emph
  {et~al.}(2020{\natexlab{c}})\citenamefont {Andreassen}, \citenamefont {Hsu},
  \citenamefont {Nachman}, \citenamefont {Suaysom},\ and\ \citenamefont
  {Suresh}}]{anders_andreassen_2020_4067673}%
  \BibitemOpen
  \bibfield  {author} {\bibinfo {author} {\bibfnamefont {A.}~\bibnamefont
  {Andreassen}}, \bibinfo {author} {\bibfnamefont {S.-C.}\ \bibnamefont {Hsu}},
  \bibinfo {author} {\bibfnamefont {B.}~\bibnamefont {Nachman}}, \bibinfo
  {author} {\bibfnamefont {N.}~\bibnamefont {Suaysom}},\ and\ \bibinfo {author}
  {\bibfnamefont {A.}~\bibnamefont {Suresh}},\ }\bibfield  {title} {\bibinfo
  {title} {Srgn: Pythia + delphes $pp \rightarrow t\bar{t}$},\ }\href
  {https://doi.org/10.5281/zenodo.4067673} {10.5281/zenodo.4067673} (\bibinfo
  {year} {2020}{\natexlab{c}})\BibitemShut {NoStop}%
\bibitem [{\citenamefont {Kasieczka}\ \emph
  {et~al.}(2019{\natexlab{b}})\citenamefont {Kasieczka}, \citenamefont
  {Nachman},\ and\ \citenamefont {Shih}}]{kasieczka_gregor_2019_4287846}%
  \BibitemOpen
  \bibfield  {author} {\bibinfo {author} {\bibfnamefont {G.}~\bibnamefont
  {Kasieczka}}, \bibinfo {author} {\bibfnamefont {B.}~\bibnamefont {Nachman}},\
  and\ \bibinfo {author} {\bibfnamefont {D.}~\bibnamefont {Shih}},\ }\bibfield
  {title} {\bibinfo {title} {{Official Datasets for LHC Olympics 2020 Anomaly
  Detection Challenge}},\ }\href {https://doi.org/10.5281/zenodo.4287846}
  {10.5281/zenodo.4287846} (\bibinfo {year} {2019}{\natexlab{b}})\BibitemShut
  {NoStop}%
\end{thebibliography}%

\end{document}